\documentclass[preprint]{elsarticle}

\usepackage{natbib}
\usepackage[labelfont=bf]{caption}
\usepackage{subcaption}
\usepackage[utf8]{inputenc}  
\usepackage[T1]{fontenc}     
\usepackage{lmodern}         
\usepackage[scaled]{helvet} 
\usepackage[english]{babel}  
\usepackage{graphicx}        
\usepackage{comment}         
\biboptions{sort&compress}   
\journal{arXiv}

\usepackage{amsmath,amsfonts,amssymb}

\usepackage{xcolor}
\usepackage{todonotes}

\graphicspath{{fig/}}
\begin{document}
\begin{frontmatter}

\title{A simplified pore-scale model for slow drainage including film flow effects}

\author{Paula Reis\fnref{label1}%
}
\author{Marcel Moura\fnref{label1}%
}
\author{Gaute Linga\fnref{label1}%
}
\author{Per Arne Rikvold\fnref{label1}%
}
\author{Eirik Grude Flekkøy\fnref{label1}%
}
\author{Knut Jørgen Måloy\fnref{label1}%
}
\fntext[label1]{PoreLab, The NJORD Centre, Department of Physics, University of Oslo, Norway%
}

\begin{abstract}
Film flow through networks of corners and capillary bridges can establish connections between seemingly isolated clusters during drainage in porous media. Coupled with drainage through the bulk of pores and throats, the flow through these networks constitutes a secondary drainage mechanism that can significantly affect fluid configurations and residual saturations. In order to investigate the prevalence of this drainage mechanism, we propose a quasi-static pore-network model based on modifying the trapped-cluster-identification algorithm in an invasion-percolation model. With the modification, wetting-phase connectivity could be provided by direct successions of pores and throats, represented by sites and bonds, as well as by chains of interconnected capillary bridges. The advancement of the fluid interface in the porous matrix was determined by the bonds' invasion thresholds, calculated based on local capillary pressure values that could be perturbed to accommodate gravitational effects. With the proposed model, experimentally verified phenomena related to slow drainage in granular porous media were reproduced, showing good qualitative agreement. 
\end{abstract}
\end{frontmatter}

\section{Introduction} 
\label{sec: Intro}

Pore-scale modeling of fluid-fluid displacements has been an active research topic for the past few decades, with applications spanning multiple fields \cite{blunt2013pore, bultreys2016imaging, Golparvar2018,zhao2019comprehensive, Chen2022pore}. In particular, extensive efforts have been directed toward the development of models related to $CO_2$ storage in subsurface geological reservoirs \cite{gao2017reactive,tahmasebi2017pore, Basirat2017pore,masoudi2021pore,payton2022pore}, oil recovery methods \cite{zhao2010pore,kallel2017pore,zhu2017pore,raeini2018generalized,su2018advances}, drying \cite{prat2002recent,surasani2008influence,panda2022pore,fei2022pore,zhao2022pore}, and water management in fuel cells \cite{mukherjee2011pore,molaeimanesh2014three,zhu2021pore,fu2022pore,guo2022pore}. The quality of predictions from such models relies on appropriately identifying and representing the primary forces driving the flow, as well as the fundamental geometrical aspects of the porous space, which can vary greatly according to the system of interest. In this work, we narrow our focus to effectively modeling the impact of film flow during slow drainage in granular porous media. With this aim, first, we identify the main flow mechanisms exhibited during slow drainage and how their prevalence can be affected by the particular geometrical features of pores within granular materials. Then, we propose an approach to incorporate these mechanisms' essential effects on a computationally cost-effective pore-scale model.  

Slow drainage flows are characterized by the displacement of a wetting fluid by a non-wetting fluid under no significant influence of viscous or inertial forces. During such flows, two main mechanisms are commonly identified, namely, a piston-like invasion of the non-wetting phase, followed by the removal of the wetting phase through networks of contiguous corners and/or capillary bridges \cite{hoogland2016drainage,hoogland2016drainageb,moura2019connectivity}. The piston-like displacement is characterized by the advancement of the fluid interface through the bulk of pore bodies and throats in fast discrete events known as Haines jumps \cite{armstrong2013interfacial,moebius2014pore,jorgen2021burst,mansouri2023interfacial}. As this mechanism unfolds, the competition between local capillary pressure ($P_c$) and pore entry pressure thresholds ($P_t$), shown in eq.\ref{eq:Pc_vs_Pt}, dictates the advancement of the invading fluid front. This mechanism is generally responsible for most of the displacement of the defending fluid and will be referred to as the primary drainage mechanism. 

\begin{equation}
    P_c>P_t=\gamma\cos{\theta}\left(\frac{1}{r_1}+\frac{1}{r_2} \right)
    \label{eq:Pc_vs_Pt}
\end{equation}

\noindent where $\gamma$ is the interfacial tension between the fluids, $\theta$ is the contact angle between the fluids and the porous material, and $r_1$ and $r_2$ are the two main interface curvature radii.

In the regions of the porous medium swept by the invading fluid front, the primary mechanism is succeeded by the displacement of the wetting fluid through film flow \cite{zhou1997hydrocarbon,tuller2001hydraulic}. At this stage, the wetting fluid can be found in the medium either in the form of isolated clusters -- structures containing one or several defending-phase-filled pores surrounded by the invading phase -- or at angular porous features passed by the interface, where $P_c$ was insufficient to mobilize the defending phase. In this unsaturated zone, the secondary drainage mechanism takes place, provided that contiguous sets of corners and capillary bridges constitute paths for wetting fluid flow \cite{flekkoy2002flow,ryazanov2009two,han2009deviation,moura2019connectivity}. Given the comparatively lower permeability of the corner/bridge flow paths, the secondary mechanism is characterized by distinctive longer time scales for drainage than the primary mechanism\cite{hoogland2016drainage,moura2019connectivity}. 

The efficiency of both drainage mechanisms is inherently tied to geometric aspects of the porous space \cite{romano2011strong,xu2014effect,vahid2022experimental}. During piston-like displacement, the pore's size distribution and spatial arrangement define the pore-invasion order, delineating the fraction of the porous medium bypassed by the front, i.e., the amount of wetting fluid retained in clusters. As for the secondary mechanism, the shape of the porous space -- along with wettability -- is the key element dictating the hydraulic conductance and connectivity of the wetting-phase film flow paths established in the medium. In the following passage, a brief review of wetting phase connectivity in granular materials is presented.

Within unsaturated granular porous media, the wetting liquid not pertaining clusters is mostly retained in the form of capillary bridges and rings \cite{Herminghaus_2005, Rieser_2015, cejas2018effect, chen2017control, chen_2018, moura2019connectivity} enclosing grain-grain contact areas. Despite their minor contribution to the wetting-phase saturation, these structures can have a significant impact on the transport properties, given that the conditions for their interconnectivity are met. In a study related to drying in granular media, \citet{cejas2018effect} investigated the role of grain packing in the stability of film networks able to transport water from the bulk of the porous medium to its surface. Their results pointed to the existence of a packing threshold in loosely packed systems where the liquid connections break and water removal from the porous medium is undermined. In a similar set of studies, \citet{chen2017control}\cite{chen_2018} investigated the effect of geometry in the formation of elongated liquid films using transparent quasi-2D micromodels of cylinders arranged between two planes. According to the spatial disposition of the cylinders, the authors could observe the formation of films consisting of chains of capillary bridges providing hydraulic connectivity between internal liquid clusters and the outer rim of the micromodels. Quasi-2D models were also adopted by \citet{moura2019connectivity} to evaluate the impact of film flow on slow drainage under varying degrees of gravitational force influence. Their experiments using modified Hele-Shaw cells filled with a single layer of monodispersed glass beads allowed the analysis of several phenomena related to film flow, such as its relative impact on the residual saturation, the size and spatial distributions of capillary bridges, and the establishment of a film-flow active zone trailing the invasion front. Interestingly, the results indicated the drainage of seemingly isolated wetting-fluid clusters through dynamic networks of capillary bridges non-coincident with the pore network invaded during piston-like displacement. 

Although significant experimental evidence points to the critical role of capillary-bridge-chain flow paths during drainage in porous media, few computational models represent their effect on the flow explicitly. A pioneering model in this direction was presented by \citet{vorhauer2015drying}, with the intent of predicting isothermal drying rates in porous media. Based on experimental observations of drying in etched-silicon micromodels, the authors proposed a 2D pore-network model (PNM) that explicitly takes into account the role of capillary bridges and rings in wetting-fluid-cluster connectivity. With this new modeling approach, better agreement with experimental results was obtained, when compared with classic PNMs that represent film-flow paths as interconnected corners of angular-shaped pores. An extension of this work was later presented by \citet{kharaghani2021three}, by changing the network topology from 2D regular square lattices to 3D regular cubic lattices. Drying rates predicted by the 3D PNM showed good quantitative agreement with experiments using monodisperse-spherical-glass-bead packs initially filled with distilled water. Another model taking into account the role of capillary bridges during drying in porous media was developed by \citet{chen_2018}, accompanying their experiments on quasi-2D porous media formed by cylinders arranged between two parallel planes. In their model, liquid films consisted of successions of capillary bridges connecting neighboring cylinders, within which flow dynamics were shaped by the effects of viscous and capillary forces. Using experimental drying rates as model inputs, their numerical results were able to qualitatively predict invasion patterns and the extent of films as the water evaporated from the porous space.

In the present work, we incorporate the capillary-bridge role in wetting-phase connectivity successfully explored in drying models on an invasion-percolation model for slow drainage under the effects of capillary and gravitational forces. Driven by the experimental findings presented by \citet{moura2019connectivity}, we use the proposed PNM to investigate  the capillary bridge size and spatial distributions, the contribution of film flow to residual saturations, and the existence of film flow active zones on quasi-2D porous media. 

\section{Methodology} 

The proposed model is based on the invasion-percolation (IP) method with trapping \cite{wilkinson1983invasion}. Designed to represent immiscible fluid-fluid displacements in porous media, this method involves the description of pores and throats as lattices of sites and bonds, to each of which an invasion threshold is assigned. Flow in the idealized porous lattices is then predicted based on the movement of the fluid interface through the least resistivity path, dictated by the bonds' invasion thresholds in the case of drainage, and the sites' invasion thresholds in the case of imbibition \cite{wilkinson1983invasion}. This modeling approach has been widely adopted in the past decades \cite{blunt2001flow} due to its ability to sensibly predict porous media flows in the capillary-dominated regime associated with low computational effort.

In the direction of investigating fundamental aspects of film flow through corners and capillary bridges in granular porous media, our modified IP model was specially tailored to reproduce slow drainage in Hele-Shaw cells filled with a single layer of monodispersed spherical beads. This choice of quasi-2D model porous media allowed us to clearly identify drainage events related to film flow and generate results that could be readily compared with available experimental data \cite{moura2019connectivity}. 

\begin{figure}[h!]
  \centering
  \includegraphics[width=0.75\textwidth]{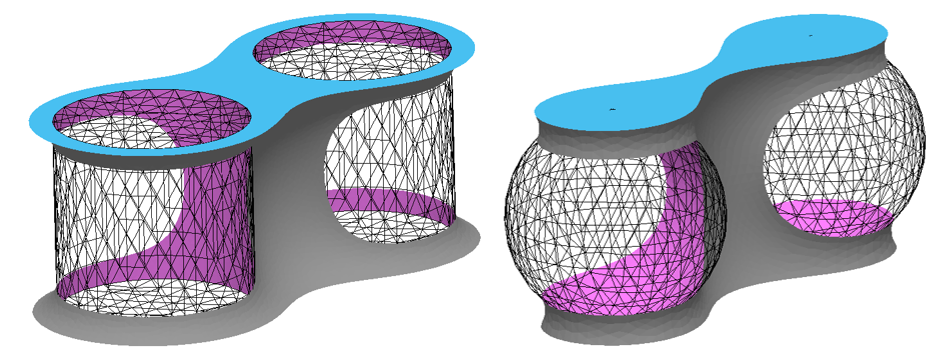}
  \caption{Examples of wetting-fluid configurations around pairs of cylinders and spheres contained between parallel planes, obtained with Surface Evolver \cite{brakke1992surface}. The blue, magenta, and gray surfaces represent the portions of fluid in contact with the planes, the structures representing grains, and the non-wetting fluid, respectively. }
  \label{fig:Rings_Grains}
\end{figure}

Insights related to fluid-fluid displacements obtained via visualization of flow in transparent quasi-2D models can be generalized to more complex media, by acknowledging how inherently 3D features of such models affect the observed flow phenomena. For the study of flow through films formed by corners and capillary bridges, an important geometric feature of quasi-2D models is the intersection of the structures representing the porous media grains, e.g. cylinders \cite{zhao2016wettability,chen2017control,maaloy2021burst,vincent2022stable} or spheres \cite{meheust2002interface,lovoll2005competition,moura2015impact,moura2019connectivity}, with the planes containing them. Following the piston-like displacement of the interface, fluid collected in these regions can play a fundamental role in wetting-phase continuity by linking capillary bridges formed between pairs of grains, as indicated in Figure \ref{fig:Rings_Grains}. 
Several recent pore-scale models incorporate aspects of these fluid configurations stemming from 3D structures in 2D lattices representing the porous space \cite{chen_2018,primkulov2018quasistatic,primkulov2021wettability,primkulov2022avalanches} and we adopt a similar approach.

In the next sections, we introduce the network representation of the porous space, the criteria for fluid interface advancement, and the role of capillary-bridge flow paths in fluid connectivity.

\subsection{Porous-media representation}
\label{sec:PM_rep}

\begin{figure}[h!]
  \centering
  \includegraphics[width=1\textwidth]{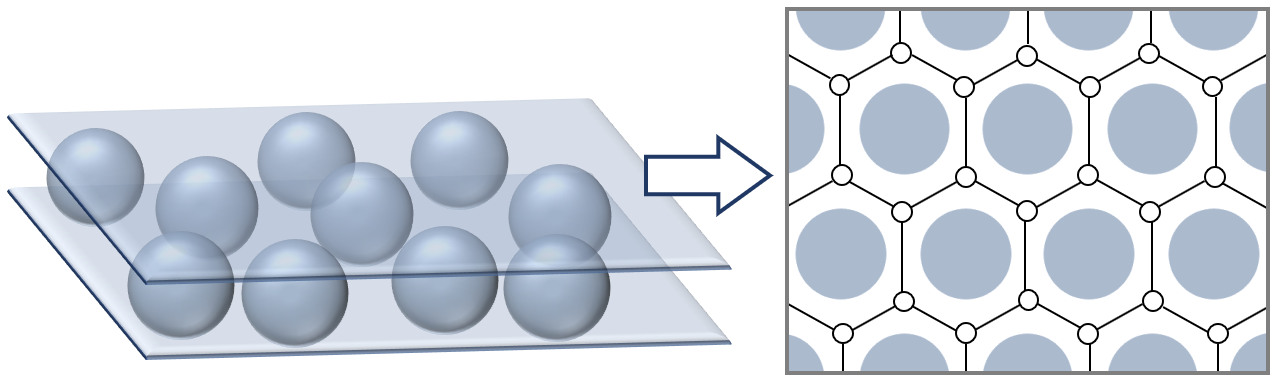}
  \caption{Sketch of the modified Hele-Shaw porous matrix used in \cite{moura2019connectivity} and its representation by a regular honeycomb lattice of nodes (open circles) and edges (solid lines).}
  \label{fig:Hele-Shaw_lattice}
\end{figure}

The description of porous media as networks of nodes (pore bodies) connected by edges (pore throats) is a common approach in the development of computationally-efficient models for multiphase flow \cite{blunt2001flow,joekar2012analysis}. In this study, we propose a 2D regular lattice representation of the quasi-2D porous media designed for the experimental investigation of film flow effects during drainage performed by \citet{moura2019connectivity}. Their porous-media models were created by randomly placing and securing spherical glass beads with a diameter of $1 mm$ between two parallel transparent plates. The resulting porous matrices could be divided into pore bodies and pore throats located at the vertices and edges of the Voronoi diagrams created using the center of the beads as seeds. Considering dense homogeneous packings of spherical beads, these porous matrices could be approximated as regular honeycomb lattices, as indicated in Figure \ref{fig:Hele-Shaw_lattice}. 

In the lattice representation, the edges portray throats with both length and height equal to the diameter of the beads, and with width related to the distance between the beads adjacent to them. The choice of width values assigned to the edges followed the probability density function of beads' separation found in the experimental models. We did not address the geometrical aspects of the lattice nodes and assumed that pores do not exert significant resistance to the drainage flows under investigation.

\subsection{Modified IP Algorithm}
\label{sec:IP_alg}

The proposed model for slow drainage can be described as an invasion-percolation model with trapping, in which the defending cluster labeling algorithm was modified to allow connectivity via capillary bridges. Its implementation exploits some basic concepts of graph theory, which will be briefly addressed next. A graph is composed of a set of nodes and a set of edges. An edge connects a pair of nodes and can be directed, if the pair of nodes is ordered, or undirected, otherwise. Pairs of nodes connected by an edge are called adjacent, and the edge is incident on the pair of nodes it connects. The number of edges incident on a node determines the node's degree. A sequence of adjacent nodes in a graph defines a path. A group of nodes is connected if there exists a path between every two nodes, and is referred to as a connected component of a graph. 

By using these concepts, the model's main idea is to create an undirected graph in which the connected components represent the wetting fluid occupation of the porous medium. As the drainage simulations start with a fully saturated porous matrix, this graph's topology initially matches the honeycomb lattice presented in Section \ref{sec:PM_rep}. The nodes corresponding to the medium's source of non-wetting fluid and sink of wetting fluid are identified and labeled as the inlet and outlet nodes, respectively. Weights are assigned to the graph's edges, corresponding to the capillary pressure threshold for the invasion of the pore throats ($P_t$). As the model takes into account the effects of capillary and gravitational forces, the calculation of $P_t$ is given by eq. \ref{eq:Pt_model}

 \begin{equation}
    P_t=2\gamma\cos{\theta}\left(\frac{1}{t_w}+\frac{1}{t_h} \right) -\Delta\rho g h 
    \label{eq:Pt_model}
\end{equation}

\noindent where $t_w$ and $t_h$ are the throat's width and height, respectively, $\Delta\rho$ is the density difference between wetting and non-wetting fluids, $g$ is the gravitational acceleration and $h$ is the elevation difference between the throat and the wetting-phase sink.

\begin{figure}[ht]
  \centering
  \includegraphics[width=1\textwidth]{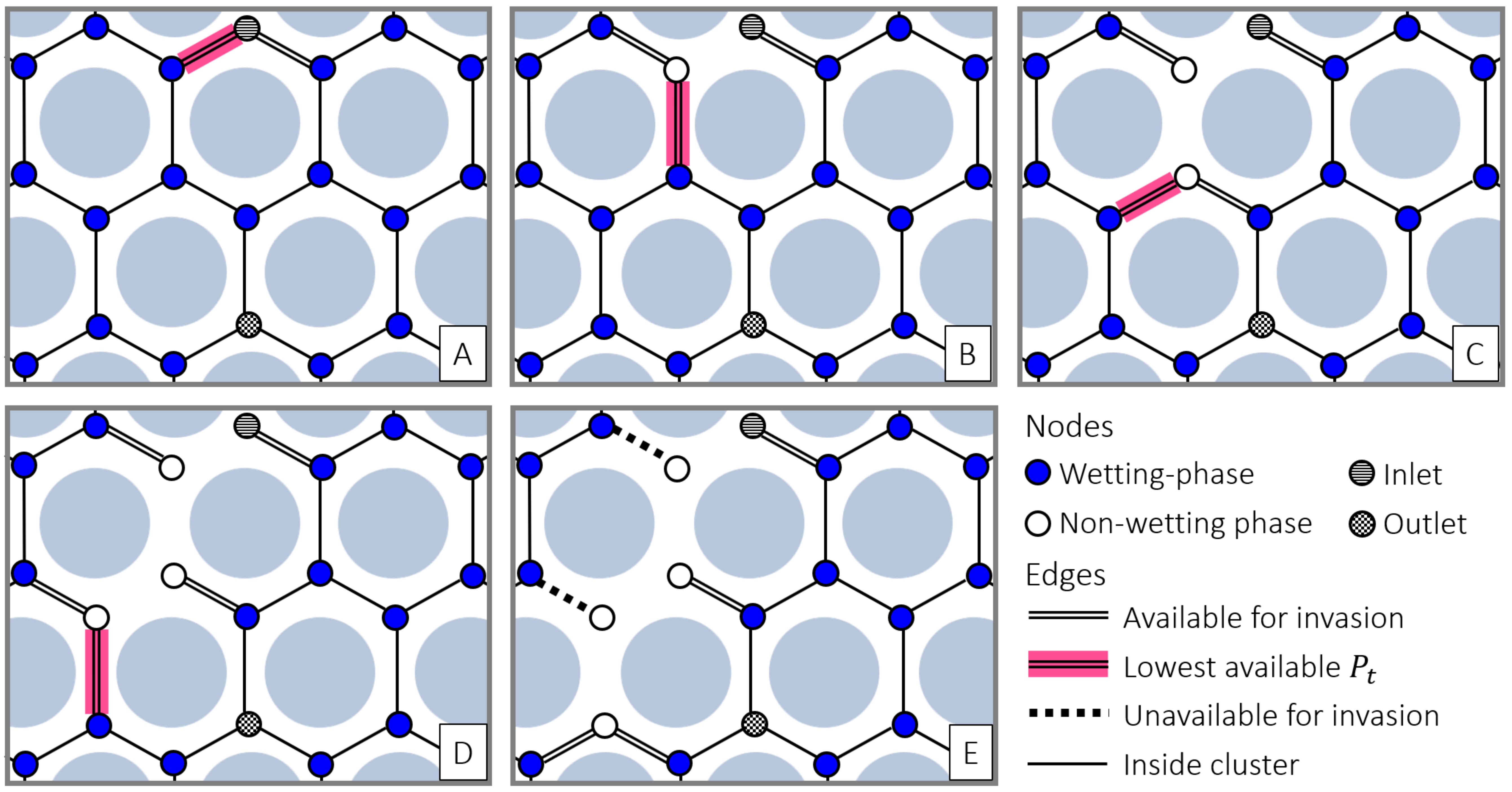}
  \caption{Representation of the modified IP model algorithm for drainage}
  \label{fig:Inv_Alg}
\end{figure}

As with IP models, our drainage simulation unfolds in discrete steps, in which the invasion of the available bond with the lowest $P_t$ takes place. This invasion is expressed by deleting the available edge with the lowest weight from the graph representing the wetting fluid occupation of the porous medium, as indicated in Figure \ref{fig:Inv_Alg}. At the beginning of the drainage simulation, only the edges incident on the inlet nodes are considered available for invasion (Figure \ref{fig:Inv_Alg}a). As an edge is deleted, the degree of the nodes on which it was incident is decreased, and these nodes' remaining edges become available for invasion (Figure \ref{fig:Inv_Alg}b-d). The decrement of a node's degree can be understood as the change in its occupation from wetting to non-wetting fluid. During the edge-deleting process, the graph can be split into multiple connected components. Connected components that do not contain outlet nodes represent wetting-fluid trapped clusters, and their edges become unavailable for invasion (Figure \ref{fig:Inv_Alg}e). Following these rules, the drainage process continues in the model until all available edges are deleted.

\subsubsection{Cluster Merging via Capillary Bridges}

\begin{figure}[ht]
  \centering
  \includegraphics[width=1\textwidth]{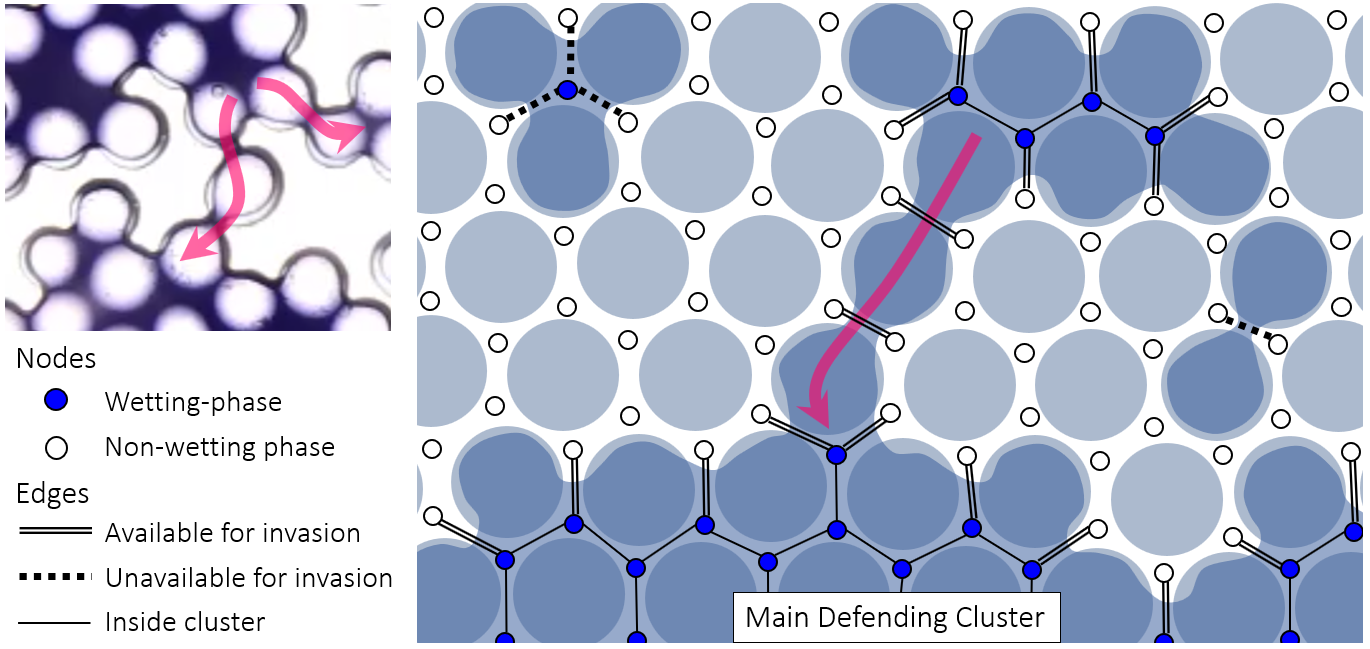}
  \caption{Sketch of Hele-Shaw porous-space representation into a 2D regular honeycomb lattice of nodes and edges.}
  \label{fig:Bridge_Conn}
\end{figure}

The drainage process illustrated in Figure \ref{fig:Inv_Alg} does not comprise the formation of capillary bridges and could be reproduced with a regular bond invasion percolation model. The cluster-labeling modification we propose in our model comes into being when capillary bridges are formed in a way that continuity is established between otherwise disconnected clusters, as suggested in Figure \ref{fig:Bridge_Conn}. 

In the upper left corner of Figure \ref{fig:Bridge_Conn}, a macro picture of a Hele-Shaw cell under drainage from \cite{moura2019connectivity} is shown. In this insert, the two arrows highlight chains formed by capillary bridges through which fluid transport is permissible. In a regular IP model, this type of connectivity is not taken into account, as no paths are formed between the nodes at the endpoints of capillary-bridge chains. This is indicated in the sketch on the right side of Figure \ref{fig:Bridge_Conn}: capillary bridges appear in the graph as edges incident on two nodes filled with the non-wetting phase, and the connection they provide for the wetting fluid does not coincide with the network representing the porous space. 

The capillary-bridge chain connectivity was then incorporated into our model by considering that edges belonging to the same hexagon in the honeycomb lattice are part of the same connected component, even if they do not directly establish a path. This rule is equivalent to acknowledging that the wetting fluid residing in the contact points between the spheres and the planes in the Hele-Shaw cell links all bridges formed around the same spheres, as illustrated in Figure \ref{fig:Rings_Grains}. With this cluster-labeling modification, our model could reproduce experimentally verified aspects of drainage events related to both primary and secondary mechanisms, as presented in Section \ref{sec:Res&Dis}

\section{Results and Discussion}
\label{sec:Res&Dis}

\begin{figure}[ht]
  \centering
  \includegraphics[width=0.75\textwidth]{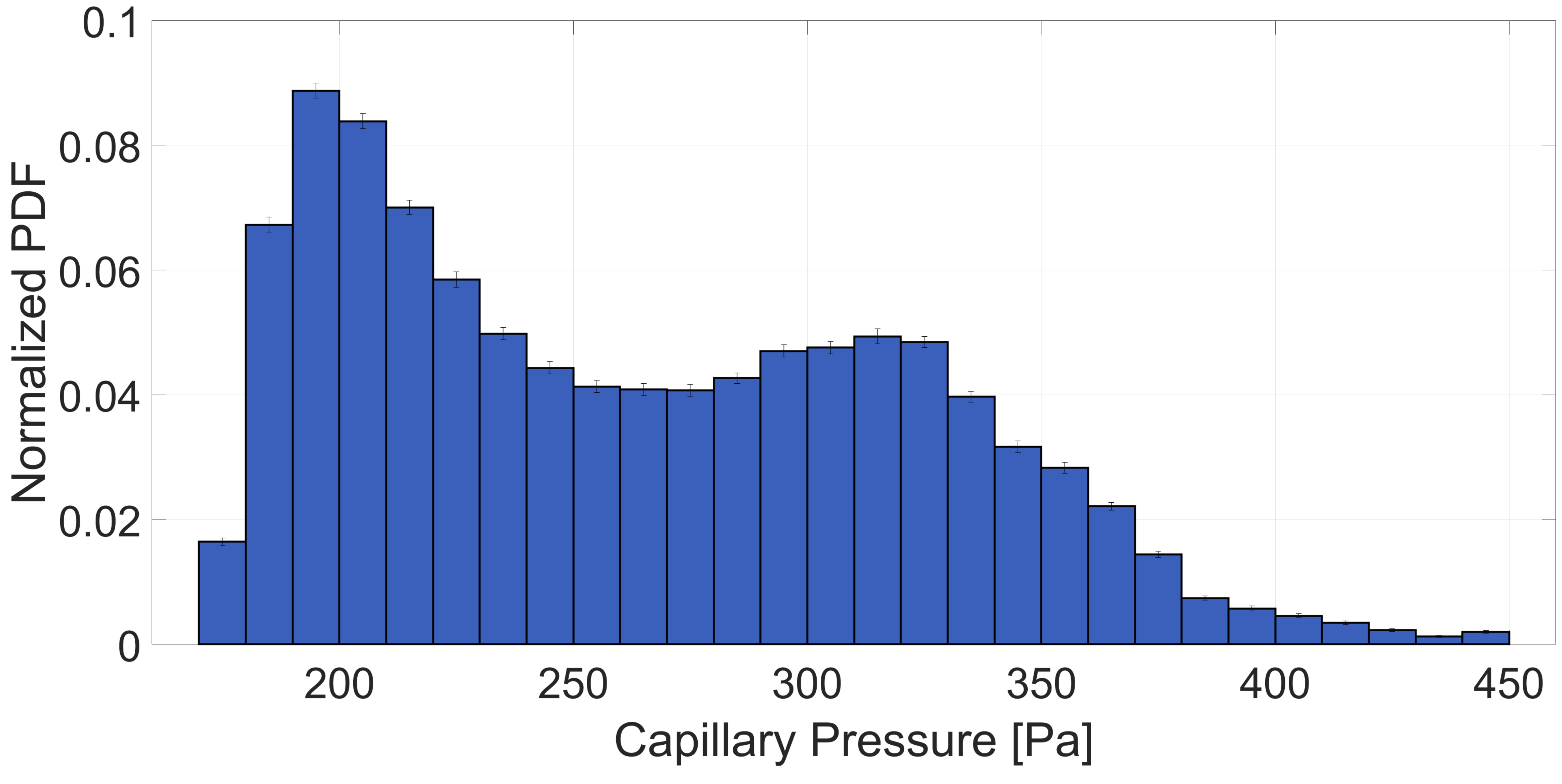}
  \caption{Probability density function of the capillary pressure of the porous matrix throats, generated to match the values reported in \citet{moura2019connectivity}.}
  \label{fig:PDF_Pc}
\end{figure}

The results presented in this section were obtained with drainage simulations on 100 random realizations of honeycomb lattices with 30703 nodes and 45903 edges, representing quasi-2D porous media of approximately 15x26 cm. The capillary pressure threshold values of the edges representing throats followed the distribution presented in Figure \ref{fig:PDF_Pc}, based on the experiments presented by \citet{moura2019connectivity}. These values were calculated with the separation between the spheres reported in \cite{moura2019connectivity} -- which varied approximately from 0.15 to 1.5 mm --, an interfacial tension between wetting and non-wetting fluids of $\gamma=64mN/m$, a density contrast between the fluids of $\Delta \rho=1203.75 kg/m^3$, and a contact angle of $\theta=0^\circ$ between the fluids and the medium. The influence of gravitational forces on the drainage was evaluated by adopting the following values of inclination of the porous media with respect to the horizontal plane: $\beta=[0, 1.25, 2.5, 5, 7.5, 10, 15, 20, 30, 45, 60, 75, 90,]^\circ$.

In the lattices, lateral boundaries ($x=0cm$ and $x=15cm$) were closed to the flow, nodes at the upper boundary ($y=26cm$) were sources of non-wetting fluid, and all nodes at the lower boundary ($y=0cm$) were connected to an external wetting-fluid sink by edges with no resistance to flow. In this way, all drainage simulations came to an end by the time the non-wetting phase percolated the lattice. 

\begin{figure}[h!]
     \centering
     
     \begin{subfigure}[b]{0.24\textwidth}
         \centering
         \includegraphics[width=\textwidth]{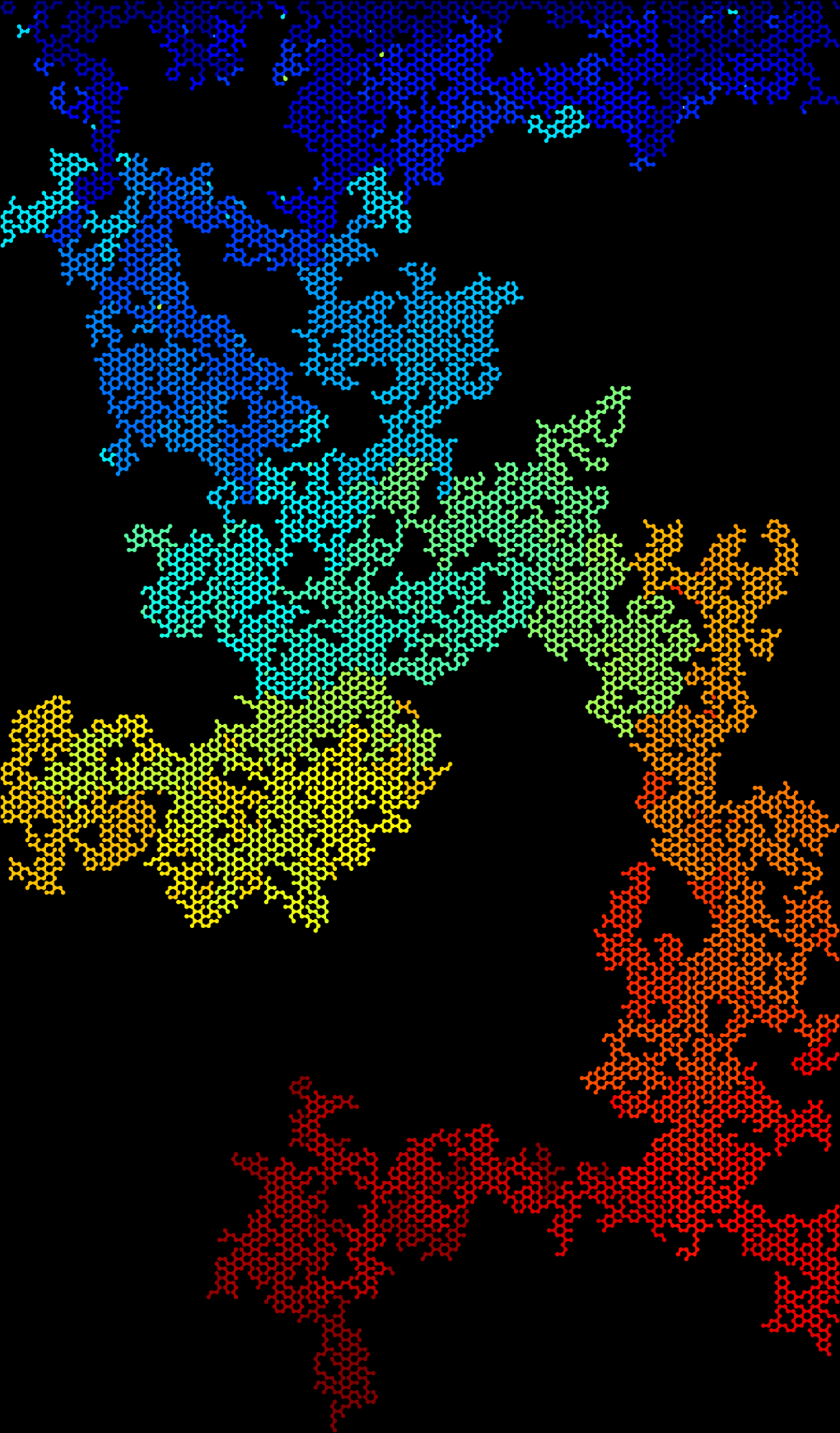}
         \caption{$\beta=0^\circ$}
         \label{fig:IT_0}
     \end{subfigure}
     \hfill
     \begin{subfigure}[b]{0.24\textwidth}
         \centering
         \includegraphics[width=\textwidth]{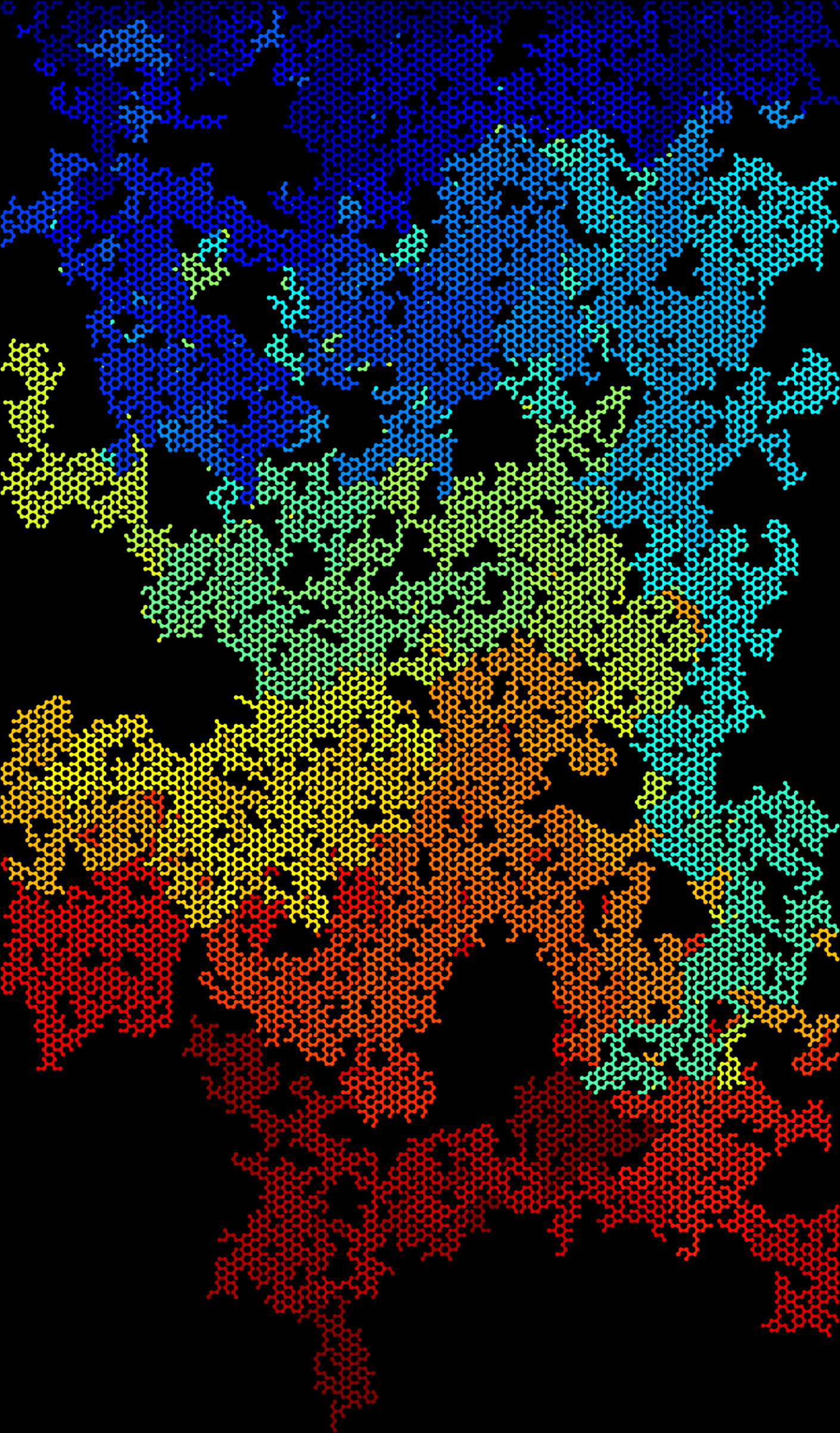}
         \caption{$\beta=1.25^\circ$}
         \label{fig:IT_1-25}
     \end{subfigure}
     \hfill
     \begin{subfigure}[b]{0.24\textwidth}
         \centering
         \includegraphics[width=\textwidth]{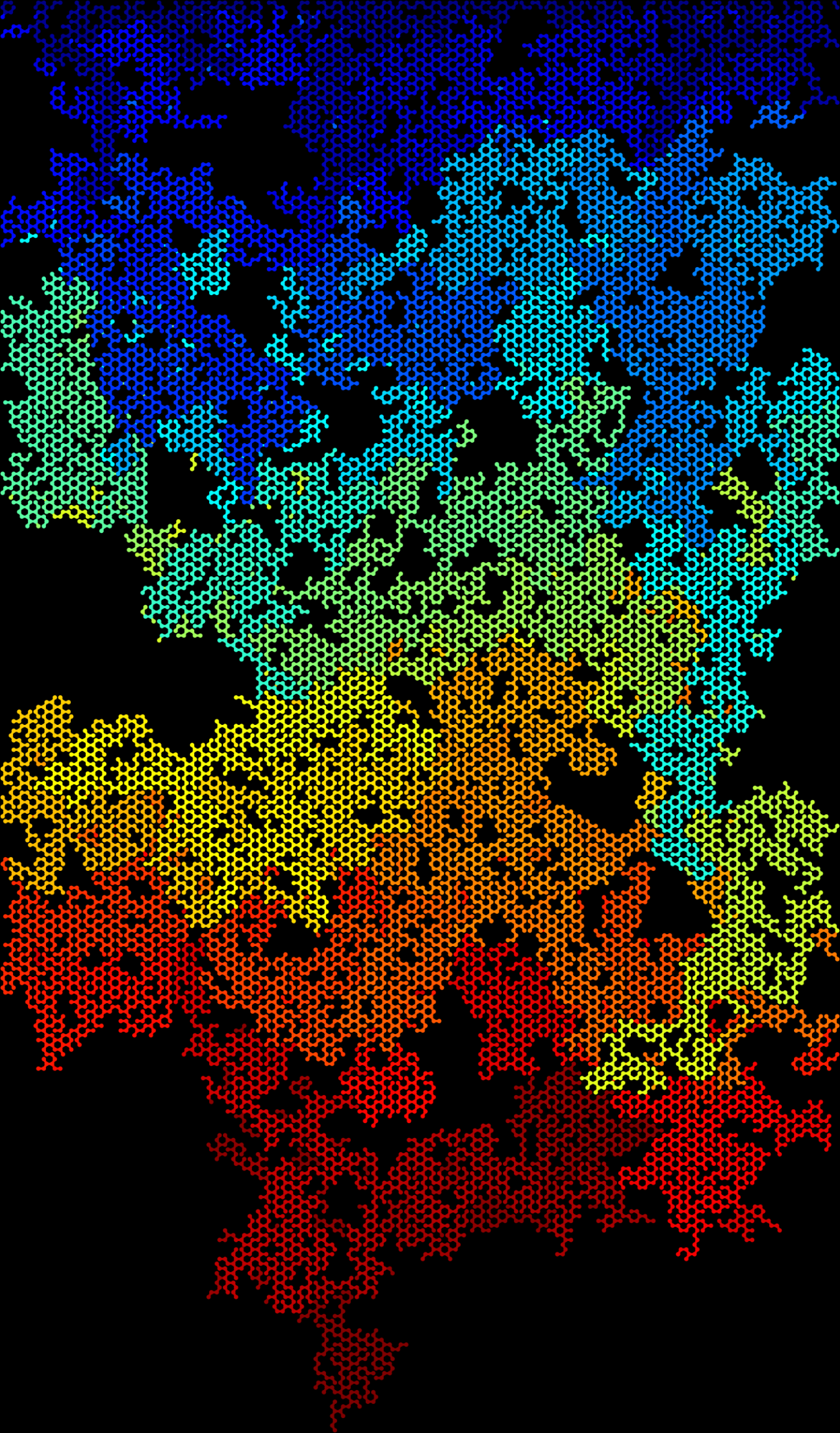}
         \caption{$\beta=2.5^\circ$}
         \label{fig:IT_2-5}
     \end{subfigure}
     \hfill
     \begin{subfigure}[b]{0.24\textwidth}
         \centering
         \includegraphics[width=\textwidth]{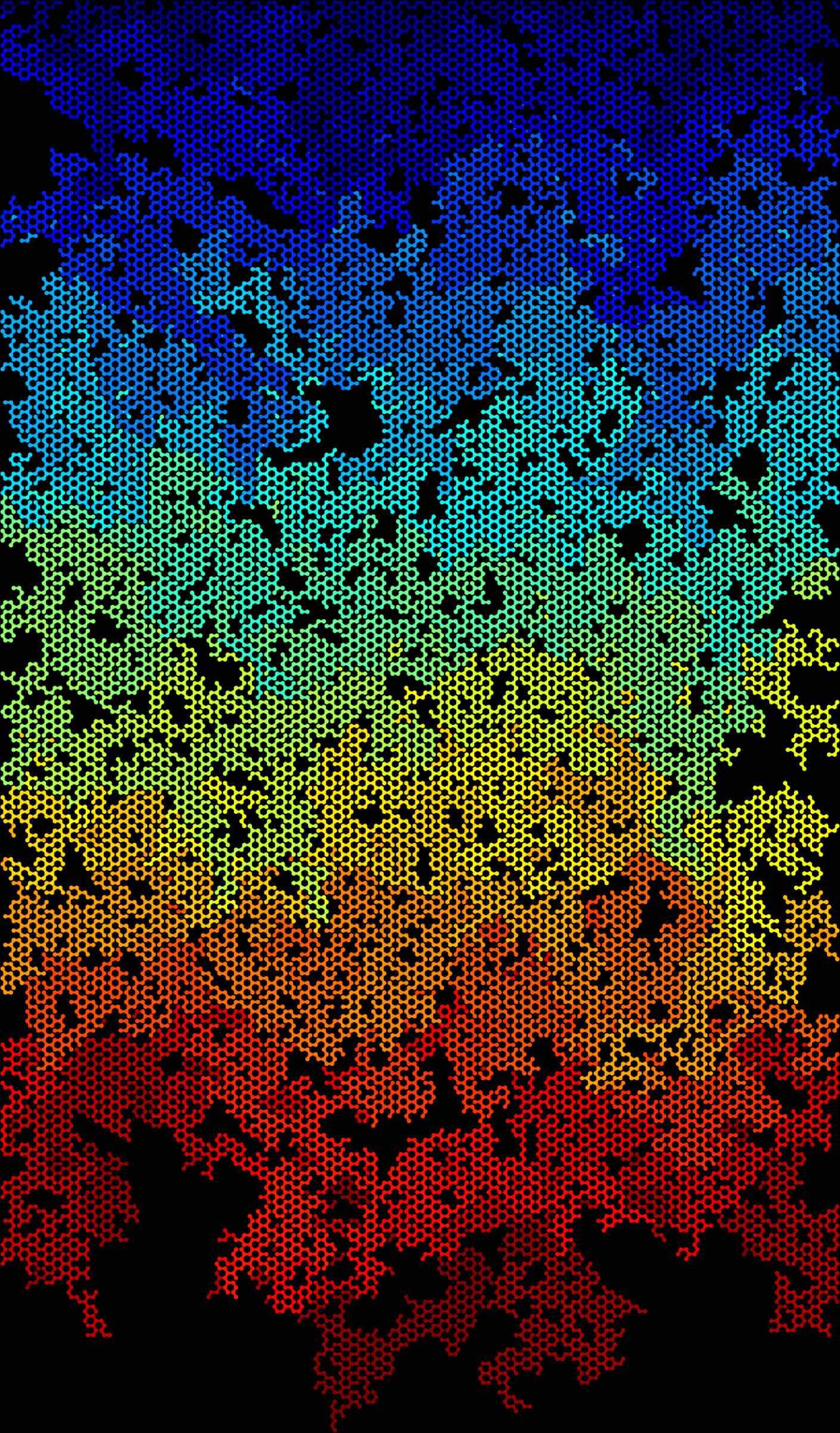}
         \caption{$\beta=7.5^\circ$}
         \label{fig:IT_7-5}
     \end{subfigure}
     \hfill
     \begin{subfigure}[b]{0.24\textwidth}
         \centering
         \includegraphics[width=\textwidth]{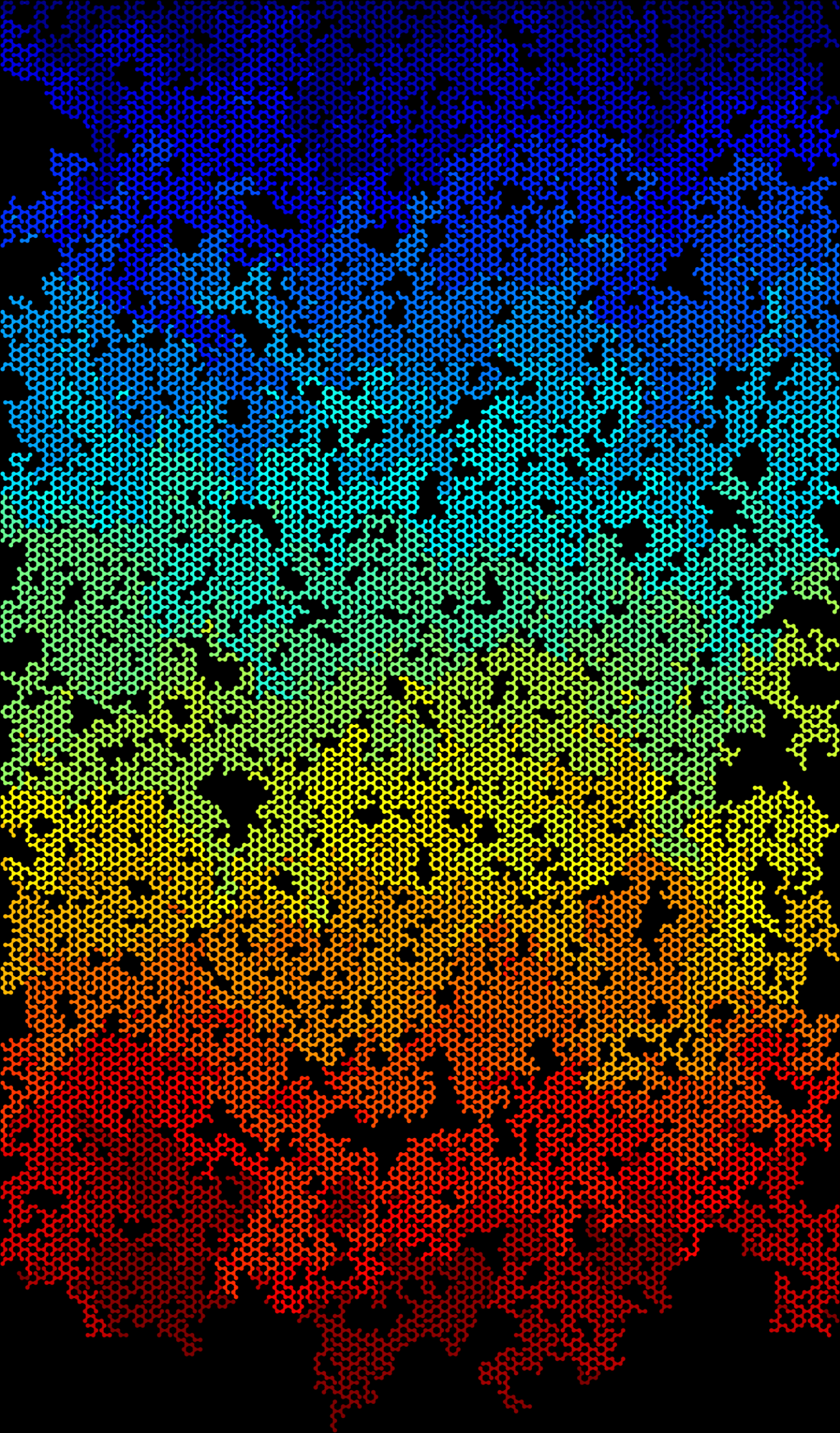}
         \caption{$\beta=10^\circ$}
         \label{fig:IT_10}
     \end{subfigure}
     \hfill
     \begin{subfigure}[b]{0.24\textwidth}
         \centering
         \includegraphics[width=\textwidth]{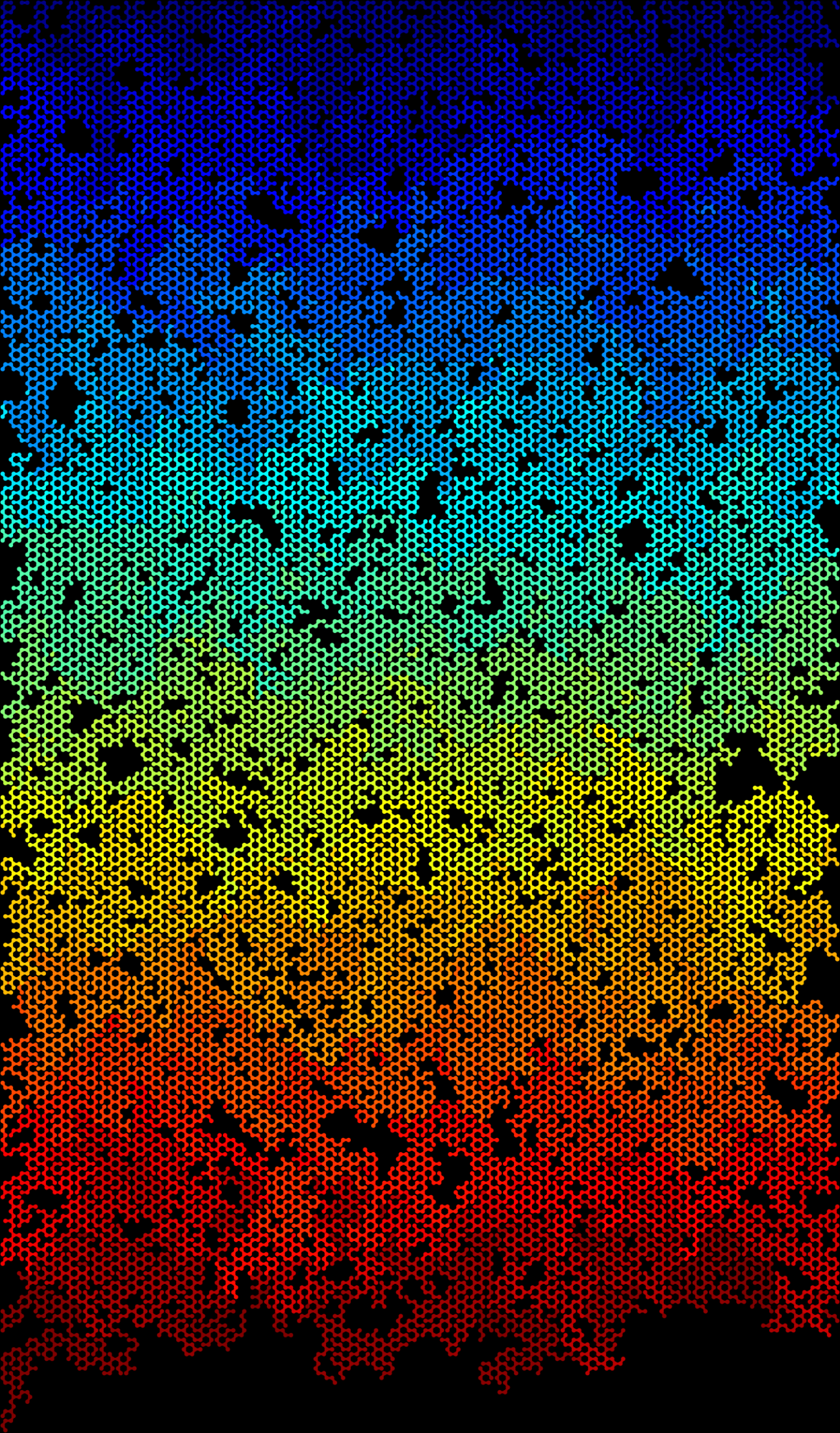}
         \caption{$\beta=20^\circ$}
         \label{fig:IT_20}
     \end{subfigure}
     \hfill
     \begin{subfigure}[b]{0.24\textwidth}
         \centering
         \includegraphics[width=\textwidth]{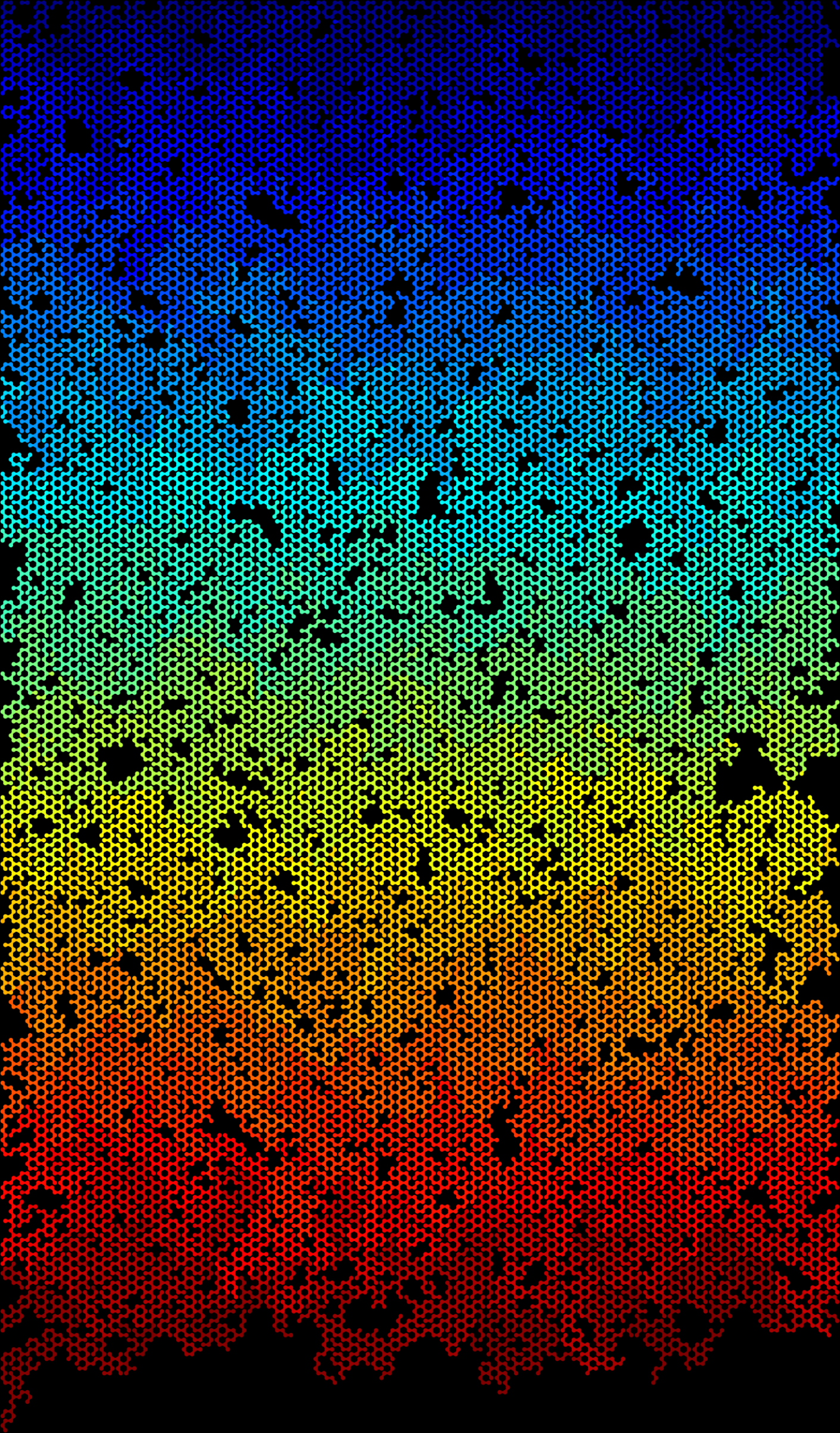}
         \caption{$\beta=30^\circ$}
         \label{fig:IT_30}
     \end{subfigure}
     \hfill
     \begin{subfigure}[b]{0.24\textwidth}
         \centering
         \includegraphics[width=\textwidth]{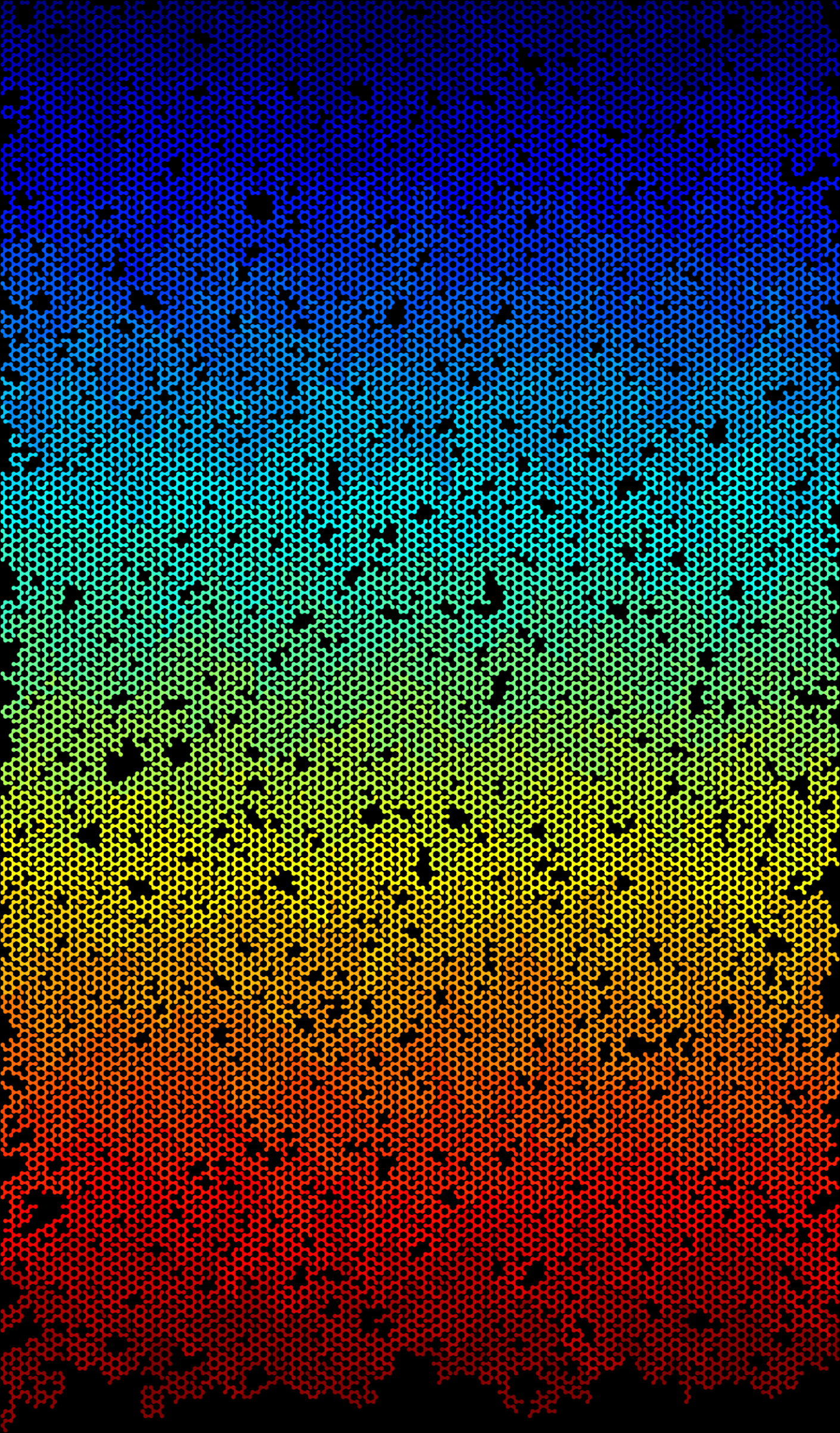}
         \caption{$\beta=60^\circ$}
         \label{fig:IT_60}
     \end{subfigure}
     
        \caption{Representation of the invasion order in the porous matrix during drainage, from dark blue (early invasion) to dark red (late invasion). }
        \label{fig:Inv_Time}
\end{figure}

Figure \ref{fig:Inv_Time} illustrates the shift in drainage dynamics predicted by the proposed model as the influence of the gravitational component increases. These results were generated using the same lattice and with inclination angles indicated in the captions. 

From \ref{fig:IT_0} to \ref{fig:IT_60}, only the nodes and edges occupied by the non-wetting phase at breakthrough are represented. The color gradient from dark blue to dark red symbolizes the order of invasion, from early to late invaded pores and throats. Black regions of the image are, therefore, correspondent to the trapped wetting fluid clusters. As expected, the unstable front characteristic of capillary fingering is progressively stabilized as $\beta$ increases, resulting in a more efficient removal of the wetting phase. The invasion dynamics presented in \ref{fig:Inv_Time} are a result of both primary and secondary drainage mechanisms, which will be discussed in more detail in the next sections.

\subsection{The impact of film flow on residual saturation}
\label{sec:FF_Sat}

\begin{figure}[h!]
     \centering
     
     \begin{subfigure}[b]{0.24\textwidth}
         \centering
         \includegraphics[width=\textwidth]{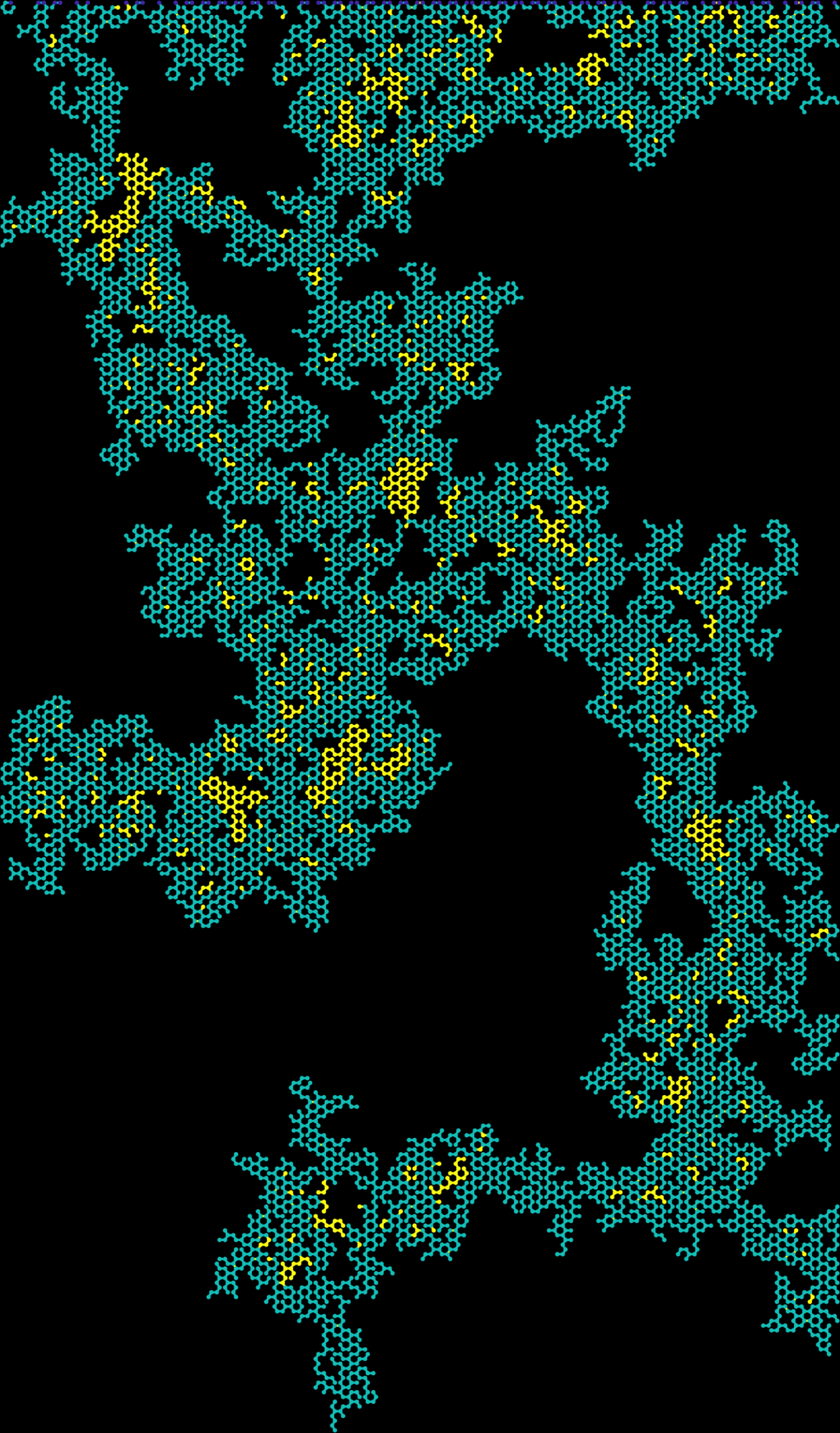}
         \caption{$\beta=0^\circ$}
         
     \end{subfigure}
     \hfill
     \begin{subfigure}[b]{0.24\textwidth}
         \centering
         \includegraphics[width=\textwidth]{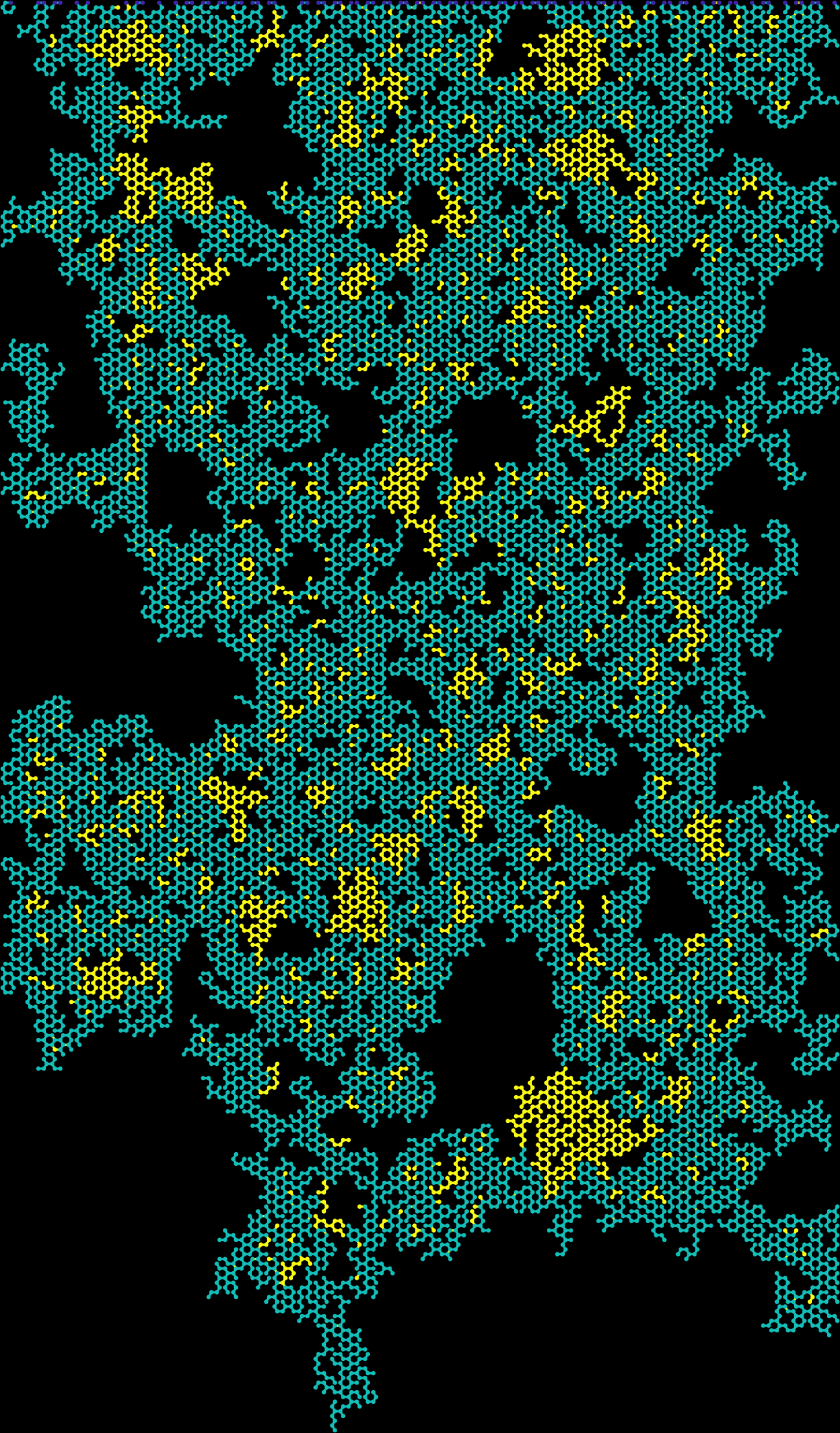}
         \caption{$\beta=1.25^\circ$}
         
     \end{subfigure}
     \hfill
     \begin{subfigure}[b]{0.24\textwidth}
         \centering
         \includegraphics[width=\textwidth]{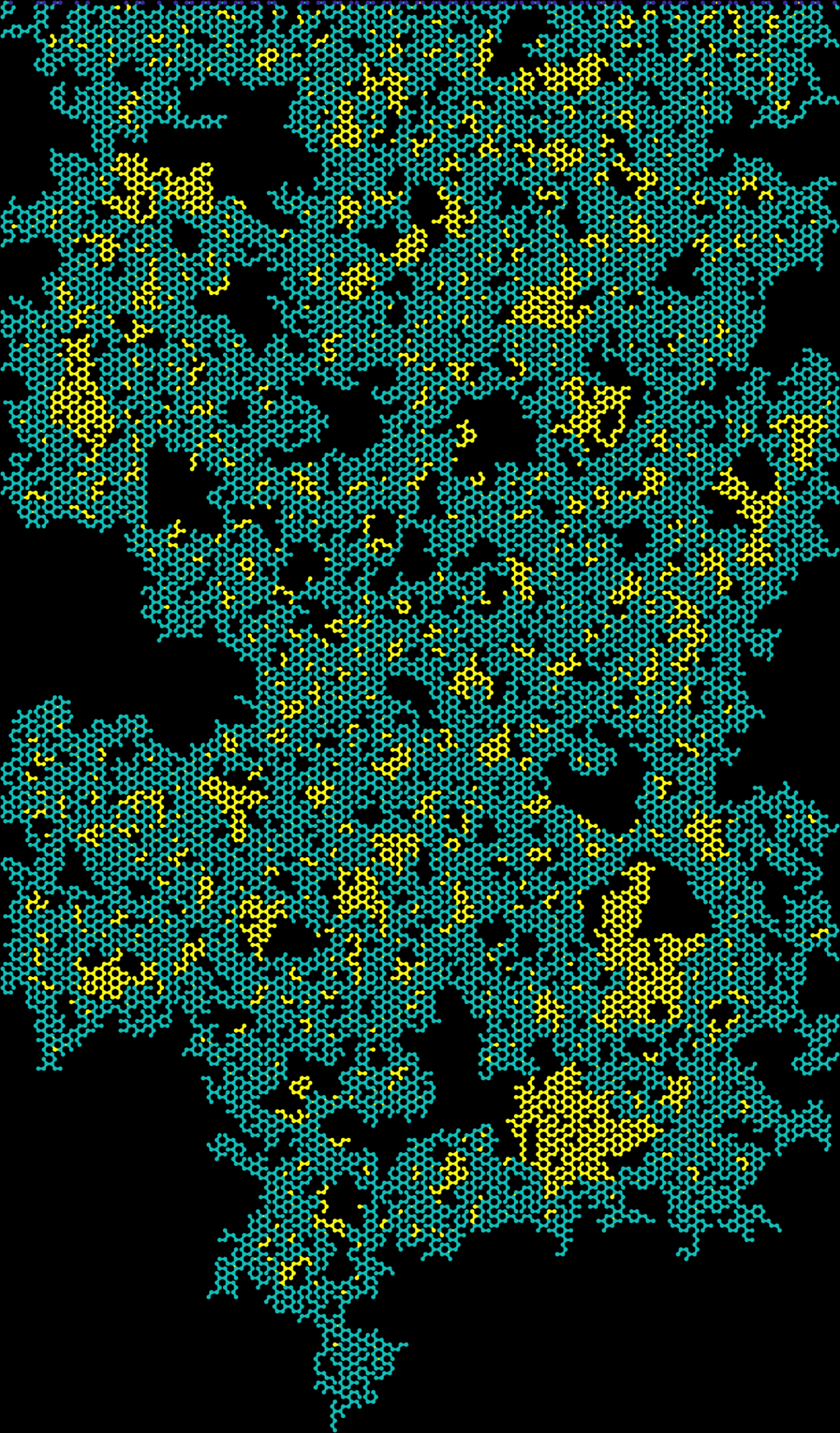}
         \caption{$\beta=2.5^\circ$}
         
     \end{subfigure}
     \hfill
   %
     \begin{subfigure}[b]{0.24\textwidth}
         \centering
         \includegraphics[width=\textwidth]{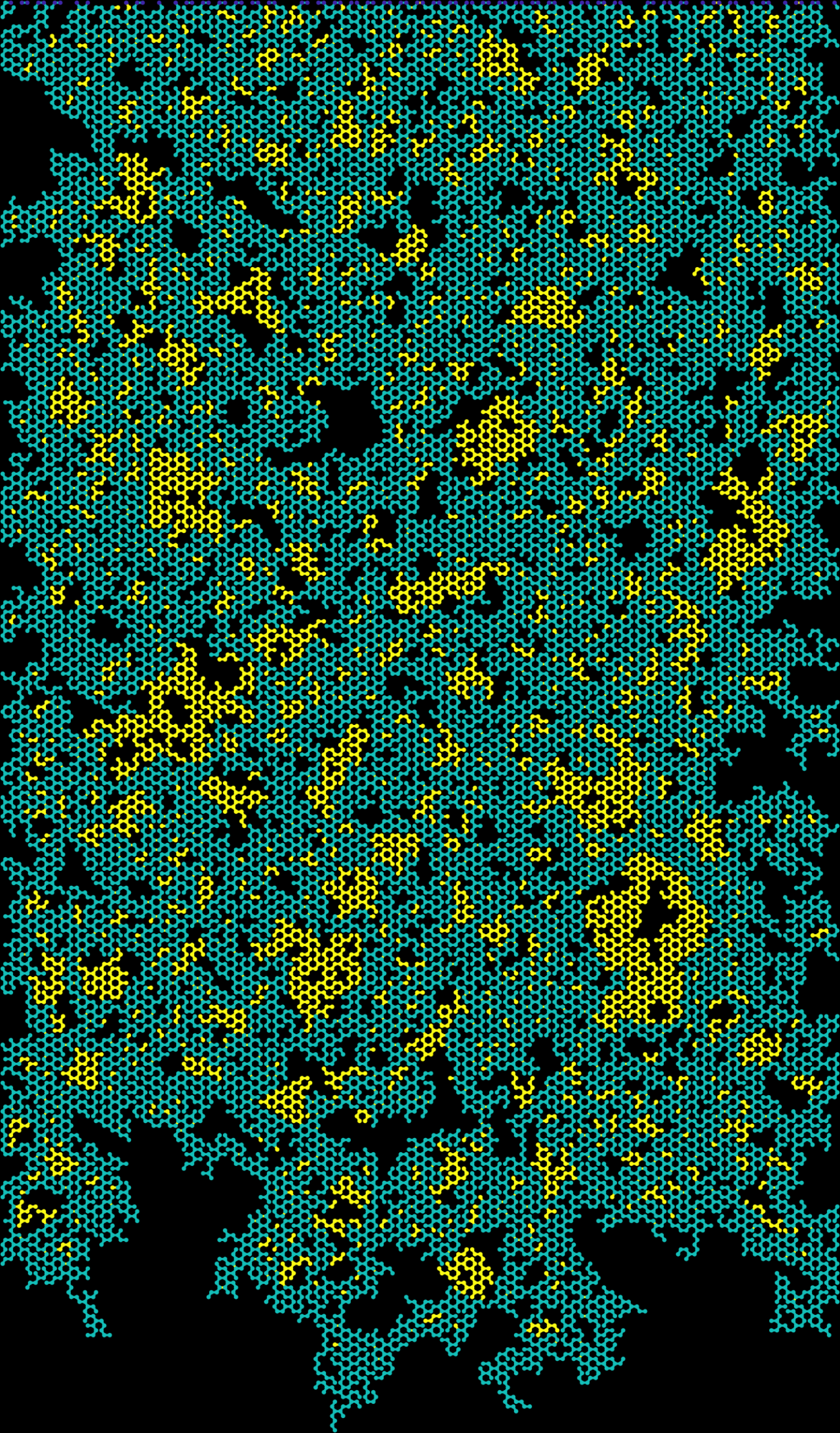}
         \caption{$\beta=7.5^\circ$}
         
     \end{subfigure}
     \hfill
     \begin{subfigure}[b]{0.24\textwidth}
         \centering
         \includegraphics[width=\textwidth]{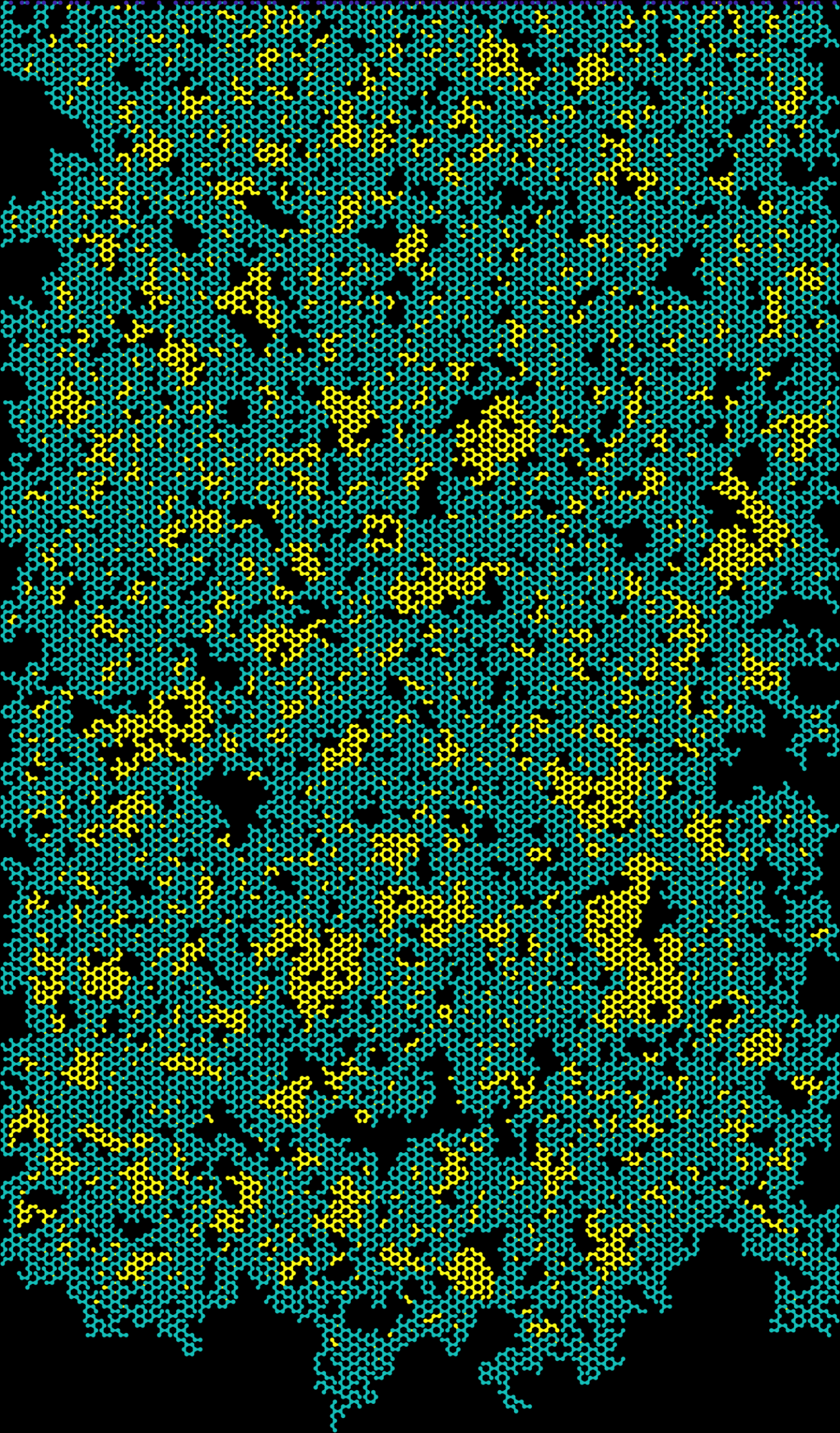}
         \caption{$\beta=10^\circ$}
         
     \end{subfigure}
     \hfill
     \begin{subfigure}[b]{0.24\textwidth}
         \centering
         \includegraphics[width=\textwidth]{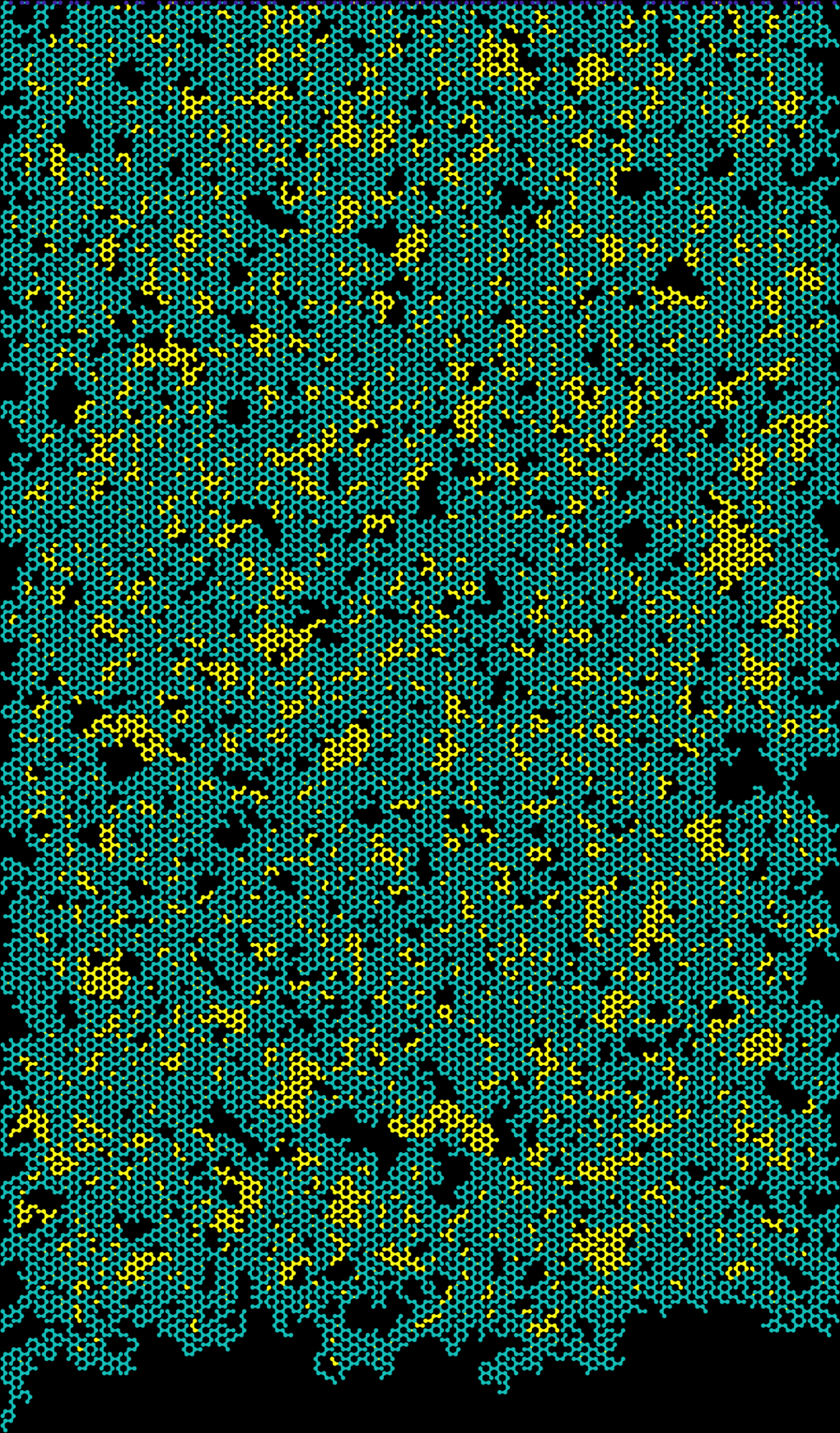}
         \caption{$\beta=20^\circ$}
         
     \end{subfigure}
     \hfill
     \begin{subfigure}[b]{0.24\textwidth}
         \centering
         \includegraphics[width=\textwidth]{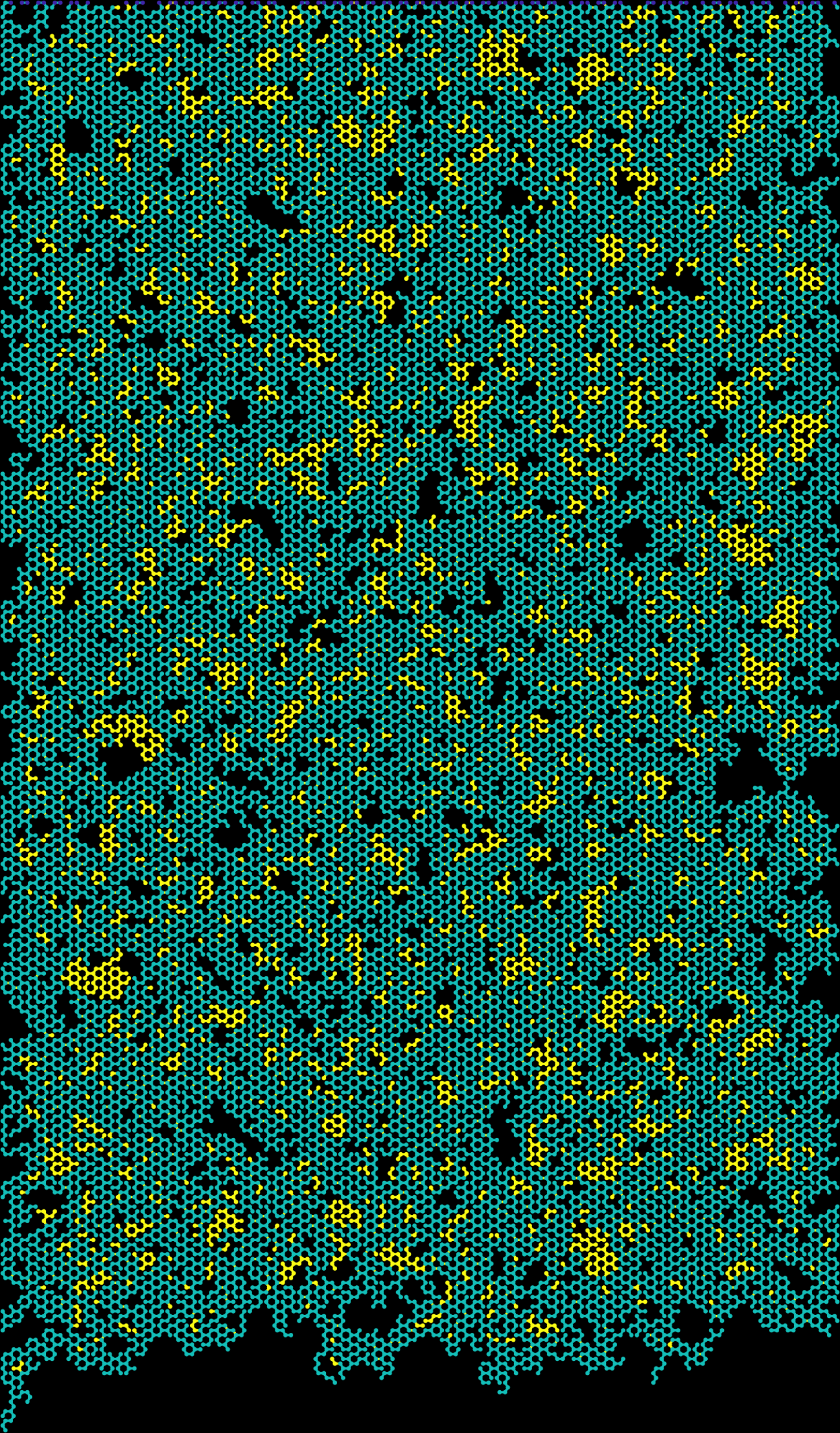}
         \caption{$\beta=30^\circ$}
         
     \end{subfigure}
     \hfill
     \begin{subfigure}[b]{0.24\textwidth}
         \centering
         \includegraphics[width=\textwidth]{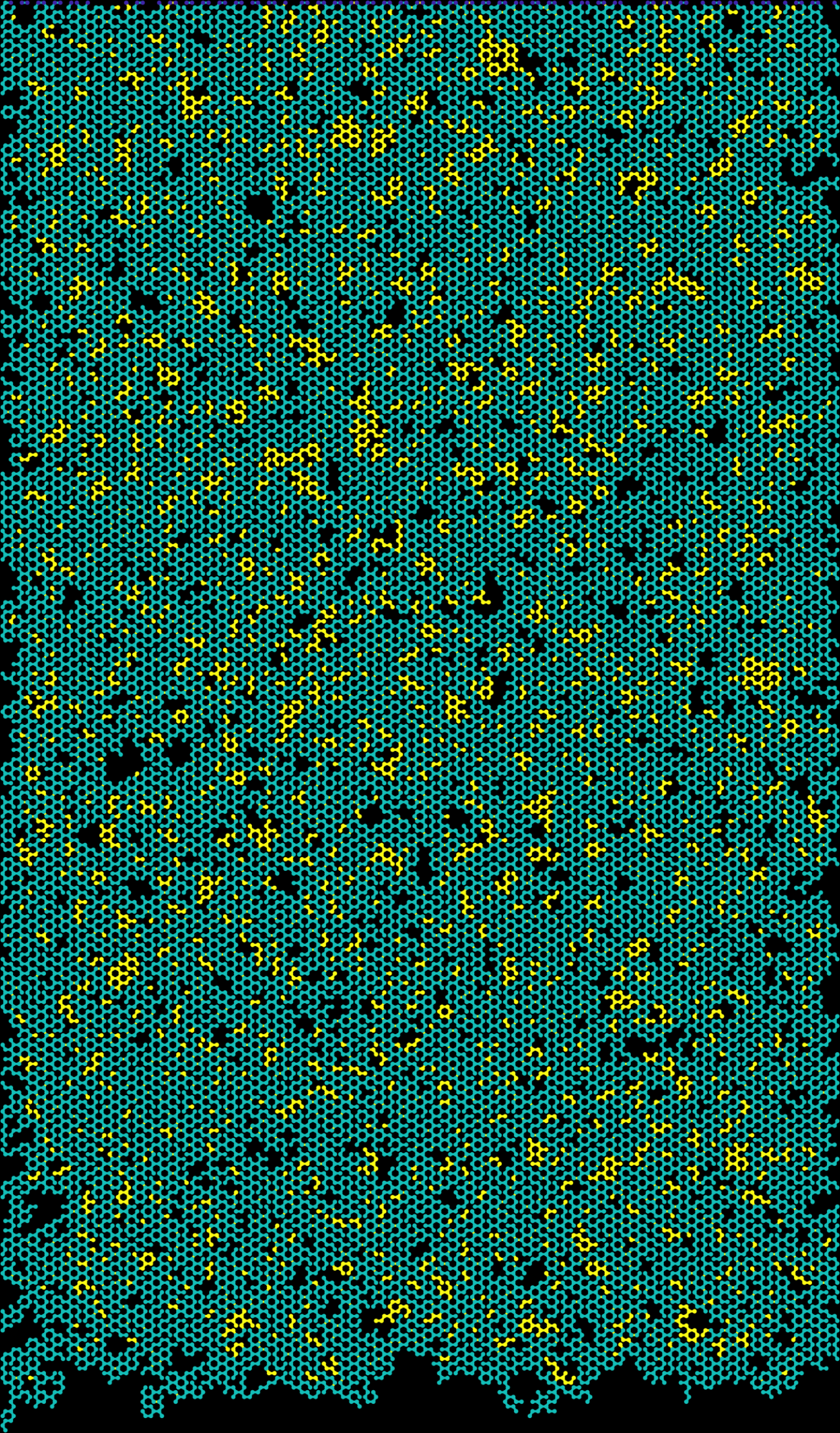}
         \caption{$\beta=60^\circ$}
         
     \end{subfigure}

        \caption{Representation of the drainage mechanism related to invasion in the porous matrix. Nodes and edges in blue were drained via the primary drainage mechanism, while nodes and edges in yellow were drained by the secondary mechanism.}
        \label{fig:Inv_Mod}
\end{figure}

As presented in Section \ref{sec: Intro}, chains of capillary bridges can establish connectivity between seemingly trapped clusters and enable their drainage, leading to a reduction in wetting-fluid residual saturations. The prevalence of primary and secondary drainage mechanisms for different degrees of gravitational forces influence is illustrated in Figure \ref{fig:Inv_Mod}.

\begin{figure}[h!]
     \centering
    \includegraphics[width=0.65\textwidth]{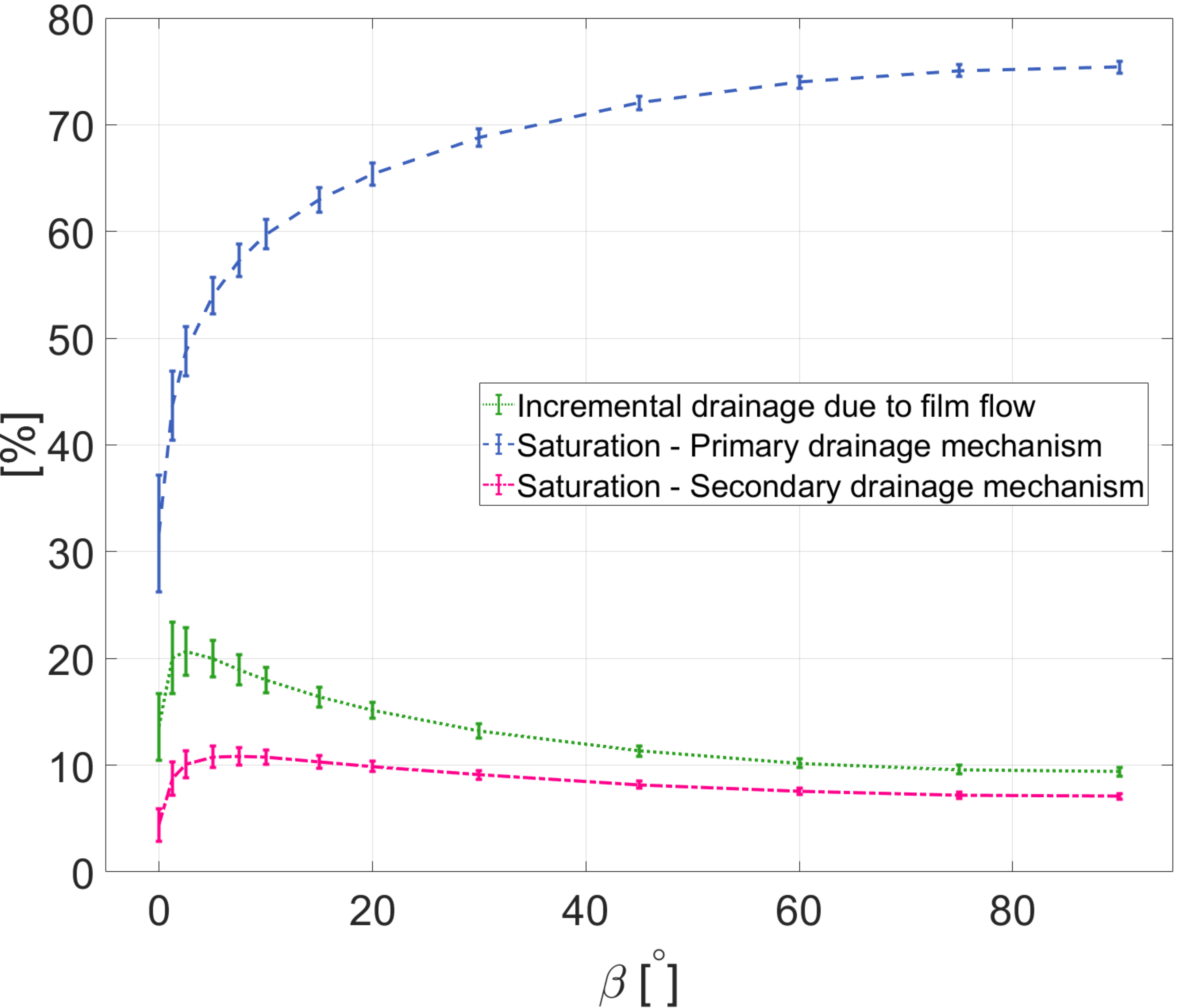}
     \caption{Contribution of primary and secondary drainage mechanisms.}
     \label{fig:Sat_Mechanims}
\end{figure}

Complementary to the information presented in Figure \ref{fig:Inv_Time}, the images in Figure \ref{fig:Inv_Mod} introduce the distinction between drained pores and throats that pertained directly to the main defending cluster (blue), to the ones that were connected to it via capillary bridges (yellow). It is clear that a significant fraction of the wetting fluid in the porous matrix could only be dislocated through the flow paths consisting of capillary bridges and corners, especially for intermediate inclination angles. The quantification of the contribution of both drainage mechanisms averaged over the different lattice realizations is presented in Figure \ref{fig:Sat_Mechanims}.

As indicated in the graph, the fraction of wetting fluid displaced due to the primary drainage mechanism increases monotonically as the influence of gravitational forces grows, while the maximum saturation drained due to the secondary mechanism is observed at $\beta=7.5^\circ$. As for the incremental drainage provided by the secondary drainage, the maximum mean value is achieved at $\beta=2.5^\circ$. This behavior may be explained due to the contradicting effects gravity can have on drainage through film flow. On one side, gravity can benefit flow from clusters placed above the average vertical position of the drainage front. Boundaries of clusters bypassed by the front are associated with higher capillary pressure thresholds for invasion than the regions in their immediate vicinity. However, as these clusters get further to the front, the negative contribution of the hydrostatic pressure to their invasion threshold may place them in a preferential position for invasion, when compared to nodes belonging to the front. This may justify the relatively low value of saturation, $4.39\pm 1.54\%$, drained due to film flow at $\beta=0^\circ$. Unlike in the other cases, the distance of a node from the front without the influence of gravity would not positively impact its chance of invasion, leading to trapping. On the other side, the positive influence of gravity on the primary mechanism is so significant that little is left for the secondary drainage at high values of $\beta$. In these cases, the efficient sweep of the porous matrix by the invasion front may, therefore, reduce the saturation drained due to film flow.

It is important to mention at this point that the absence of viscous effects in the proposed model may affect the analysis of drainage through film flow based on the data obtained with the proposed model. While we propose to investigate slow drainage, which is associated with negligible influence of viscous forces, there is a distinction between the effect of viscosity on the flow through the bulk of pores and throats and on the flow through chains of capillary bridges. The higher impact of viscous forces on the latter penalizes the secondary drainage mechanism, and we expect that the contribution of film flow to drainage presented here to be overestimated to some extent. The incorporation of viscous effects in the model could also lead to a shift in the values of $\beta$ associated with maximum drainage via film flow, as the negative influence of viscous forces on this mechanism should be proportionally more relevant for low inclination. Finally, viscous effects are partially responsible for the different time scales associated with primary and secondary drainage mechanisms identified experimentally in \cite{moura2019connectivity}.The joint analysis of Figures \ref{fig:Inv_Time} and \ref{fig:Inv_Mod} allow us to identify some contrast in the invasion time of clusters drained due to film flow and their surroundings, drained through the bulk of pores and throats, but this distinction should be more prominent. Still, the results shown in Figure \ref{fig:Sat_Mechanims} exhibit a reasonable agreement with experimental data obtained by \cite{moura2019connectivity}. In that study, an increase of $8.3\%$ in non-wetting phase saturation was reported for a drainage experiment with $\beta=20^\circ$, an up to $10\%$ in analyzed subsets of the porous matrix for a case with $\beta=30^\circ$. The discrepancy between experimental and simulated results is in line with the expected due to the lack of representation of viscous effects in the model, and we intend to address this shortcoming in a future work.

\subsection{The film flow active zone}
\label{sec:AZ}

In this section, we analyze the formation region trailing the invasion front where drainage events related to film flow are more likely to occur. First identified experimentally by \citet{moura2019connectivity}, this region was termed "active zone", and its detailed characterization can strengthen our understanding of the secondary drainage mechanism. 

The extent of the active zone is directly related to the size and spatial distribution of the portions of the porous matrix bypassed by the invasion front. First, there must be a series of capillary bridges positioned in a contiguous form in order to form a chain. This imposes a constraint on how wide the active zone can be, as long sequences of connected capillary bridges are less likely to occur than short ones. Second, the frequency and size of wetting-phase clusters behind the front contribute to the width of the active zone, as clusters connected to the front by capillary bridges can then lead to the direct connection of further regions in the porous matrix. Thus, the existence of large wetting-phase clusters can broaden the zone where secondary drainage is possible, even if long chains of capillary bridges are not formed. 

\begin{figure}[h!]
     \centering
     
     \begin{subfigure}[b]{0.24\textwidth}
         \centering
         \includegraphics[width=\textwidth]{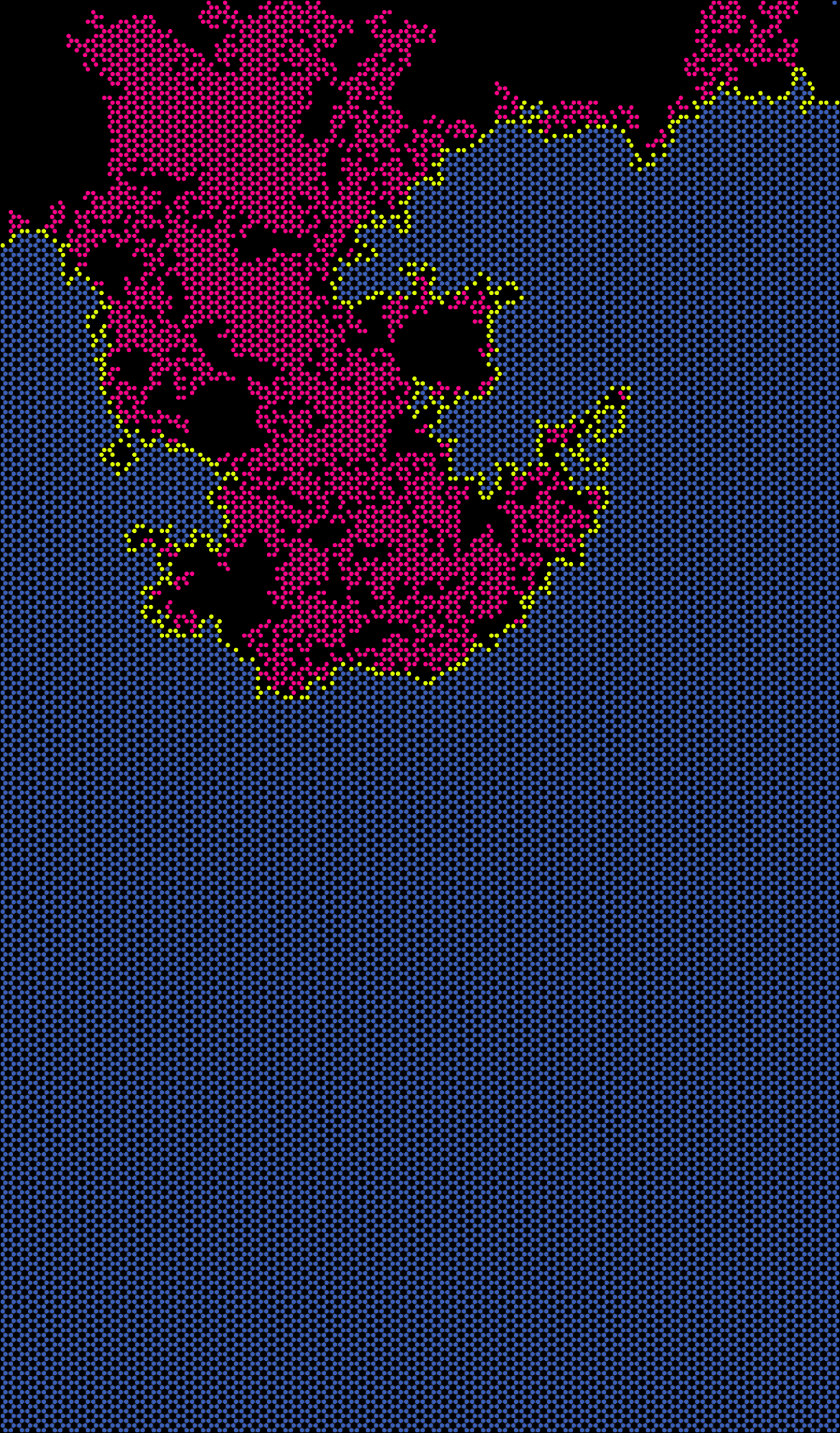}
         \caption{$\beta=0^\circ$}
         \label{fig:}
     \end{subfigure}
     \hfill
     \begin{subfigure}[b]{0.24\textwidth}
         \centering
         \includegraphics[width=\textwidth]{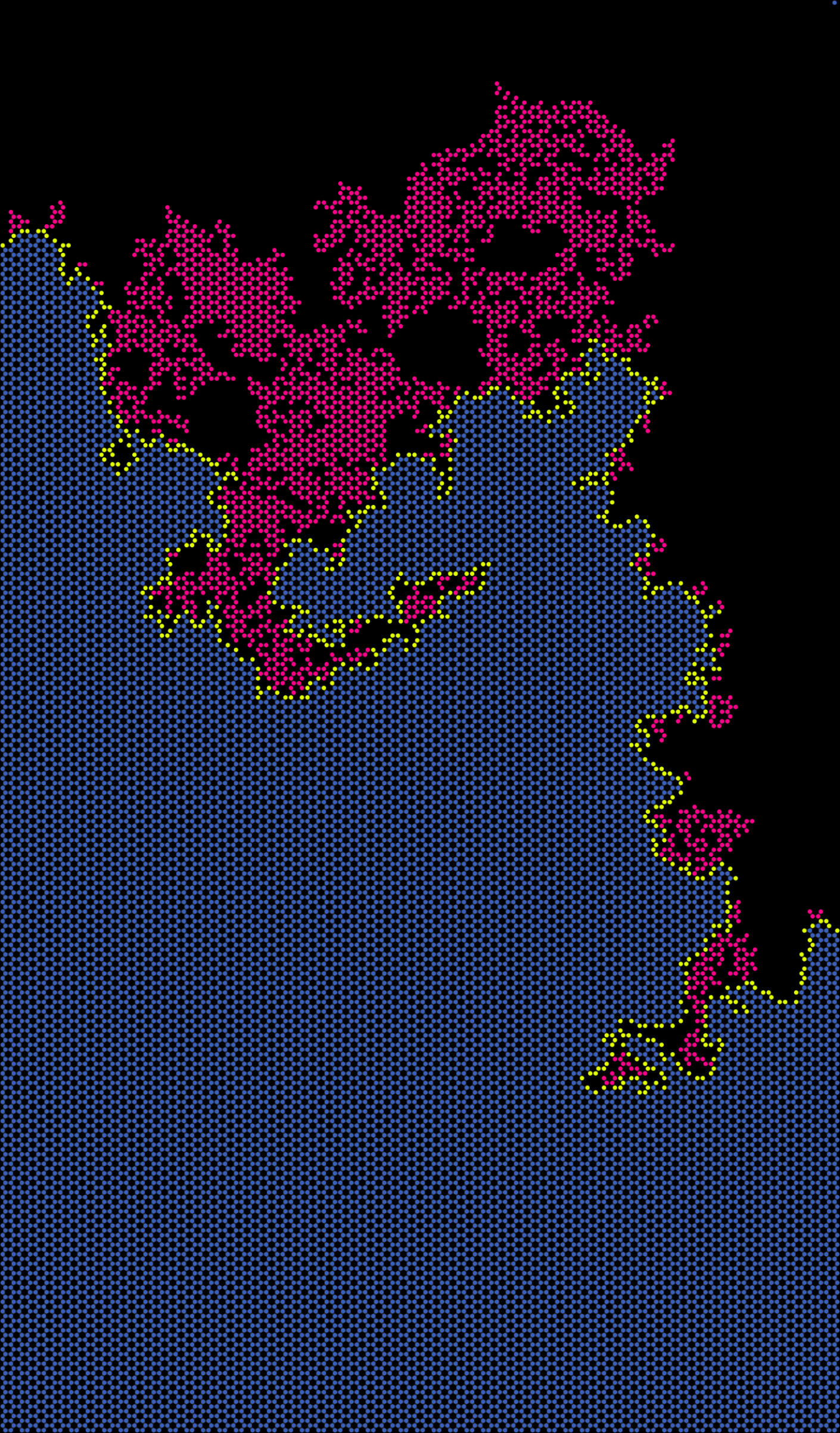}
         \caption{$\beta=1.25^\circ$}
         
     \end{subfigure}
     \hfill
     \begin{subfigure}[b]{0.24\textwidth}
         \centering
         \includegraphics[width=\textwidth]{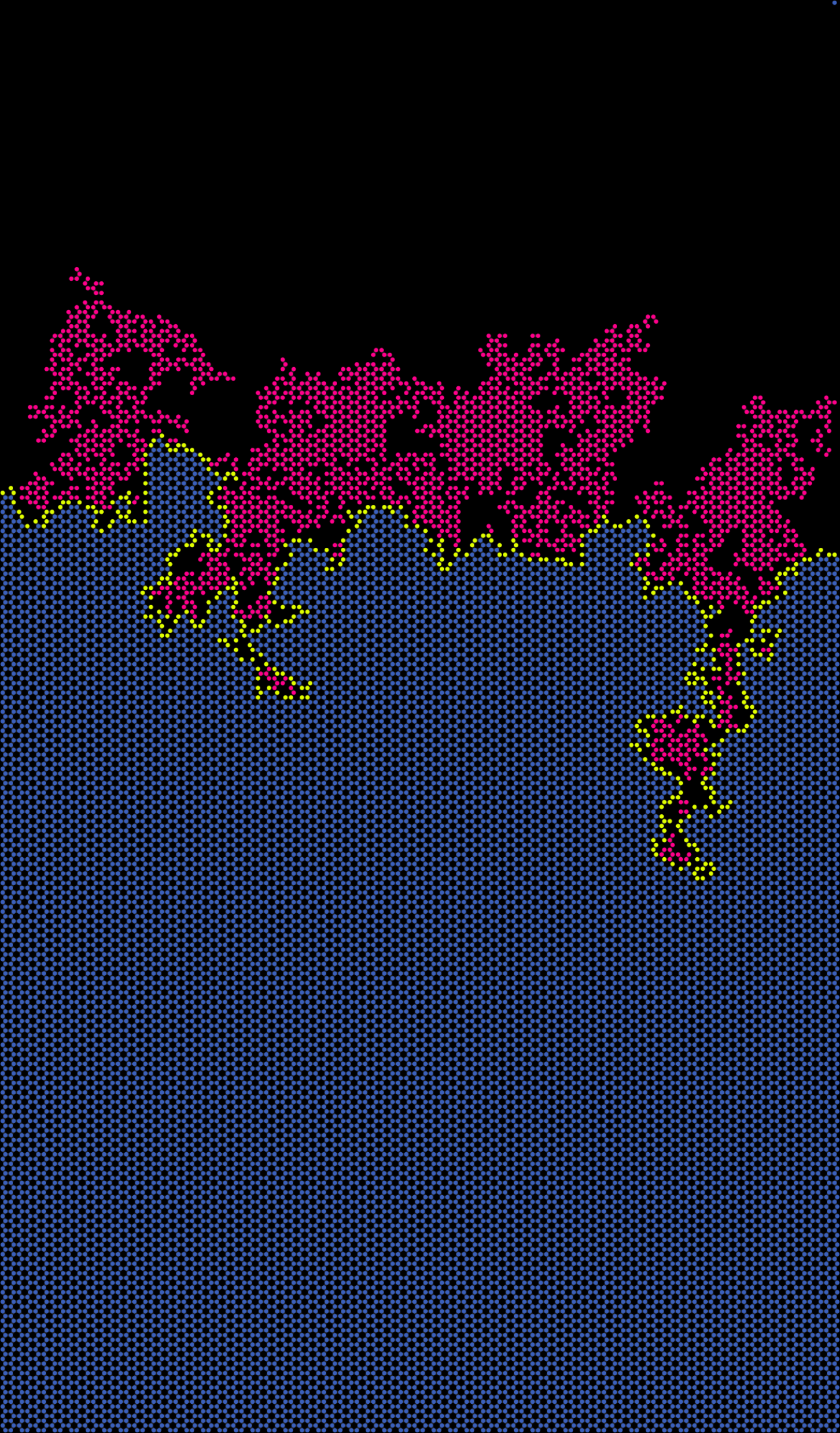}
         \caption{$\beta=2.5^\circ$}
         
     \end{subfigure}
     \hfill
     %
     \begin{subfigure}[b]{0.24\textwidth}
         \centering
         \includegraphics[width=\textwidth]{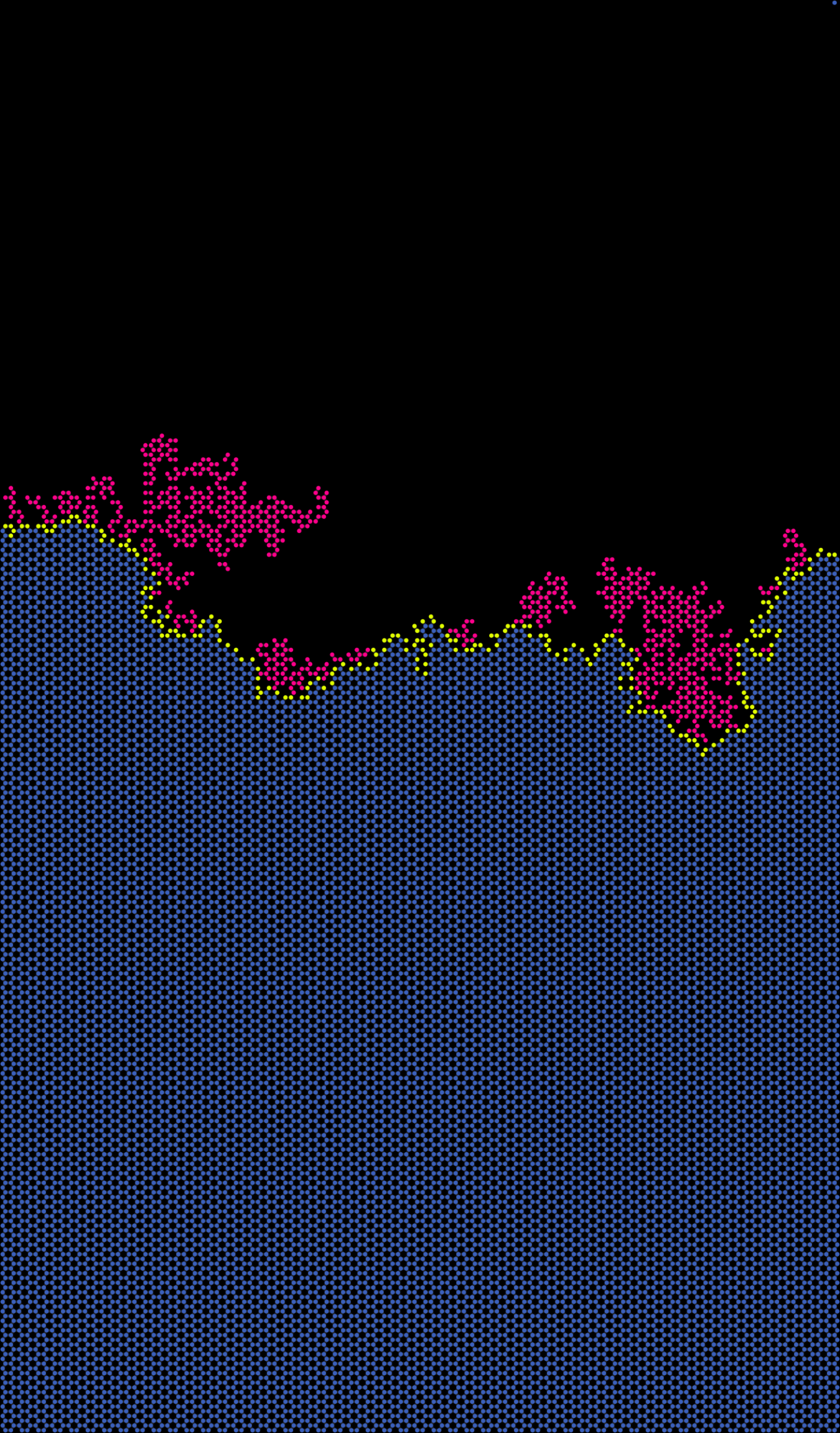}
         \caption{$\beta=7.5^\circ$}
         
     \end{subfigure}
     \hfill
     \begin{subfigure}[b]{0.24\textwidth}
         \centering
         \includegraphics[width=\textwidth]{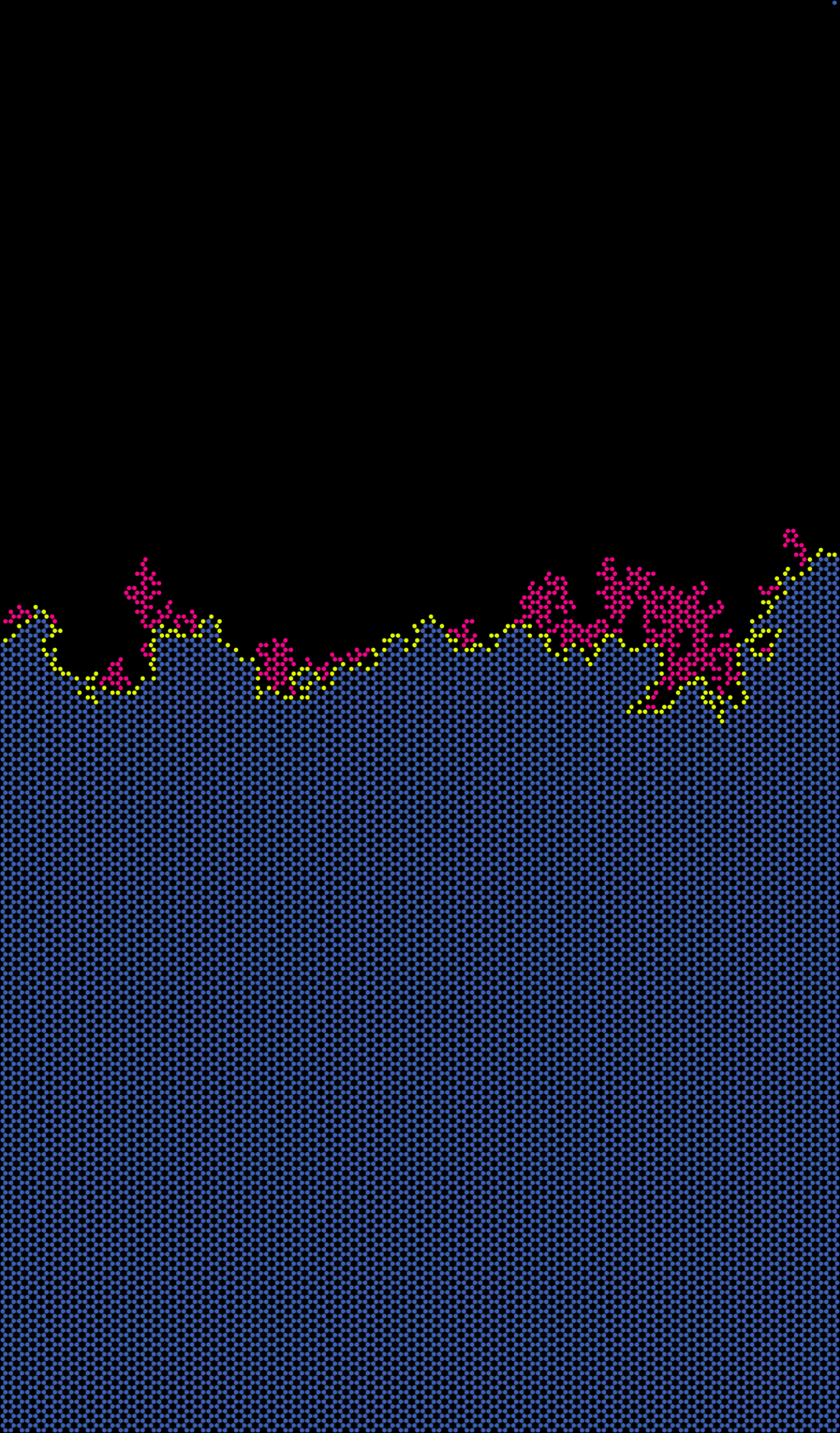}
         \caption{$\beta=10^\circ$}
         
     \end{subfigure}
     \hfill
          \begin{subfigure}[b]{0.24\textwidth}
         \centering
         \includegraphics[width=\textwidth]{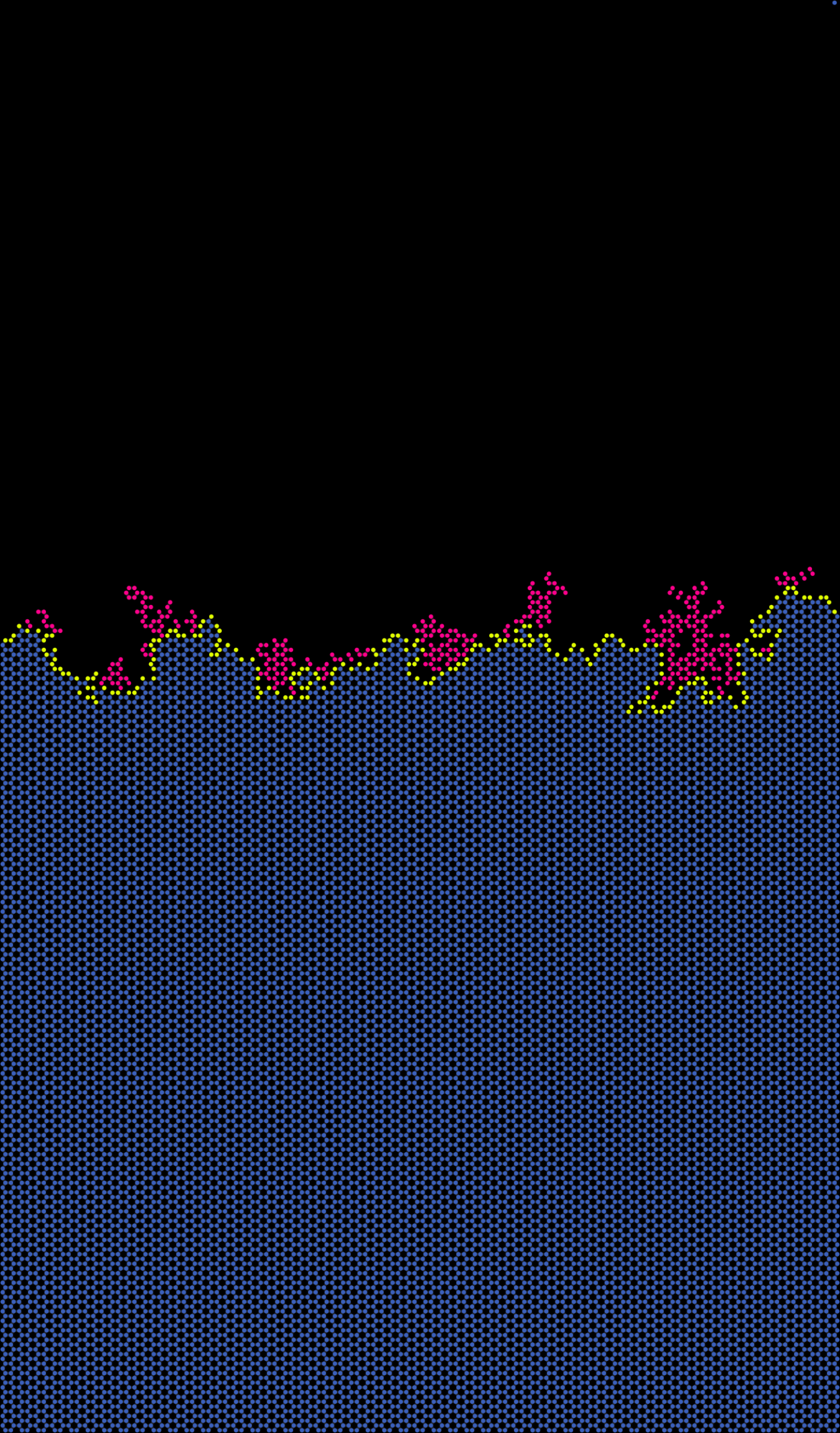}
         \caption{$\beta=20^\circ$}
         
     \end{subfigure}
     \hfill
     \begin{subfigure}[b]{0.24\textwidth}
         \centering
         \includegraphics[width=\textwidth]{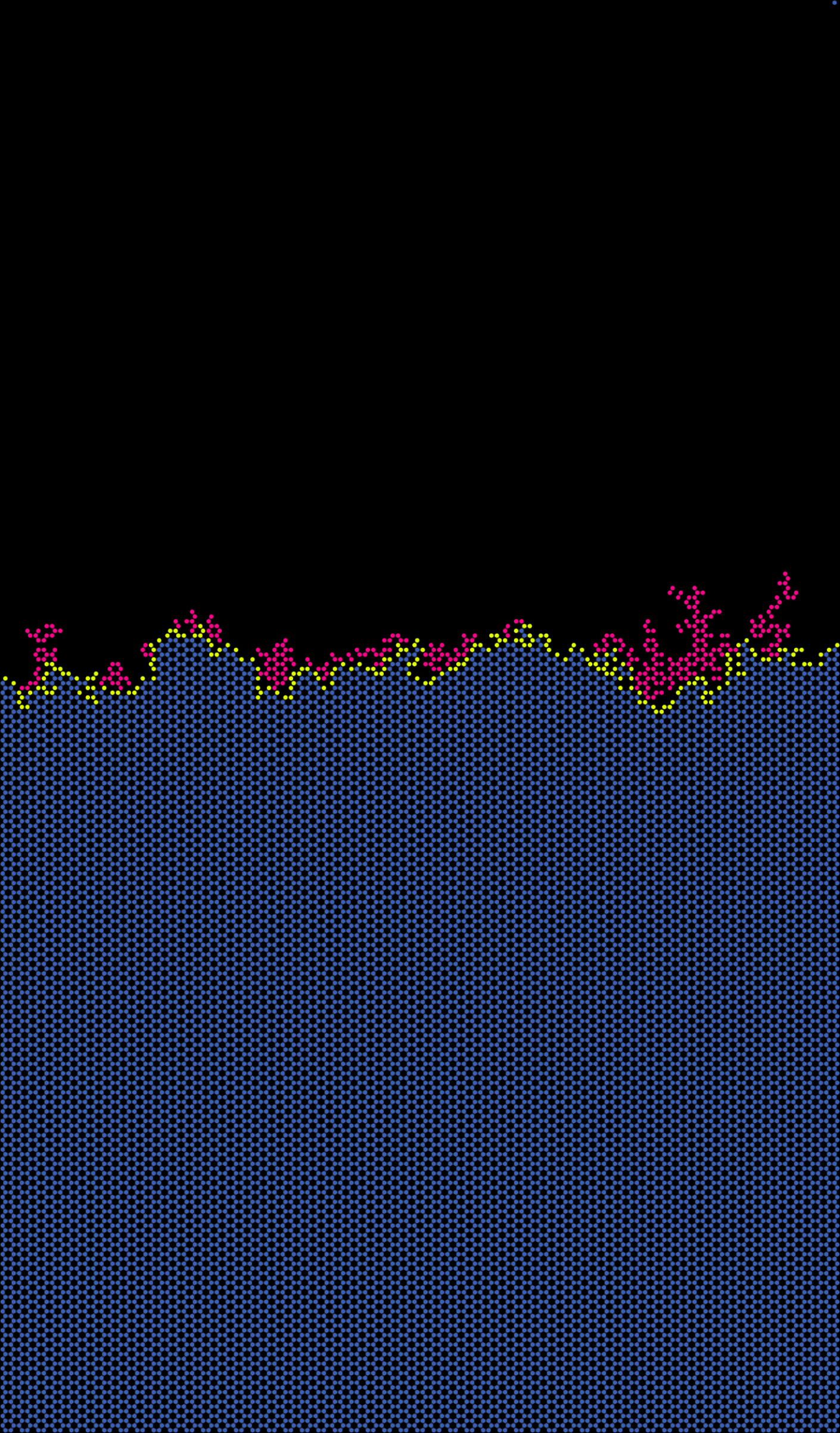}
         \caption{$\beta=30^\circ$}
         
     \end{subfigure}
     \hfill
     \begin{subfigure}[b]{0.24\textwidth}
         \centering
         \includegraphics[width=\textwidth]{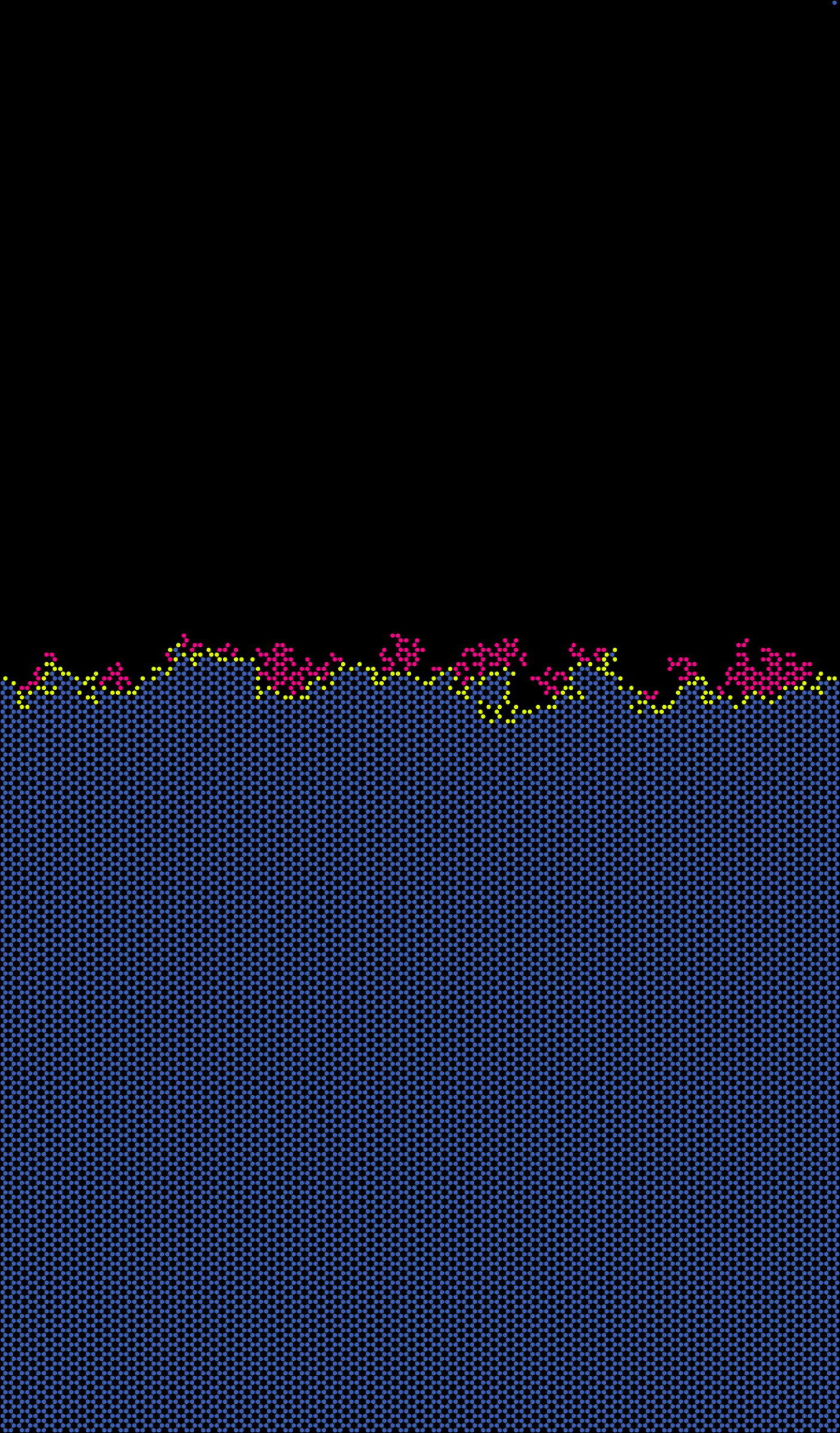}
         \caption{$\beta=60^\circ$}
         
     \end{subfigure}
     
        \caption{Representation of nodes connected to the front via films (pink), nodes at the front (yellow), and nodes inside the main defending cluster (blue) halfway through the drainage simulations.}
        \label{fig:Node_Conn_Front}
\end{figure}

Figure
\ref{fig:Node_Conn_Front} illustrated the nodes connected to the main defending cluster via films, halfway through drainage, under different degrees of influence of gravitational forces. There is a clear indication from these images that the active zone gets progressively more compact as $\beta$ increases. For the cases where the gravitational component is zero or relatively small, $0^\circ\leq\beta\leq 2.5^\circ$, the frequency of large wetting fluid clusters behind the front -- as noticeable in Figures \ref{fig:Inv_Time} and \ref{fig:Inv_Mod} -- lead to very extensive portions of the porous matrix connected to the front via films. As the inclination of the porous matrix becomes steeper, there seems to be a short range of angles where the active zone rapidly shrinks, $2.5^\circ\leq\beta<10^\circ$, followed by a long inclination interval, $10^\circ\leq\beta\leq90^\circ$, where a narrow active zone is curtailed slowly. 

\begin{figure}[h!]
  \centering
\begin{subfigure}[b]{0.49\textwidth}
         \centering
         \includegraphics[width=\textwidth]{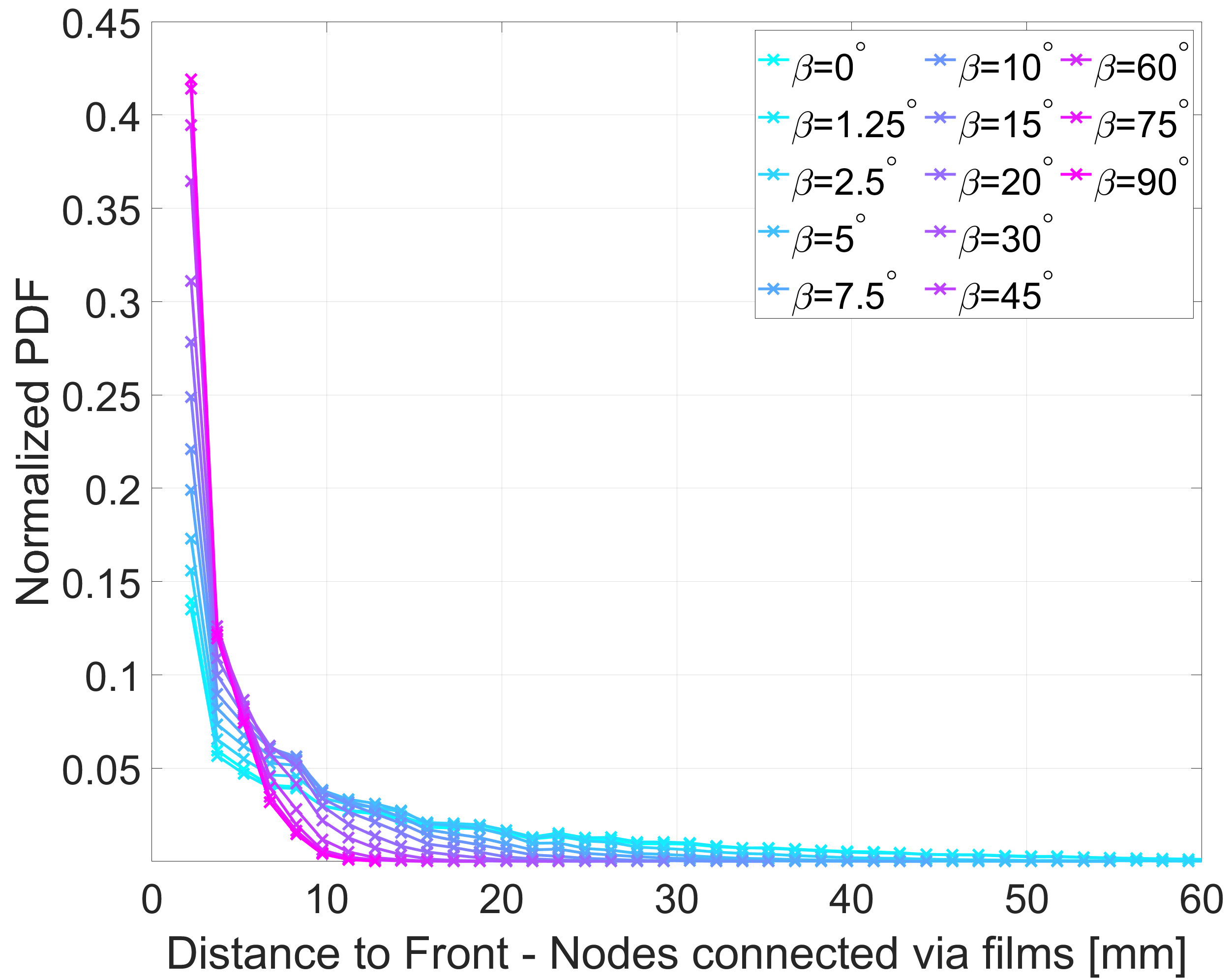}
         \caption{}
         \label{fig:pdf_conn}
     \end{subfigure}
     \hfill
     \begin{subfigure}[b]{0.49\textwidth}
         \centering
         \includegraphics[width=\textwidth]{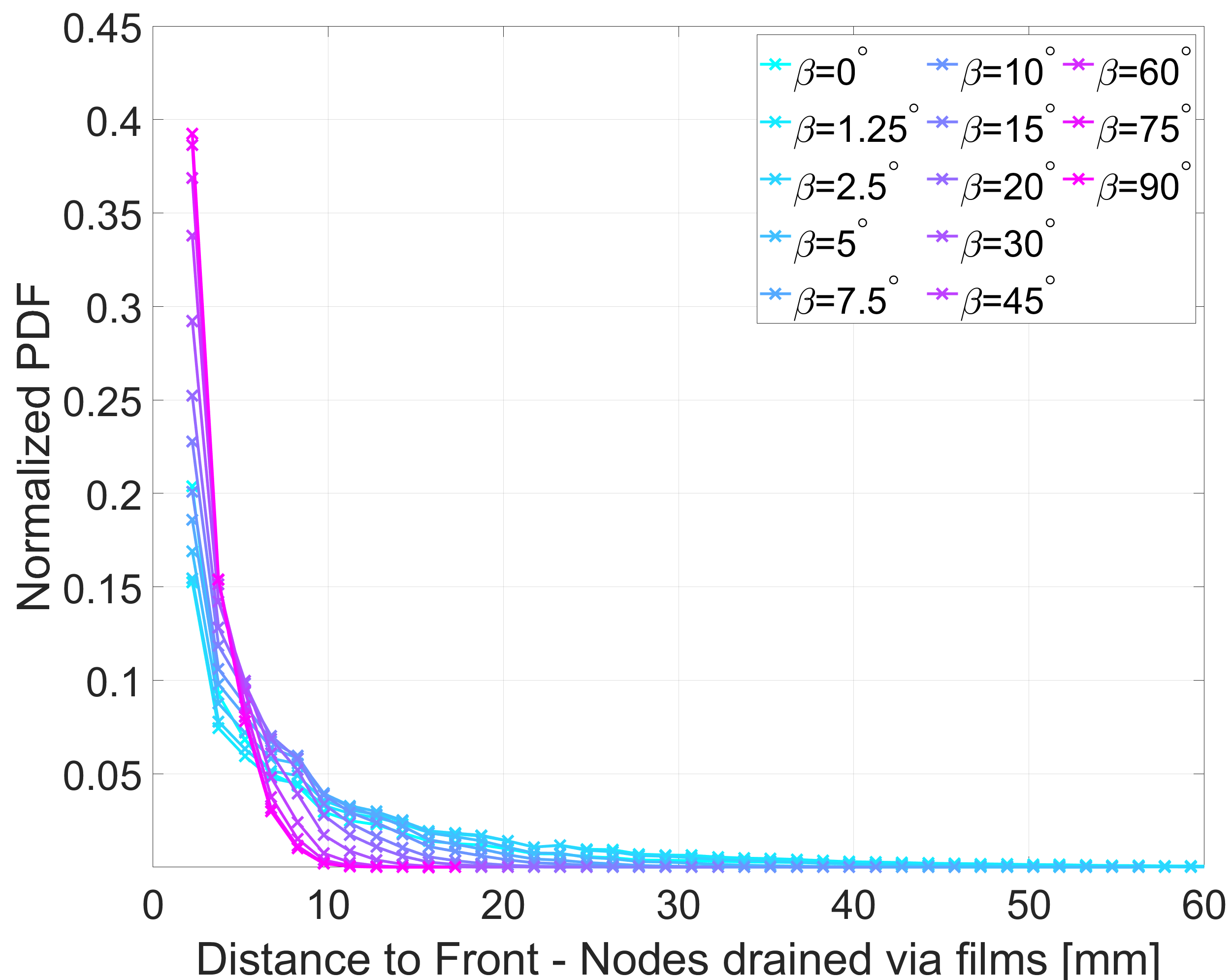}
         \caption{}
         \label{fig:pdf_drained}
     \end{subfigure}

\caption{Normalized PDF of the distance to the front of nodes connected via film flow (a), and of the nodes effectively drained via film flow (b).}
  \label{fig:PDF_DFF}
\end{figure}

The normalized probability density function (PDF) of distance from the front to the nodes connected and drained via film flow is shown in Figure \ref{fig:pdf_conn}. As suggested before, the probability of finding nodes pertaining to the active zone closer to the front increases significantly as gravitational forces become more effective during drainage. The same trend was observed experimentally in \cite{moura2019connectivity} when comparing the PDF of the distance from the invasion front of film flow events in experiments with $\beta=10^\circ, 20^\circ$ and $30^\circ$. Interestingly, the normalized PDFs presented in \ref{fig:pdf_conn} and \ref{fig:pdf_drained} are remarkably similar. This indicates that, among all nodes that could be drained by film flow at a given moment, there is no clear preference that further nodes would be invaded instead of nearer ones. 

\begin{figure}[h!]
  \centering
  \includegraphics[width=0.5\textwidth]{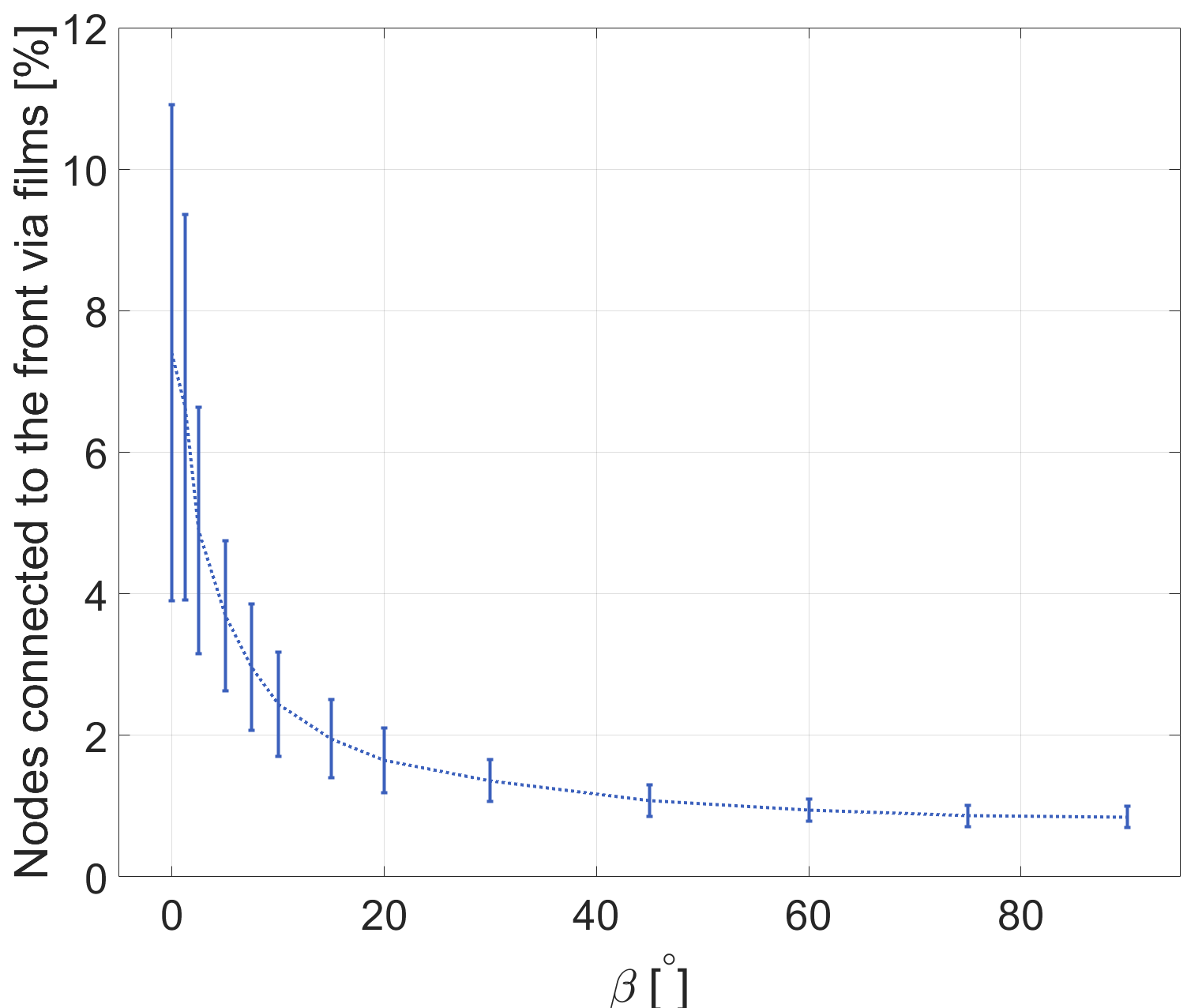}
  \caption{Average fraction of the total number nodes belonging to the lattice connected to the invasion front during the drainage under different levels of influence of gravitational forces}
  \label{fig:frac_conn}
\end{figure}

Another interesting fact is that wider active zones are not necessarily linked to more significant effects of film flow events in drainage. Figure \ref{fig:frac_conn} indicates the average fraction of all nodes belonging to the lattice connected to the main defending cluster via films, at any point during drainage. The joint analysis of data presented in Figures \ref{fig:frac_conn} and \ref{fig:Sat_Mechanims} indicates that the large amount of nodes belonging to the networks through which the secondary drainage mechanism takes place at very low inclination angles do no translate into greater transport of wetting fluid to the main defending cluster.

\begin{figure}[h!]
     \centering
     
     \begin{subfigure}[b]{0.24\textwidth}
         \centering
         \includegraphics[width=\textwidth]{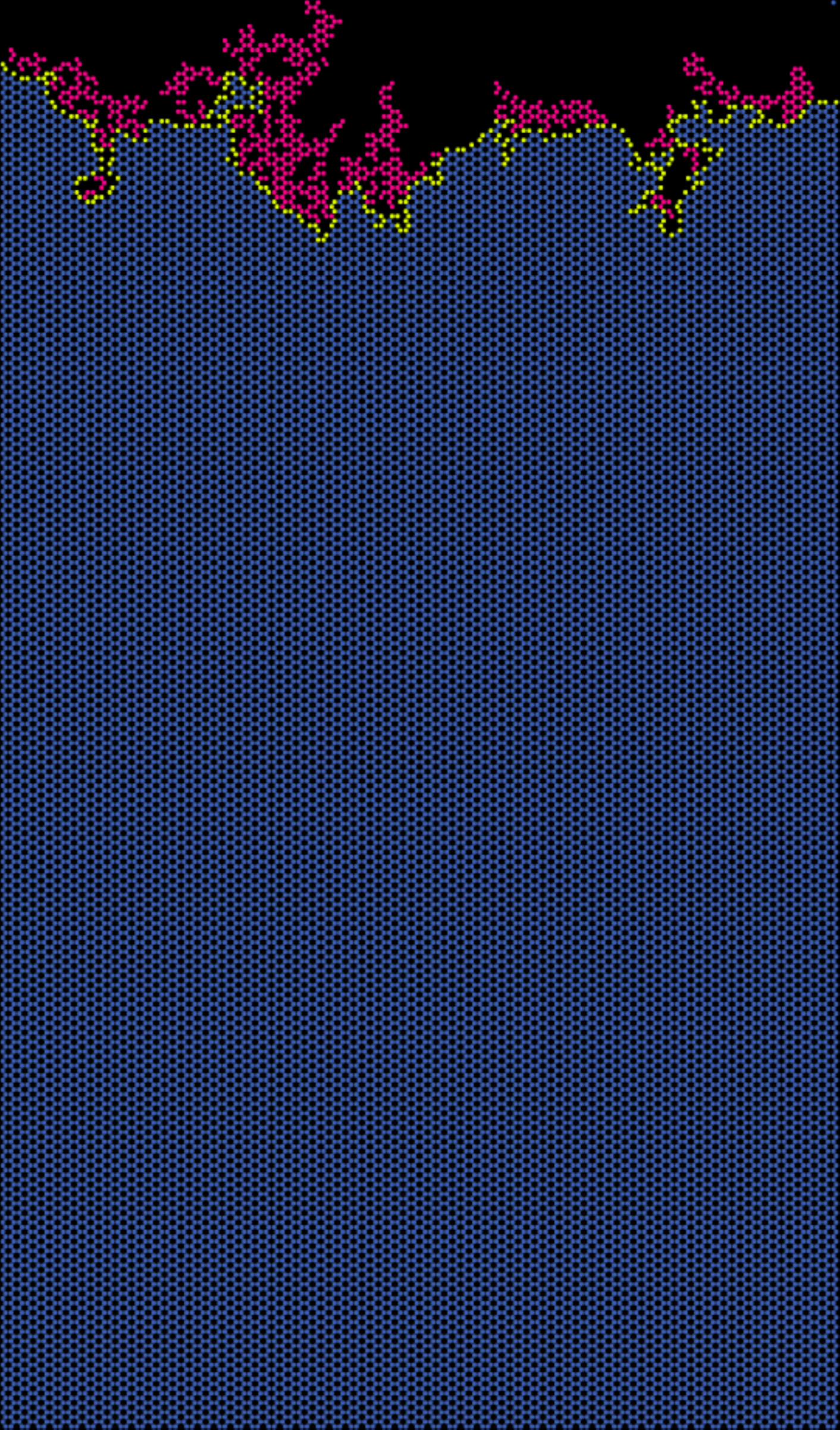}
         \caption{}
     \end{subfigure}
     \hfill
     \begin{subfigure}[b]{0.24\textwidth}
         \centering
         \includegraphics[width=\textwidth]{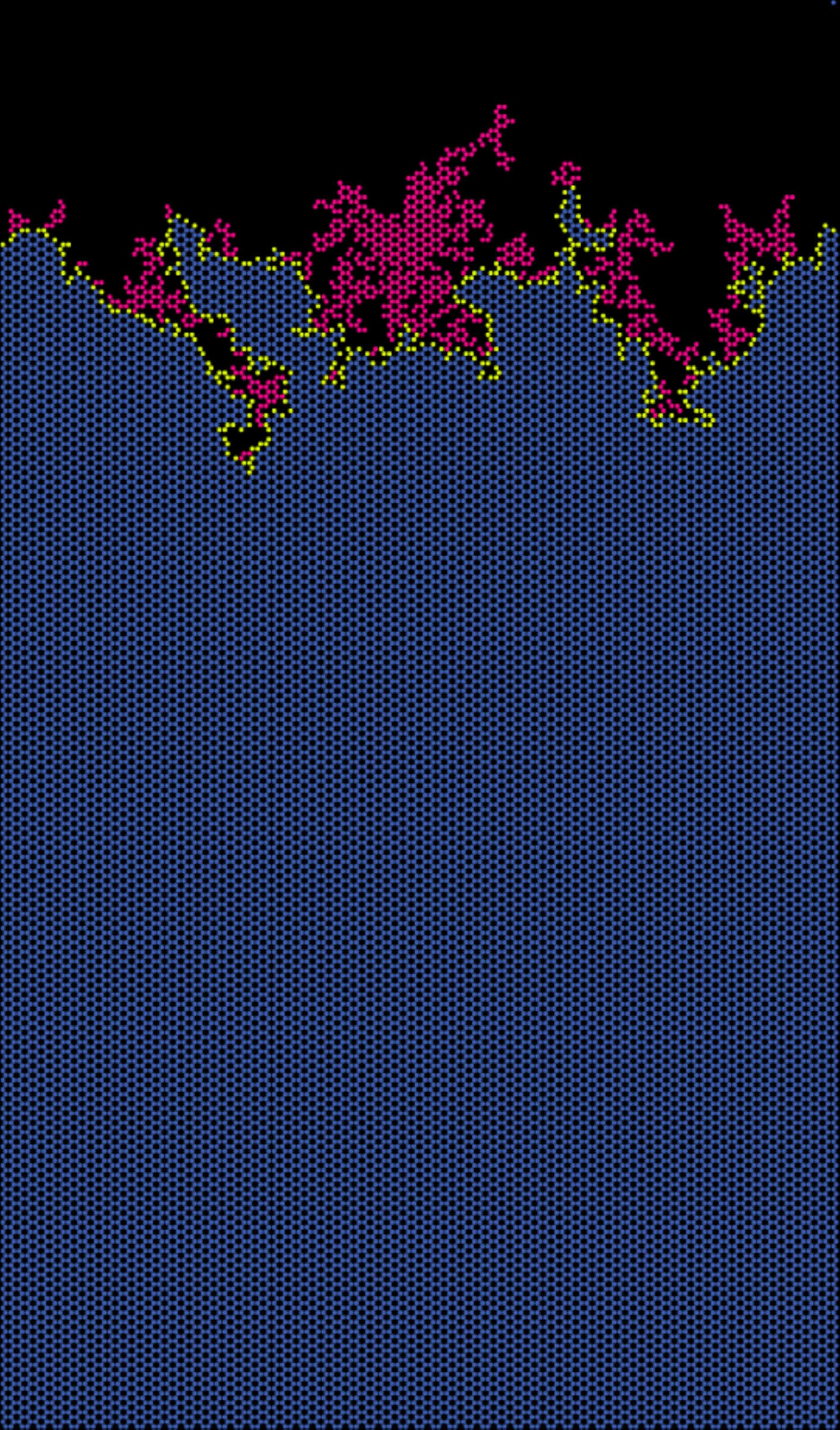}
         \caption{}
     \end{subfigure}
     \hfill
     \begin{subfigure}[b]{0.24\textwidth}
         \centering
         \includegraphics[width=\textwidth]{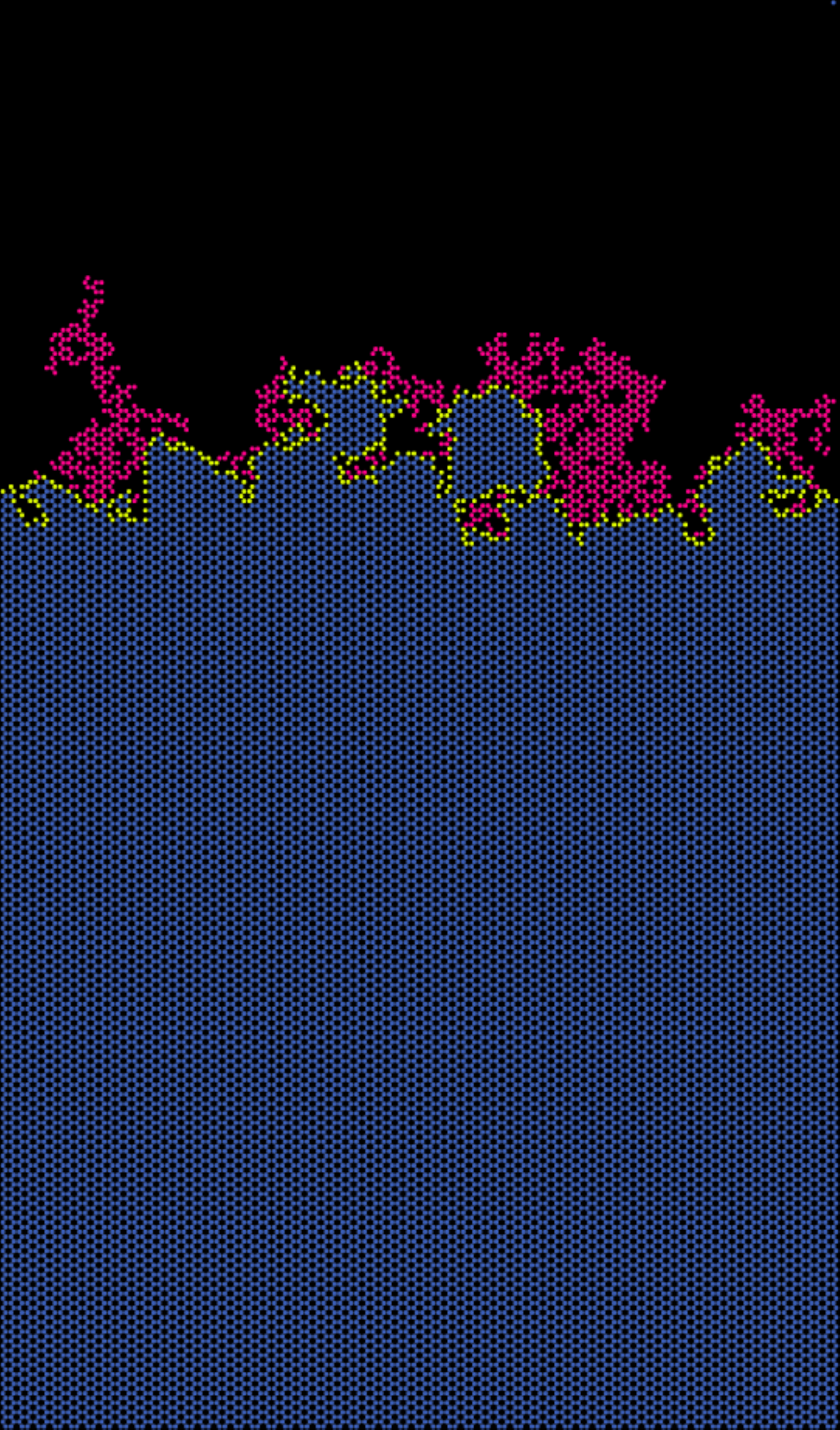}
         \caption{}
     \end{subfigure}
     \hfill
     \begin{subfigure}[b]{0.24\textwidth}
         \centering
         \includegraphics[width=\textwidth]{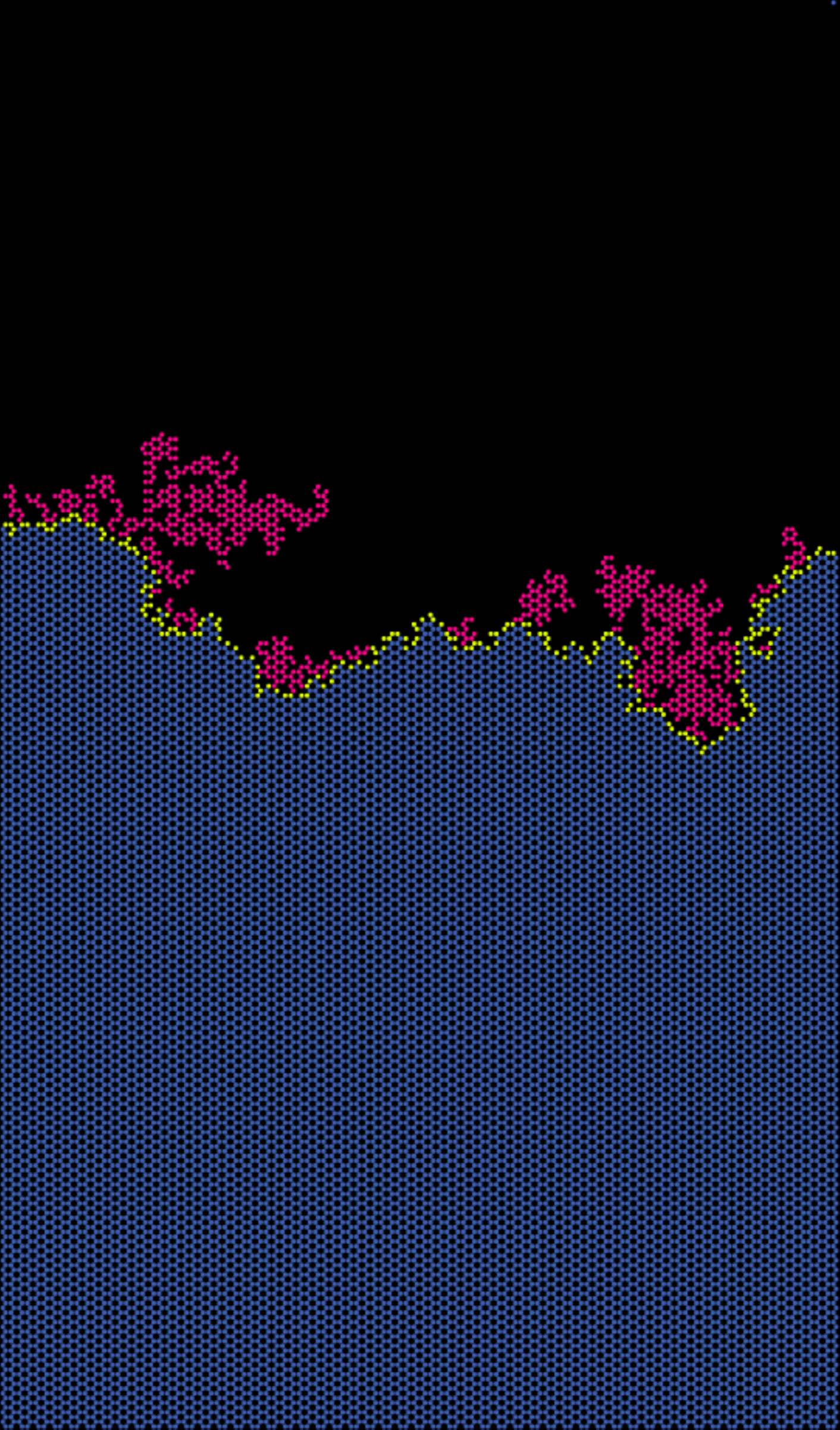}
         \caption{}
         \label{fig:Zoomed_in_Next}
     \end{subfigure}
     \hfill
     \begin{subfigure}[b]{0.24\textwidth}
         \centering
         \includegraphics[width=\textwidth]{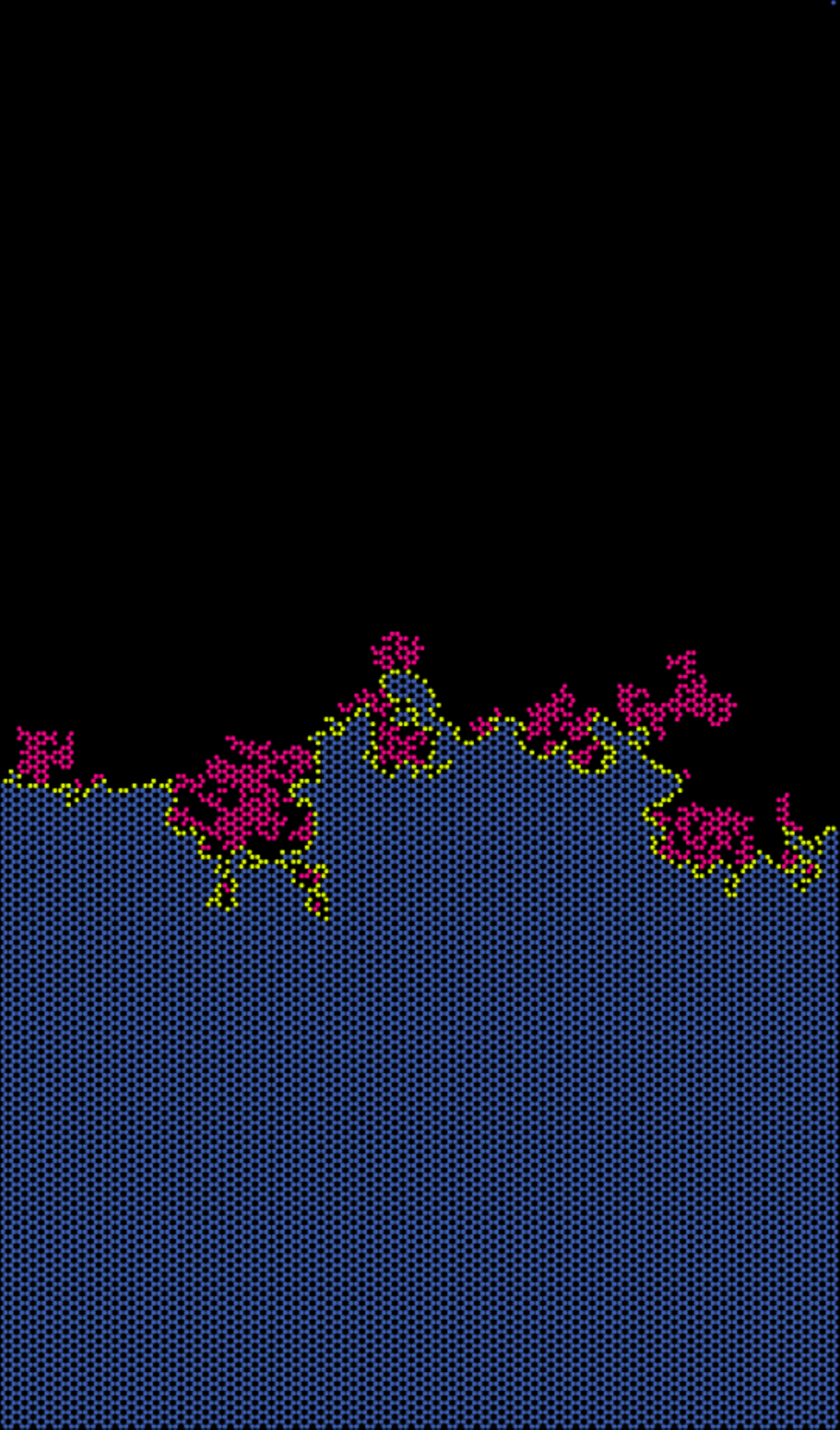}
         \caption{}
     \end{subfigure}
     \hfill
     \begin{subfigure}[b]{0.24\textwidth}
         \centering
         \includegraphics[width=\textwidth]{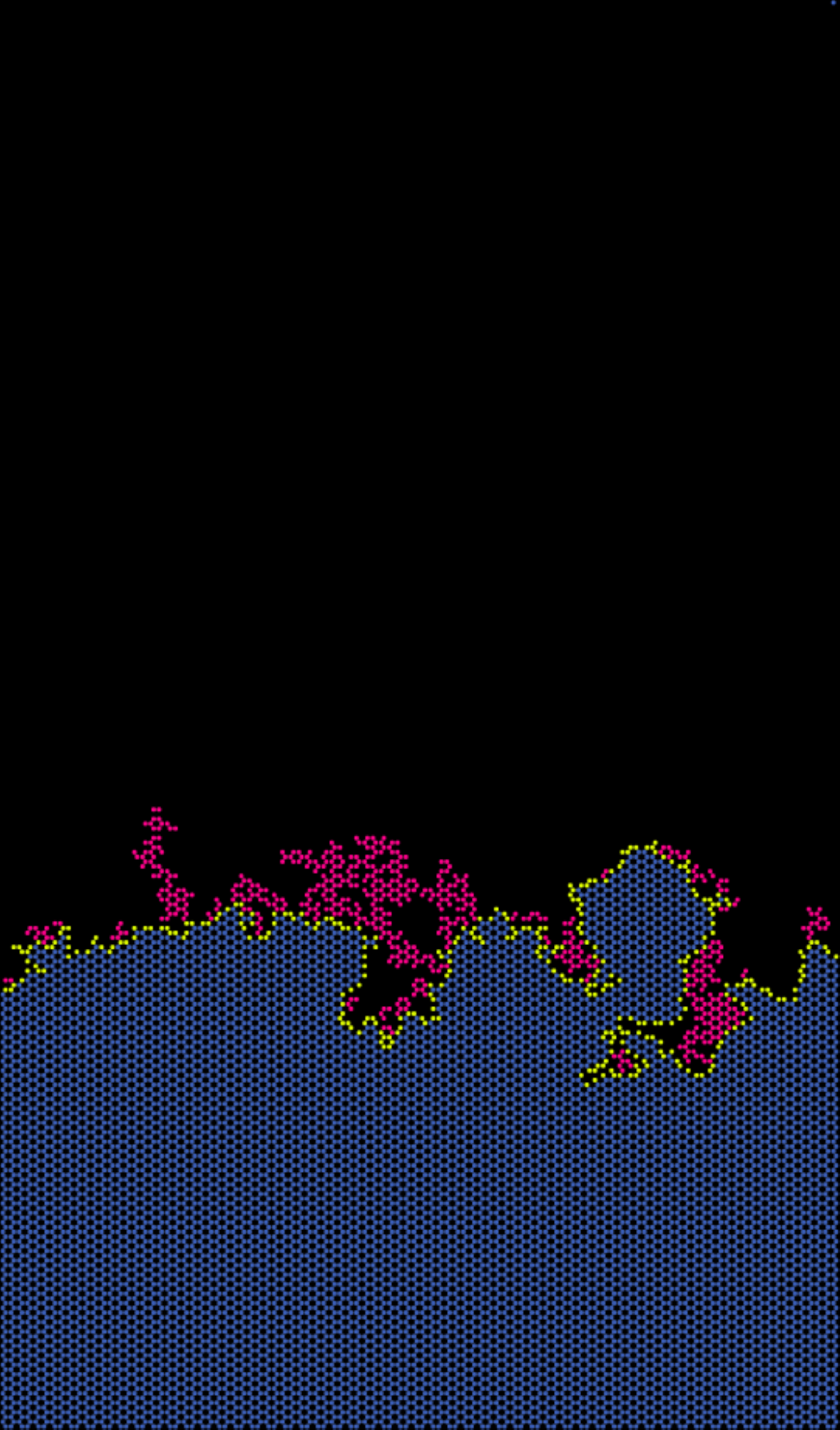}
         \caption{}
     \end{subfigure}
     \hfill
     \begin{subfigure}[b]{0.24\textwidth}
         \centering
         \includegraphics[width=\textwidth]{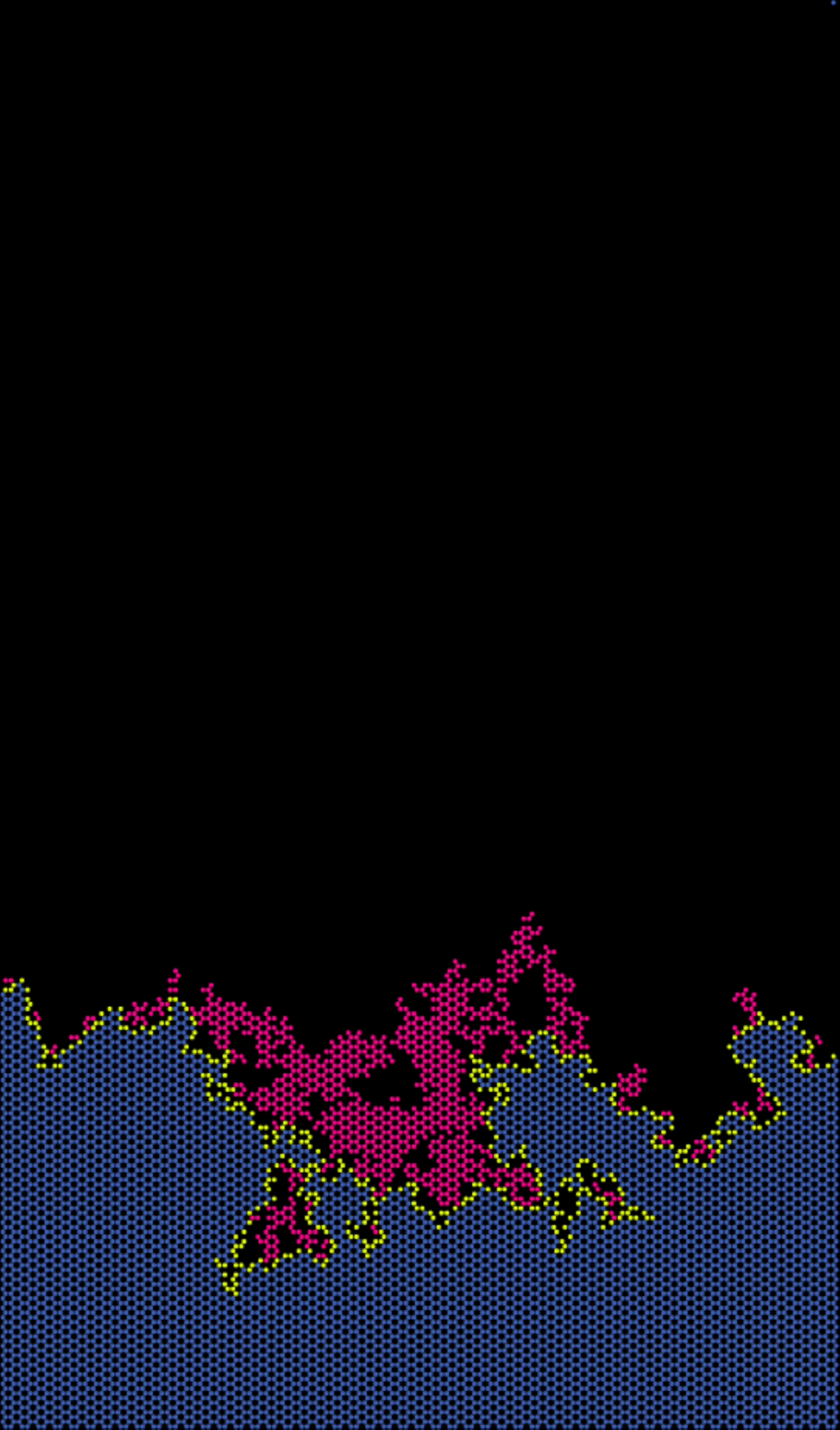}
         \caption{}
     \end{subfigure}
     \hfill
     \begin{subfigure}[b]{0.24\textwidth}
         \centering
         \includegraphics[width=\textwidth]{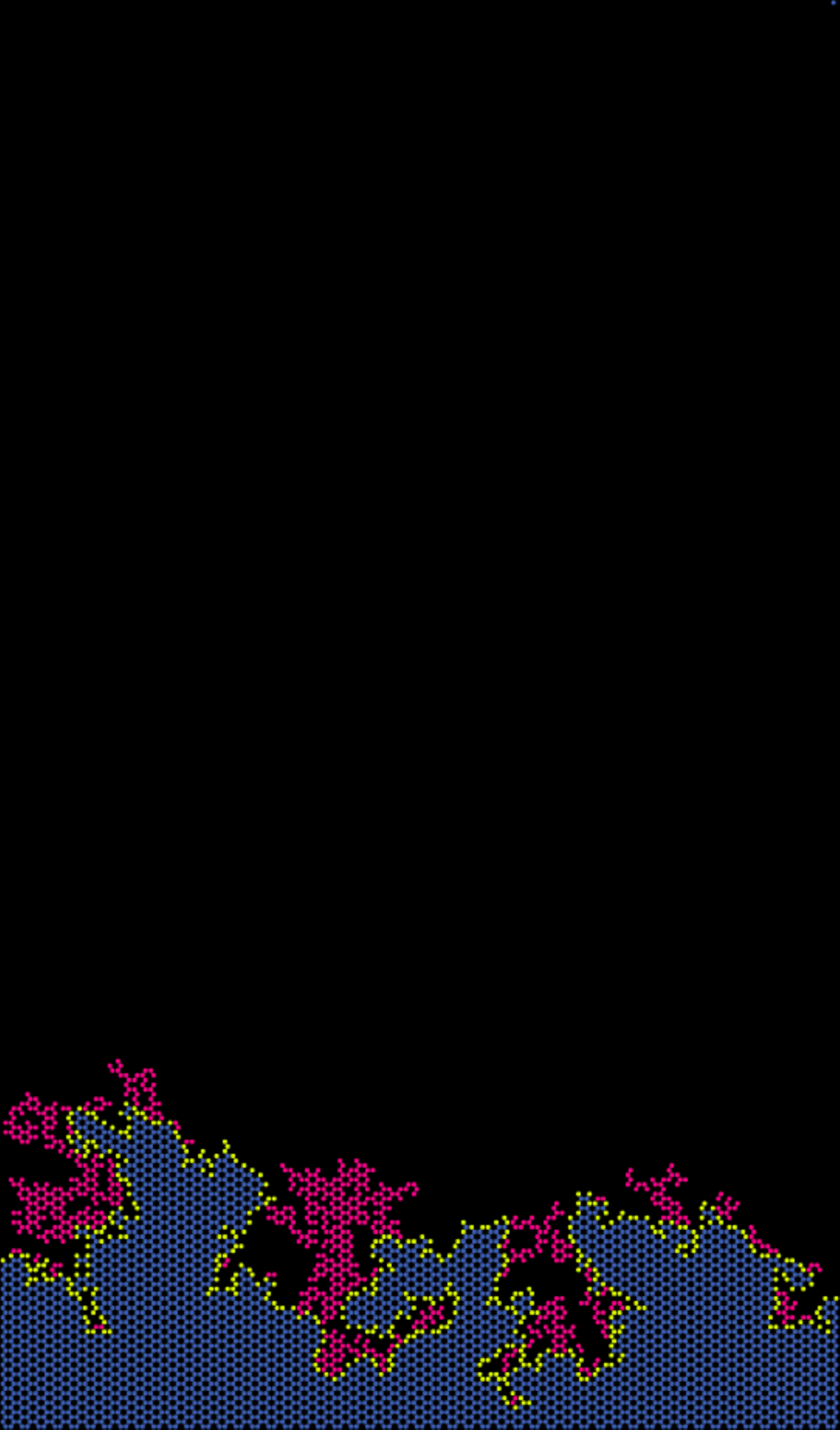}
         \caption{}
     \end{subfigure}
        \caption{Representation of nodes connected to the front via films for $\beta=7.5^\circ$ at six stages during drainage}
        \label{fig:Node_Conn_Front_Seq_7.5}
\end{figure}

The characterization of the active zone using the size of the secondary-drainage-mechanism networks should also take into account the fact that the topology of such networks is under constant evolution as the porous matrix is drained. This phenomenon was acknowledged in \cite{moura2019connectivity}, when the authors described the networks associated with primary drainage as "static" -- formed by the lattice of pores and throats representing the porous space --, and the ones associated with film flow events as "dynamic" -- formed by unsteady subsets of the lattice following the invasion front. The extent of the fluctuation in the dynamic networks' sizes during drainage can be inferred by the error bars in Figure \ref{fig:frac_conn}, and an insight into their topological variations is presented in Figure \ref{fig:Node_Conn_Front_Seq_7.5}

\begin{figure}[h!]
  \centering
  \includegraphics[width=\textwidth]{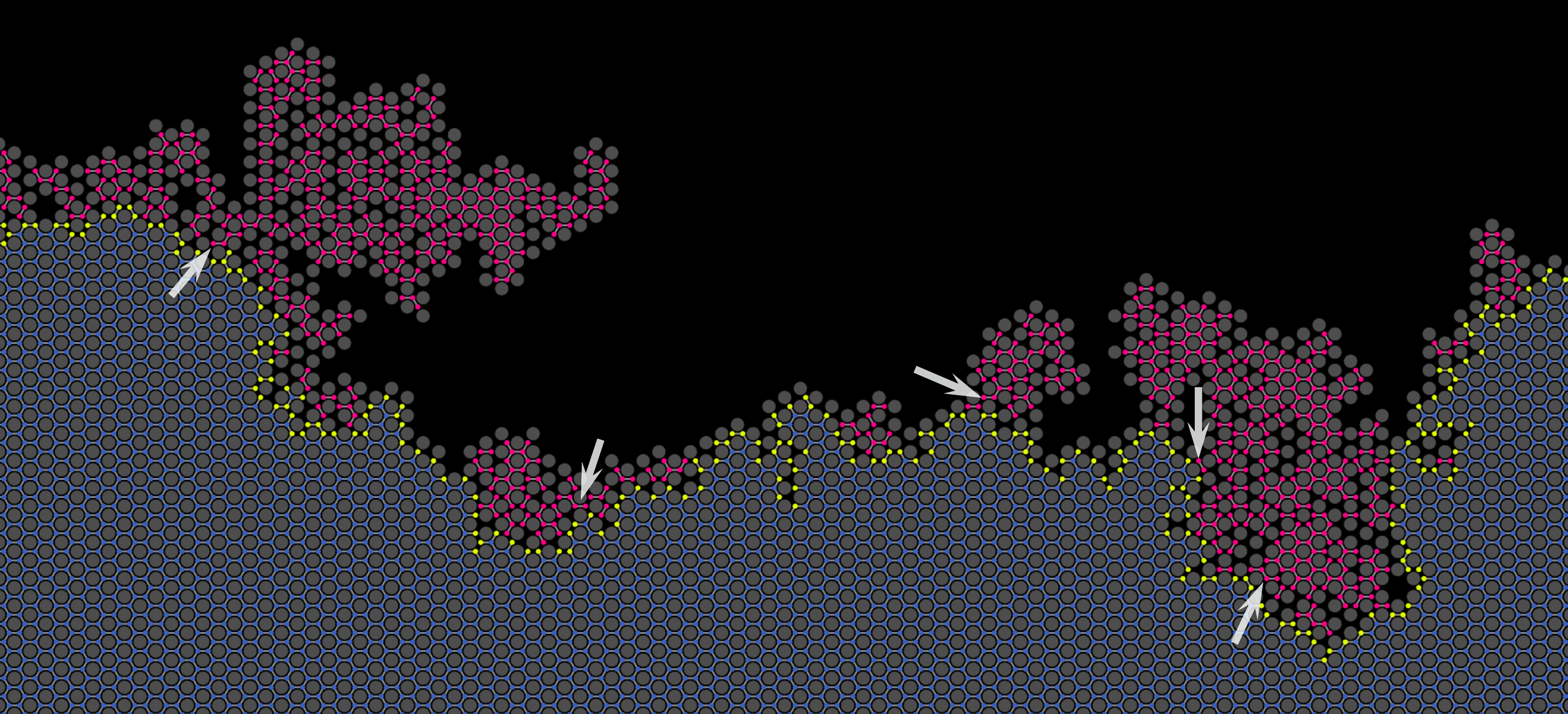}
  \caption{Detail of the secondary drainage network shown in \ref{fig:Zoomed_in_Next}. Illustration of the spherical beads surrounded by the networks is included for clarity.}
  \label{fig:SecNet_detail}
\end{figure}

This figure illustrates the secondary drainage networks at eight equidistant stages of a drainage simulation with $\beta=7.5^\circ$. We can observe that even though their width is somewhat constant, their topology and size can vary significantly. A reason pointed out in \citet{moura2019connectivity} for the frequent substantial transformations in these dynamic networks is the potentially global effect of single capillary bridges in the wetting-phase connectivity. As shown in detail in Figure \ref{fig:SecNet_detail}, which represents the region around the invasion front in Figure \ref{fig:Zoomed_in_Next}, the five capillary bridges indicated by white arrows are responsible for most of the communication between the dynamic film-flow networks with the main defending cluster. Any drainage event in the throats hosting these bridges, or in their immediate vicinity, could lead to instant disconnection of large portions of the secondary drainage networks. In order to expand our understanding of the likelihood of film flow events in a given porous matrix, an analysis of the capillary bridge frequency in the used lattices, as well as the probability of their occurrence given a throat's size is presented in the next Section.

\subsection{The frequency and size distribution of capillary bridges}

\begin{figure}[h!]
  \centering
\begin{subfigure}[b]{0.485\textwidth}
         \centering
         \includegraphics[width=\textwidth]{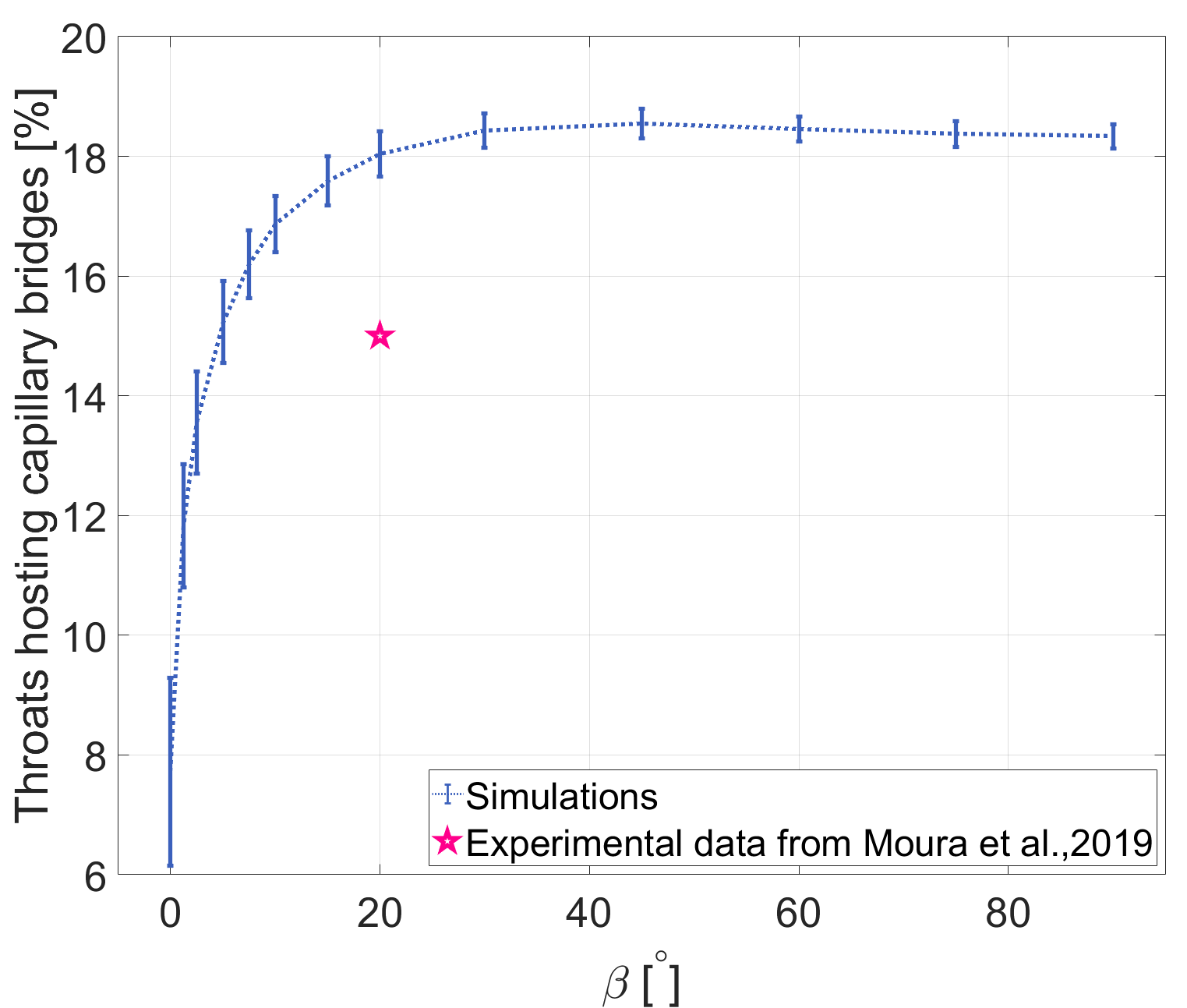}
         \caption{}
         \label{fig:Throat_CB_frac}
     \end{subfigure}
     \hfill
     \begin{subfigure}[b]{0.505\textwidth}
         \centering
         \includegraphics[width=\textwidth]{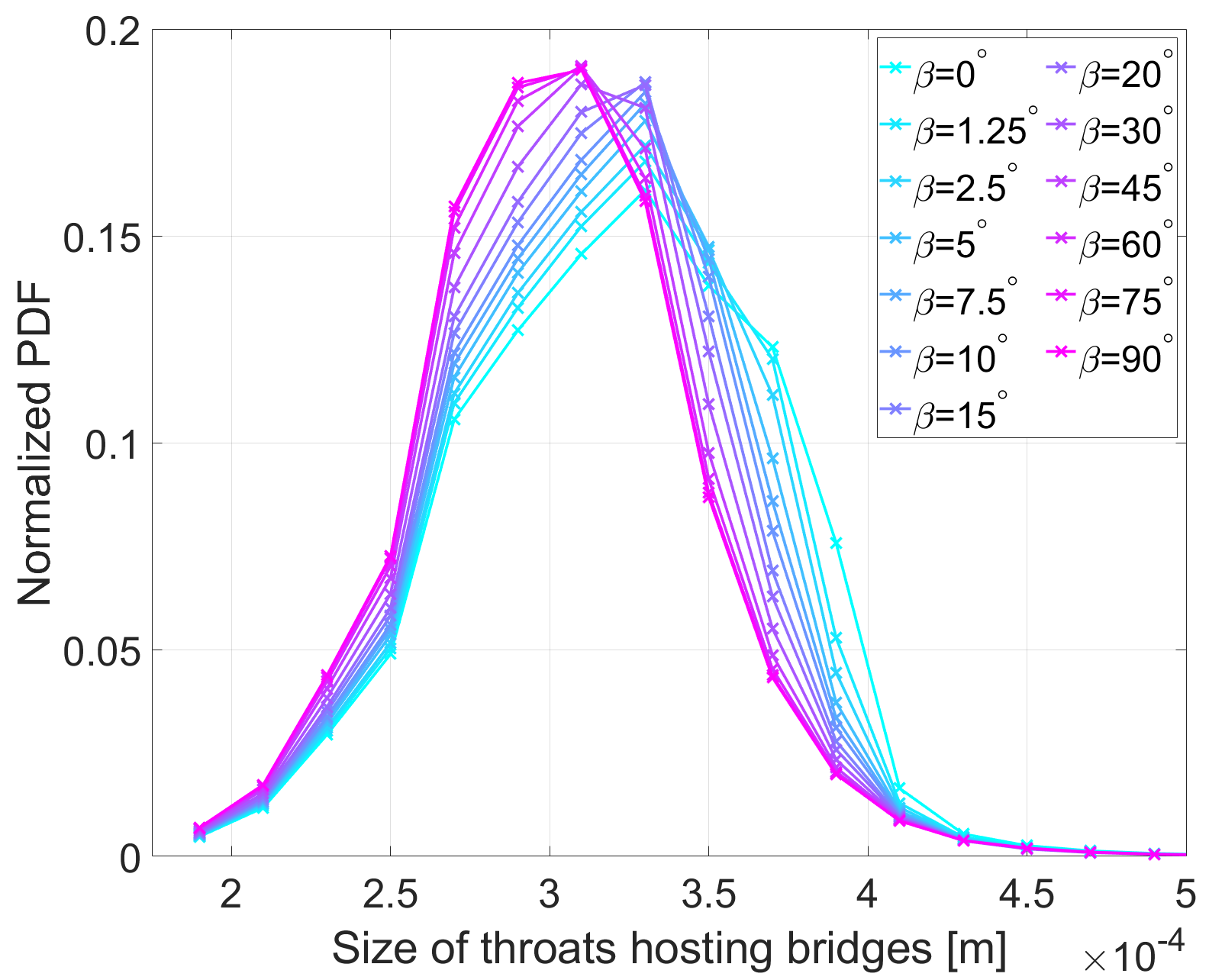}
         \caption{}
         \label{fig:PDF_size_throat_cb}
     \end{subfigure}
  \label{fig:Freq_and_Size_Throat_CB}
  \caption{Frequency of throats containing capillary bridges (a) and their size distributions (b).}
\end{figure}

Given the central role of capillary bridges in film flow events in granular porous media, in this section we examine how likely it is for a capillary bridge to be formed in the quasi-2D porous matrices under investigation, and where they are mostly expected to occur. Figure \ref{fig:Throat_CB_frac} shows the fraction of throats hosting capillary bridges, among all throats in the lattices, at the moment of the invading fluid breakthrough. We can see that from $\beta=0^\circ$ to $\beta=20^\circ$ there is a sharp increase in the occurrence of bridges, which henceforth stabilizes at approximately $18.5\%$. The comparison of these results with the ones presented in Figures \ref{fig:Sat_Mechanims} and \ref{fig:frac_conn} indicates that the pervasiveness of capillary bridges alone cannot be correlated with a greater capacity of connecting, nor draining, extensive regions in the lattice to the main defending cluster. In fact, the increase in formation of capillary bridges at higher inclination angles can be linked to the enhanced displacement of wetting-fluid via the primary drainage mechanism, leaving smaller -- and most likely disconnected -- clusters behind the front. An increase in the frequency of capillary bridges as inclination angles get steeper was also observed experimentally by \citet{moura2019connectivity}, when comparing cases with $\beta=20^\circ$ and $30^\circ$. In that study, a fraction of $15\%$ of throats occupied by bridges at $\beta=20^\circ$ was also reported, which is approximately $17\%$ lower than the value predicted by the model.

Complementary to the information presented in Figure \ref{fig:Throat_CB_frac}, Figure \ref{fig:PDF_size_throat_cb} presents the PDFs of the sizes of throats ($t_w$) in which capillary bridges were formed during drainage at different inclination angles. In this graph we can see that, even though the porous matrices comprised throats with sizes up to 1.5 mm, all the bridges were found in throats smaller than 0.45 mm. It is also noticeable that the same range of throat sizes hosting capillary bridges was found for all tested $\beta$ values. As the inclination angles get steeper, however, there is a tendency of the bridges to concentrate at the lower end of this range. 

\begin{figure}[h!]
  \centering
  \includegraphics[width=0.75\textwidth]{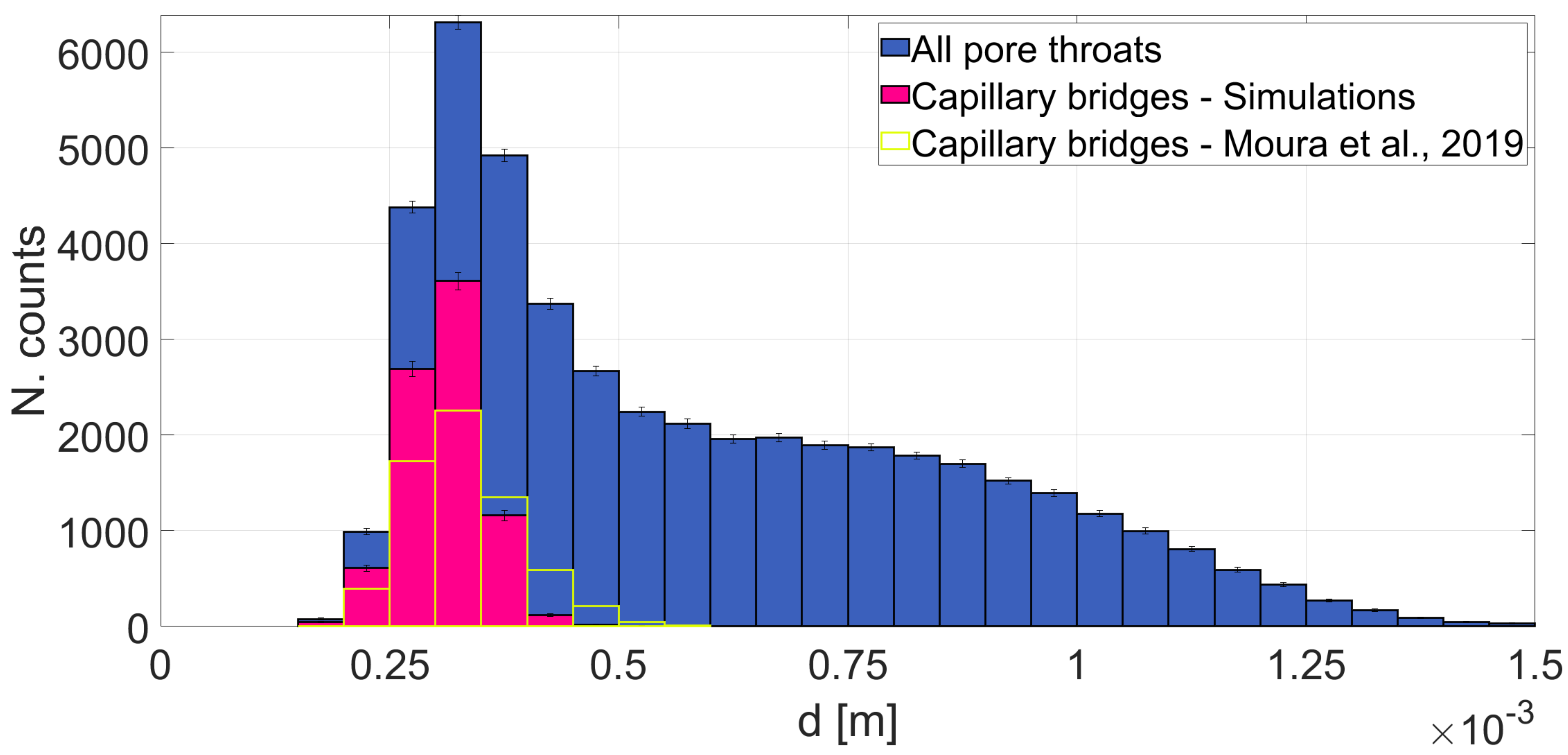}
  \caption{Probability density function of the sizes of throats containing capillary bridges experimentally and with the proposed model, for $\beta=20^\circ$.}
  \label{fig:PDF_bridges_20}
\end{figure}

This heterogeneous occupation of throats by bridges was also identified experimentally in \cite{moura2019connectivity}, as shown in Figure \ref{fig:PDF_bridges_20} (where the number of counts of the experimental results were rescaled to match the size of the lattices used in the simulations). For the drainage experiment conducted at $\beta=20^\circ$, the reported maximum size of throat hosting a bridge was approximately 0.5mm, only $10\%$ larger than the size predicted by the model. In that study, this maximum size was justified as a result of two main factors: the Plateau-Rayleigh instability in the bridges -- which would lead to the snap-off of bridges accommodated between spheres set at wider separations -- and the nature of the drainage process -- which leads to the invasion of larger throats during the primary drainage mechanism. Given the low frequency of snap-off events reported experimentally, and the fact that our model does not incorporate a criterion for capillary bridge snap-off due to hydrodynamic instabilities, we have a strong indication that the nature of the drainage process is the predominant factor associated with the size of throats expected to host bridges.

\section{Conclusions}

In this study, a simple quasi-static pore-network model for slow drainage in granular porous media was presented. The model was based on the bond-invasion-percolation method, and incorporated a modification in the trapped-cluster-identification algorithm in order to acknowledge  wetting-phase connectivity provided by chains of capillary bridges. Porous media were represented in the model as regular honeycomb lattices of nodes and edges, symbolizing the pores and throats found in Hele-Shaw cells filled with a single layer of monodispersed spherical beads. This choice of idealized quasi-2D porous media allowed us to clearly characterize drainage events due to film flow, and the obtained insights about the phenomenon can mostly be generalized to drainage in 3D granular materials.

With the proposed model, the prevalence of primary and secondary drainage mechanisms under different levels of influence of gravitational forces was investigated, and the results were compared with experimental data from \citet{moura2019connectivity}. Despite the model's simple representation of the porous space and drainage dynamics, qualitative agreement with experimental data was obtained in the analyses regarding the impact of film flow on residual saturations, the establishment of a film flow active zone and the occurrence of capillary bridges in the medium. With the demonstrated ability to represent fundamental aspects of complex physical phenomena, the proposed model could work as a complementary tool to experiments, in the quest for a comprehensive understanding of drainage in granular porous media.


\section*{References}
\bibliography{paper}

\begin{thebibliography}{60}
\providecommand{\natexlab}[1]{#1}
\providecommand{\url}[1]{\texttt{#1}}
\expandafter\ifx\csname urlstyle\endcsname\relax
  \providecommand{\doi}[1]{doi: #1}\else
  \providecommand{\doi}{doi: \begingroup \urlstyle{rm}\Url}\fi

\bibitem[Blunt et~al.(2013)Blunt, Bijeljic, Dong, Gharbi, Iglauer, Mostaghimi,
  Paluszny, and Pentland]{blunt2013pore}
Martin~J Blunt, Branko Bijeljic, Hu~Dong, Oussama Gharbi, Stefan Iglauer,
  Peyman Mostaghimi, Adriana Paluszny, and Christopher Pentland.
\newblock Pore-scale imaging and modelling.
\newblock \emph{Advances in Water resources}, 51:\penalty0 197--216, 2013.

\bibitem[Bultreys et~al.(2016)Bultreys, De~Boever, and
  Cnudde]{bultreys2016imaging}
Tom Bultreys, Wesley De~Boever, and Veerle Cnudde.
\newblock Imaging and image-based fluid transport modeling at the pore scale in
  geological materials: A practical introduction to the current
  state-of-the-art.
\newblock \emph{Earth-Science Reviews}, 155:\penalty0 93--128, 2016.

\bibitem[Golparvar et~al.(2018)Golparvar, Zhou, Wu, Ma, and Yu]{Golparvar2018}
Amir Golparvar, Yingfang Zhou, Kejian Wu, Jingsheng Ma, and Zhixin Yu.
\newblock A comprehensive review of pore scale modeling methodologies for
  multiphase flow in porous media.
\newblock \emph{Advances in Geo-Energy Research}, 2\penalty0 (4):\penalty0
  418--440, 2018.

\bibitem[Zhao et~al.(2019)Zhao, MacMinn, Primkulov, Chen, Valocchi, Zhao, Kang,
  Bruning, McClure, Miller, et~al.]{zhao2019comprehensive}
Benzhong Zhao, Christopher~W MacMinn, Bauyrzhan~K Primkulov, Yu~Chen, Albert~J
  Valocchi, Jianlin Zhao, Qinjun Kang, Kelsey Bruning, James~E McClure, Cass~T
  Miller, et~al.
\newblock Comprehensive comparison of pore-scale models for multiphase flow in
  porous media.
\newblock \emph{Proceedings of the National Academy of Sciences}, 116\penalty0
  (28):\penalty0 13799--13806, 2019.

\bibitem[Chen et~al.(2022)Chen, He, Zhao, Kang, Li, Carmeliet, Shikazono, and
  Tao]{Chen2022pore}
Li~Chen, An~He, Jianlin Zhao, Qinjun Kang, Zeng-Yao Li, Jan Carmeliet, Naoki
  Shikazono, and Wen-Quan Tao.
\newblock Pore-scale modeling of complex transport phenomena in porous media.
\newblock \emph{Progress in Energy and Combustion Science}, 88:\penalty0
  100968, 2022.
\newblock ISSN 0360-1285.

\bibitem[Gao et~al.(2017)Gao, Xing, Tian, Pearce, Sedek, Golding, and
  Rudolph]{gao2017reactive}
Jinfang Gao, Huilin Xing, Zhiwei Tian, Julie~K Pearce, Mohamed Sedek, Suzanne~D
  Golding, and Victor Rudolph.
\newblock Reactive transport in porous media for co2 sequestration: Pore scale
  modeling using the lattice boltzmann method.
\newblock \emph{Computers \& Geosciences}, 98:\penalty0 9--20, 2017.

\bibitem[Tahmasebi et~al.(2017)Tahmasebi, Sahimi, Kohanpur, and
  Valocchi]{tahmasebi2017pore}
Pejman Tahmasebi, Muhammad Sahimi, Amir~H Kohanpur, and Albert Valocchi.
\newblock Pore-scale simulation of flow of co2 and brine in reconstructed and
  actual 3d rock cores.
\newblock \emph{Journal of Petroleum Science and Engineering}, 155:\penalty0
  21--33, 2017.

\bibitem[Basirat et~al.(2017)Basirat, Yang, and Niemi]{Basirat2017pore}
Farzad Basirat, Zhibing Yang, and Auli Niemi.
\newblock Pore-scale modeling of wettability effects on co2–brine
  displacement during geological storage.
\newblock \emph{Advances in Water Resources}, 109:\penalty0 181--195, 2017.
\newblock ISSN 0309-1708.

\bibitem[Masoudi et~al.(2021)Masoudi, Fazeli, Miri, and
  Hellevang]{masoudi2021pore}
Mohammad Masoudi, Hossein Fazeli, Rohaldin Miri, and Helge Hellevang.
\newblock Pore scale modeling and evaluation of clogging behavior of salt
  crystal aggregates in co2-rich phase during carbon storage.
\newblock \emph{International Journal of Greenhouse Gas Control}, 111:\penalty0
  103475, 2021.

\bibitem[Payton et~al.(2022)Payton, Sun, Chiarella, and
  Kingdon]{payton2022pore}
Ryan~L Payton, Yizhuo Sun, Domenico Chiarella, and Andrew Kingdon.
\newblock Pore scale numerical modelling of geological carbon storage through
  mineral trapping using true pore geometries.
\newblock \emph{Transport in Porous Media}, 141\penalty0 (3):\penalty0
  667--693, 2022.

\bibitem[Zhao et~al.(2010)Zhao, Blunt, and Yao]{zhao2010pore}
Xiucai Zhao, Martin~J Blunt, and Jun Yao.
\newblock Pore-scale modeling: Effects of wettability on waterflood oil
  recovery.
\newblock \emph{Journal of Petroleum Science and Engineering}, 71\penalty0
  (3-4):\penalty0 169--178, 2010.

\bibitem[Kallel et~al.(2017)Kallel, Van~Dijke, Sorbie, and
  Wood]{kallel2017pore}
Wissem Kallel, Marinus Izaak~Jan Van~Dijke, Kenneth~Stuart Sorbie, and Rachel
  Wood.
\newblock Pore-scale modeling of wettability alteration during primary
  drainage.
\newblock \emph{Water Resources Research}, 53\penalty0 (3):\penalty0
  1891--1907, 2017.

\bibitem[Zhu et~al.(2017)Zhu, Yao, Li, Sun, and Zhang]{zhu2017pore}
Guangpu Zhu, Jun Yao, Aifen Li, Hai Sun, and Lei Zhang.
\newblock Pore-scale investigation of carbon dioxide-enhanced oil recovery.
\newblock \emph{Energy \& Fuels}, 31\penalty0 (5):\penalty0 5324--5332, 2017.

\bibitem[Raeini et~al.(2018)Raeini, Bijeljic, and Blunt]{raeini2018generalized}
Ali~Q Raeini, Branko Bijeljic, and Martin~J Blunt.
\newblock Generalized network modeling of capillary-dominated two-phase flow.
\newblock \emph{Physical Review E}, 97\penalty0 (2):\penalty0 023308, 2018.

\bibitem[Su et~al.(2018)Su, Wang, Gu, Zhang, and Chen]{su2018advances}
Junwei Su, Le~Wang, Zhaolin Gu, Yunwei Zhang, and Chungang Chen.
\newblock Advances in pore-scale simulation of oil reservoirs.
\newblock \emph{Energies}, 11\penalty0 (5):\penalty0 1132, 2018.

\bibitem[Prat(2002)]{prat2002recent}
Marc Prat.
\newblock Recent advances in pore-scale models for drying of porous media.
\newblock \emph{Chemical engineering journal}, 86\penalty0 (1-2):\penalty0
  153--164, 2002.

\bibitem[Surasani et~al.(2008)Surasani, Metzger, and
  Tsotsas]{surasani2008influence}
VK~Surasani, T~Metzger, and E~Tsotsas.
\newblock Influence of heating mode on drying behavior of capillary porous
  media: Pore scale modeling.
\newblock \emph{Chemical engineering science}, 63\penalty0 (21):\penalty0
  5218--5228, 2008.

\bibitem[Panda et~al.(2022)Panda, Bhaskaran, Paliwal, Kharaghani, Tsotsas, and
  Surasani]{panda2022pore}
Debashis Panda, Supriya Bhaskaran, Shubhani Paliwal, Abdolreza Kharaghani,
  Evangelos Tsotsas, and Vikranth~Kumar Surasani.
\newblock Pore-scale physics of drying porous media revealed by lattice
  boltzmann simulations.
\newblock \emph{Drying Technology}, 40\penalty0 (6):\penalty0 1114--1129, 2022.

\bibitem[Fei et~al.(2022)Fei, Qin, Zhao, Derome, and Carmeliet]{fei2022pore}
Linlin Fei, Feifei Qin, Jianlin Zhao, Dominique Derome, and Jan Carmeliet.
\newblock Pore-scale study on convective drying of porous media.
\newblock \emph{Langmuir}, 38\penalty0 (19):\penalty0 6023--6035, 2022.

\bibitem[Zhao et~al.(2022)Zhao, Qin, Kang, Derome, and Carmeliet]{zhao2022pore}
Jianlin Zhao, Feifei Qin, Qinjun Kang, Dominique Derome, and Jan Carmeliet.
\newblock Pore-scale simulation of drying in porous media using a hybrid
  lattice boltzmann: pore network model.
\newblock \emph{Drying Technology}, 40\penalty0 (4):\penalty0 719--734, 2022.

\bibitem[Mukherjee et~al.(2011)Mukherjee, Kang, and Wang]{mukherjee2011pore}
Partha~P Mukherjee, Qinjun Kang, and Chao-Yang Wang.
\newblock Pore-scale modeling of two-phase transport in polymer electrolyte
  fuel cells—progress and perspective.
\newblock \emph{Energy \& Environmental Science}, 4\penalty0 (2):\penalty0
  346--369, 2011.

\bibitem[Molaeimanesh and Akbari(2014)]{molaeimanesh2014three}
GR~Molaeimanesh and MH~Akbari.
\newblock A three-dimensional pore-scale model of the cathode electrode in
  polymer-electrolyte membrane fuel cell by lattice boltzmann method.
\newblock \emph{Journal of Power Sources}, 258:\penalty0 89--97, 2014.

\bibitem[Zhu et~al.(2021)Zhu, Zhang, Xiao, Bazylak, Gao, and Sui]{zhu2021pore}
Lijun Zhu, Heng Zhang, Liusheng Xiao, Aimy Bazylak, Xin Gao, and Pang-Chieh
  Sui.
\newblock Pore-scale modeling of gas diffusion layers: Effects of compression
  on transport properties.
\newblock \emph{Journal of Power Sources}, 496:\penalty0 229822, 2021.

\bibitem[Fu et~al.(2022)Fu, Zhang, Zhu, Ye, Sui, Djilali, and Liao]{fu2022pore}
Ya-lu Fu, Biao Zhang, Xun Zhu, Ding-ding Ye, Pang-Chieh Sui, Ned Djilali, and
  Qiang Liao.
\newblock Pore-scale modeling of mass transport in the air-breathing cathode of
  membraneless microfluidic fuel cells.
\newblock \emph{International Journal of Heat and Mass Transfer}, 188:\penalty0
  122590, 2022.

\bibitem[Guo et~al.(2022)Guo, Chen, Zhang, Peng, and Tao]{guo2022pore}
Lingyi Guo, Li~Chen, Ruiyuan Zhang, Ming Peng, and Wen-Quan Tao.
\newblock Pore-scale simulation of two-phase flow and oxygen reactive transport
  in gas diffusion layer of proton exchange membrane fuel cells: Effects of
  nonuniform wettability and porosity.
\newblock \emph{Energy}, 253:\penalty0 124101, 2022.

\bibitem[Hoogland et~al.(2016{\natexlab{a}})Hoogland, Lehmann, Mokso, and
  Or]{hoogland2016drainage}
Frouke Hoogland, Peter Lehmann, Rajmund Mokso, and Dani Or.
\newblock Drainage mechanisms in porous media: From piston-like invasion to
  formation of corner flow networks.
\newblock \emph{Water Resources Research}, 52\penalty0 (11):\penalty0
  8413--8436, 2016{\natexlab{a}}.

\bibitem[Hoogland et~al.(2016{\natexlab{b}})Hoogland, Lehmann, and
  Or]{hoogland2016drainageb}
Frouke Hoogland, Peter Lehmann, and Dani Or.
\newblock Drainage dynamics controlled by corner flow: Application of the foam
  drainage equation.
\newblock \emph{Water Resources Research}, 52\penalty0 (11):\penalty0
  8402--8412, 2016{\natexlab{b}}.

\bibitem[Moura et~al.(2019)Moura, Flekk{\o}y, M{\aa}l{\o}y, Sch{\"a}fer, and
  Toussaint]{moura2019connectivity}
Marcel Moura, Eirik~Grude Flekk{\o}y, Knut~J{\o}rgen M{\aa}l{\o}y, Gerhard
  Sch{\"a}fer, and Renaud Toussaint.
\newblock Connectivity enhancement due to film flow in porous media.
\newblock \emph{Physical Review Fluids}, 4\penalty0 (9):\penalty0 094102, 2019.

\bibitem[Armstrong and Berg(2013)]{armstrong2013interfacial}
Ryan~T Armstrong and Steffen Berg.
\newblock Interfacial velocities and capillary pressure gradients during haines
  jumps.
\newblock \emph{Physical Review E}, 88\penalty0 (4):\penalty0 043010, 2013.

\bibitem[Moebius and Or(2014)]{moebius2014pore}
Franziska Moebius and Dani Or.
\newblock Pore scale dynamics underlying the motion of drainage fronts in
  porous media.
\newblock \emph{Water Resources Research}, 50\penalty0 (11):\penalty0
  8441--8457, 2014.

\bibitem[J{\o}rgen~M{\aa}l{\o}y et~al.(2021)J{\o}rgen~M{\aa}l{\o}y, Moura,
  Hansen, Grude~Flekk{\o}y, and Toussaint]{jorgen2021burst}
Knut J{\o}rgen~M{\aa}l{\o}y, Marcel Moura, Alex Hansen, Eirik Grude~Flekk{\o}y,
  and Renaud Toussaint.
\newblock Burst dynamics, up-scaling and dissipation of slow drainage in porous
  media.
\newblock \emph{arXiv e-prints}, pages arXiv--2110, 2021.

\bibitem[Mansouri-Boroujeni et~al.(2023)Mansouri-Boroujeni, Soulaine, Azaroual,
  and Roman]{mansouri2023interfacial}
Mahdi Mansouri-Boroujeni, Cyprien Soulaine, Mohamed Azaroual, and Sophie Roman.
\newblock How interfacial dynamics controls drainage pore-invasion patterns in
  porous media.
\newblock \emph{Advances in Water Resources}, 171:\penalty0 104353, 2023.

\bibitem[Zhou et~al.(1997)Zhou, Blunt, and Orr~Jr]{zhou1997hydrocarbon}
Dengen Zhou, Martin Blunt, and FM~Orr~Jr.
\newblock Hydrocarbon drainage along corners of noncircular capillaries.
\newblock \emph{Journal of colloid and interface science}, 187\penalty0
  (1):\penalty0 11--21, 1997.

\bibitem[Tuller and Or(2001)]{tuller2001hydraulic}
Markus Tuller and Dani Or.
\newblock Hydraulic conductivity of variably saturated porous media: Film and
  corner flow in angular pore space.
\newblock \emph{Water resources research}, 37\penalty0 (5):\penalty0
  1257--1276, 2001.

\bibitem[Flekk{\o}y et~al.(2002)Flekk{\o}y, Schmittbuhl, L{\o}vholt, Oxaal,
  M{\aa}l{\o}y, and Aagaard]{flekkoy2002flow}
Eirik~G Flekk{\o}y, Jean Schmittbuhl, Finn L{\o}vholt, Unni Oxaal,
  Knut~J{\o}rgen M{\aa}l{\o}y, and Per Aagaard.
\newblock Flow paths in wetting unsaturated flow: Experiments and simulations.
\newblock \emph{Physical review E}, 65\penalty0 (3):\penalty0 036312, 2002.

\bibitem[Ryazanov et~al.(2009)Ryazanov, Van~Dijke, and Sorbie]{ryazanov2009two}
Andrey~V Ryazanov, Marinus Izaak~Jan Van~Dijke, and Kenneth~Stuart Sorbie.
\newblock Two-phase pore-network modelling: existence of oil layers during
  water invasion.
\newblock \emph{Transport in Porous Media}, 80:\penalty0 79--99, 2009.

\bibitem[Han et~al.(2009)Han, Youssef, Rosenberg, Fleury, and
  Levitz]{han2009deviation}
M~Han, S~Youssef, E~Rosenberg, M~Fleury, and P~Levitz.
\newblock Deviation from archie’s law in partially saturated porous media:
  Wetting film versus disconnectedness of the conducting phase.
\newblock \emph{Physical Review E}, 79\penalty0 (3):\penalty0 031127, 2009.

\bibitem[Romano et~al.(2011)Romano, Chabert, Cuenca, and
  Bodiguel]{romano2011strong}
Marta Romano, Max Chabert, Amandine Cuenca, and Hugues Bodiguel.
\newblock Strong influence of geometrical heterogeneity on drainage in porous
  media.
\newblock \emph{Physical Review E}, 84\penalty0 (6):\penalty0 065302, 2011.

\bibitem[Xu et~al.(2014)Xu, Ok, Xiao, Neeves, and Yin]{xu2014effect}
Wei Xu, Jeong~Tae Ok, Feng Xiao, Keith~B Neeves, and Xiaolong Yin.
\newblock Effect of pore geometry and interfacial tension on water-oil
  displacement efficiency in oil-wet microfluidic porous media analogs.
\newblock \emph{Physics of Fluids}, 26\penalty0 (9):\penalty0 093102, 2014.

\bibitem[Vahid~Dastjerdi et~al.(2022)Vahid~Dastjerdi, Karadimitriou,
  Hassanizadeh, and Steeb]{vahid2022experimental}
Samaneh Vahid~Dastjerdi, Nikolaos Karadimitriou, S~Majid Hassanizadeh, and
  Holger Steeb.
\newblock Experimental evaluation of fluid connectivity in two-phase flow in
  porous media during drainage.
\newblock \emph{Water Resources Research}, 58\penalty0 (11):\penalty0
  e2022WR033451, 2022.

\bibitem[Herminghaus(2005)]{Herminghaus_2005}
S.~Herminghaus.
\newblock Dynamics of wet granular matter.
\newblock \emph{Advances in Physics}, 54\penalty0 (3):\penalty0 221--261, 2005.
\newblock \doi{10.1080/00018730500167855}.

\bibitem[Rieser et~al.(2015)Rieser, Arratia, Yodh, Gollub, and
  Durian]{Rieser_2015}
Jennifer~M. Rieser, P.~E. Arratia, A.~G. Yodh, J.~P. Gollub, and D.~J. Durian.
\newblock Tunable capillary-induced attraction between vertical cylinders.
\newblock \emph{Langmuir}, 31\penalty0 (8):\penalty0 2421--2429, 2015.
\newblock \doi{10.1021/la5046139}.
\newblock PMID: 25646573.

\bibitem[Cejas et~al.(2018)Cejas, Hough, Fr{\'e}tigny, and
  Dreyfus]{cejas2018effect}
Cesare~M Cejas, Lawrence~A Hough, Christian Fr{\'e}tigny, and R{\'e}mi Dreyfus.
\newblock Effect of geometry on the dewetting of granular chains by
  evaporation.
\newblock \emph{Soft Matter}, 14\penalty0 (34):\penalty0 6994--7002, 2018.

\bibitem[Chen et~al.(2017)Chen, Duru, Joseph, Geoffroy, and
  Prat]{chen2017control}
Chen Chen, Paul Duru, Pierre Joseph, Sandrine Geoffroy, and Marc Prat.
\newblock Control of evaporation by geometry in capillary structures. from
  confined pillar arrays in a gap radial gradient to phyllotaxy-inspired
  geometry.
\newblock \emph{Scientific reports}, 7\penalty0 (1):\penalty0 15110, 2017.

\bibitem[Chen et~al.(2018)Chen, Joseph, Geoffroy, Prat, and Duru]{chen_2018}
C.~Chen, P.~Joseph, S.~Geoffroy, M.~Prat, and P.~Duru.
\newblock Evaporation with the formation of chains of liquid bridges.
\newblock \emph{Journal of Fluid Mechanics}, 837:\penalty0 703–728, 2018.
\newblock \doi{10.1017/jfm.2017.827}.

\bibitem[Vorhauer et~al.(2015)Vorhauer, Wang, Kharaghani, Tsotsas, and
  Prat]{vorhauer2015drying}
Nicole Vorhauer, YJ~Wang, A~Kharaghani, Evangelos Tsotsas, and Marc Prat.
\newblock Drying with formation of capillary rings in a model porous medium.
\newblock \emph{Transport in Porous Media}, 110:\penalty0 197--223, 2015.

\bibitem[Kharaghani et~al.(2021)Kharaghani, Mahmood, Wang, and
  Tsotsas]{kharaghani2021three}
Abdolreza Kharaghani, Hafiz~Tariq Mahmood, Yujing Wang, and Evangelos Tsotsas.
\newblock Three-dimensional visualization and modeling of capillary liquid
  rings observed during drying of dense particle packings.
\newblock \emph{International Journal of Heat and Mass Transfer}, 177:\penalty0
  121505, 2021.

\bibitem[Wilkinson and Willemsen(1983)]{wilkinson1983invasion}
David Wilkinson and Jorge~F Willemsen.
\newblock Invasion percolation: a new form of percolation theory.
\newblock \emph{Journal of physics A: Mathematical and general}, 16\penalty0
  (14):\penalty0 3365, 1983.

\bibitem[Blunt(2001)]{blunt2001flow}
Martin~J Blunt.
\newblock Flow in porous media—pore-network models and multiphase flow.
\newblock \emph{Current opinion in colloid \& interface science}, 6\penalty0
  (3):\penalty0 197--207, 2001.

\bibitem[Zhao et~al.(2016)Zhao, MacMinn, and Juanes]{zhao2016wettability}
Benzhong Zhao, Christopher~W MacMinn, and Ruben Juanes.
\newblock Wettability control on multiphase flow in patterned microfluidics.
\newblock \emph{Proceedings of the National Academy of Sciences}, 113\penalty0
  (37):\penalty0 10251--10256, 2016.

\bibitem[M{\aa}l{\o}y et~al.(2021)M{\aa}l{\o}y, Moura, Hansen, Flekk{\o}y, and
  Toussaint]{maaloy2021burst}
Knut~J{\o}rgen M{\aa}l{\o}y, Marcel Moura, Alex Hansen, Eirik~Grude Flekk{\o}y,
  and Renaud Toussaint.
\newblock Burst dynamics, upscaling and dissipation of slow drainage in porous
  media.
\newblock \emph{Frontiers in Physics}, page 718, 2021.

\bibitem[Vincent-Dospital et~al.(2022)Vincent-Dospital, Moura, Toussaint, and
  M{\aa}l{\o}y]{vincent2022stable}
Tom Vincent-Dospital, Marcel Moura, Renaud Toussaint, and Knut~J{\o}rgen
  M{\aa}l{\o}y.
\newblock Stable and unstable capillary fingering in porous media with a
  gradient in grain size.
\newblock \emph{Communications Physics}, 5\penalty0 (1):\penalty0 306, 2022.

\bibitem[M{\'e}heust et~al.(2002)M{\'e}heust, L{\o}voll, M{\aa}l{\o}y, and
  Schmittbuhl]{meheust2002interface}
Yves M{\'e}heust, Grunde L{\o}voll, Knut~J{\o}rgen M{\aa}l{\o}y, and Jean
  Schmittbuhl.
\newblock Interface scaling in a two-dimensional porous medium under combined
  viscous, gravity, and capillary effects.
\newblock \emph{Physical Review E}, 66\penalty0 (5):\penalty0 051603, 2002.

\bibitem[L{\o}voll et~al.(2005)L{\o}voll, M{\'e}heust, M{\aa}l{\o}y, Aker, and
  Schmittbuhl]{lovoll2005competition}
Grunde L{\o}voll, Yves M{\'e}heust, Knut~J{\o}rgen M{\aa}l{\o}y, Eyvind Aker,
  and Jean Schmittbuhl.
\newblock Competition of gravity, capillary and viscous forces during drainage
  in a two-dimensional porous medium, a pore scale study.
\newblock \emph{Energy}, 30\penalty0 (6):\penalty0 861--872, 2005.

\bibitem[Moura et~al.(2015)Moura, Fiorentino, M{\aa}l{\o}y, Sch{\"a}fer, and
  Toussaint]{moura2015impact}
M~Moura, E-A Fiorentino, Knut~J{\o}rgen M{\aa}l{\o}y, Gerhard Sch{\"a}fer, and
  Renaud Toussaint.
\newblock Impact of sample geometry on the measurement of pressure-saturation
  curves: Experiments and simulations.
\newblock \emph{Water Resources Research}, 51\penalty0 (11):\penalty0
  8900--8926, 2015.

\bibitem[Primkulov et~al.(2018)Primkulov, Talman, Khaleghi, Shokri,
  Chalaturnyk, Zhao, MacMinn, and Juanes]{primkulov2018quasistatic}
Bauyrzhan~K Primkulov, Stephen Talman, Keivan Khaleghi, Alireza~Rangriz Shokri,
  Rick Chalaturnyk, Benzhong Zhao, Christopher~W MacMinn, and Ruben Juanes.
\newblock Quasistatic fluid-fluid displacement in porous media:
  Invasion-percolation through a wetting transition.
\newblock \emph{Physical Review Fluids}, 3\penalty0 (10):\penalty0 104001,
  2018.

\bibitem[Primkulov et~al.(2021)Primkulov, Pahlavan, Fu, Zhao, MacMinn, and
  Juanes]{primkulov2021wettability}
Bauyrzhan~K Primkulov, Amir~A Pahlavan, Xiaojing Fu, Benzhong Zhao,
  Christopher~W MacMinn, and Ruben Juanes.
\newblock Wettability and lenormand's diagram.
\newblock \emph{Journal of Fluid Mechanics}, 923:\penalty0 A34, 2021.

\bibitem[Primkulov et~al.(2022)Primkulov, Zhao, MacMinn, and
  Juanes]{primkulov2022avalanches}
Bauyrzhan~K Primkulov, Benzhong Zhao, Christopher~W MacMinn, and Ruben Juanes.
\newblock Avalanches in strong imbibition.
\newblock \emph{Communications Physics}, 5\penalty0 (1):\penalty0 52, 2022.

\bibitem[Brakke(1992)]{brakke1992surface}
Kenneth~A Brakke.
\newblock The surface evolver.
\newblock \emph{Experimental mathematics}, 1\penalty0 (2):\penalty0 141--165,
  1992.

\bibitem[Joekar-Niasar and Hassanizadeh(2012)]{joekar2012analysis}
V~Joekar-Niasar and SM~Hassanizadeh.
\newblock Analysis of fundamentals of two-phase flow in porous media using
  dynamic pore-network models: A review.
\newblock \emph{Critical reviews in environmental science and technology},
  42\penalty0 (18):\penalty0 1895--1976, 2012.

\end{thebibliography}
\end{document}